\documentclass[a4paper]{article}
\pdfoutput=1

\usepackage[utf8]{inputenc}
\usepackage{amsmath}
\usepackage{jheppub}
\usepackage{amsfonts}
\usepackage{amssymb}
\usepackage{mathtools}
\usepackage{graphicx}
\usepackage{relsize}
\usepackage{pdfpages}
\usepackage{subfig}
\usepackage{multicol}
\usepackage{subfig}
\usepackage{pdfpages}
\usepackage{float}

\newcommand{\D}[1]{(\mathbf{1}_{#1},0)}
\newcommand{\ta}[1]{\mathbf{T}^{a}_{#1}} 
\newcommand{\tb}[1]{\mathbf{T}^{b}_{#1}}
\newcommand{\tc}[1]{\mathbf{T}^{c}_{#1}}
\newcommand{\td}[1]{\mathbf{T}^{d}_{#1}}
\newcommand{\tf}[1]{\mathbf{T}^{f}_{#1}} 
\newcommand{\tg}[1]{\mathbf{T}^{g}_{#1}}
\newcommand{\te}[1]{\mathbf{T}^{e}_{#1}}
\newcommand{\tj}[1]{\mathbf{T}^{j}_{#1}}
\newcommand{\thh}[1]{\mathbf{T}^{h}_{#1}}
\newcommand{\tkk}[1]{\mathbf{T}^{k}_{#1}}
\newcommand{\tm}[1]{\mathbf{T}^{m}_{#1}}
\newcommand{\tl}[1]{\mathbf{T}^{l}_{#1}}
\newcommand{\tn}[1]{\mathbf{T}^{n}_{#1}}
\newcommand{\tv}[1]{\mathbf{T}^{v}_{#1}}
\newcommand{\tu}[1]{\mathbf{T}^{u}_{#1}}
\newcommand{\tp}[1]{\mathbf{T}^{p}_{#1}}
\newcommand{\tq}[1]{\mathbf{T}^{q}_{#1}}
\newcommand{\fabc}{f^{abc}}
\newcommand{\fabg}{f^{abg}}
\newcommand{\fcej}{f^{cej}}
\newcommand{\fedh}{f^{edh}}
\newcommand{\fcdg}{f^{cdg}}
\newcommand{\faeh}{f^{aeh}}
\newcommand{\febh}{f^{ebh}}
\newcommand{\fdef}{f^{def}}
\newcommand{\ffaj}{f^{faj}}
\newcommand{\faeg}{f^{aeg}}
\newcommand{\fdag}{f^{dag}}
\newcommand{\fbdh}{f^{bdh}}
\newcommand{\fdej}{f^{dej}}
\newcommand{\fcgh}{f^{cgh}}
\newcommand{\fdeg}{f^{deg}}
\newcommand{\fdjk}{f^{djk}}
\newcommand{\fbdk}{f^{bdk}}
\newcommand{\fgal}{f^{gal}}
\newcommand{\fhdm}{f^{hdm}}
\newcommand{\fceg}{f^{ceg}}
\newcommand{\fdbj}{f^{dbj}}
\newcommand{\fadh}{f^{adh}}
\newcommand{\fedk}{f^{edk}}
\newcommand{\fach}{f^{ach}}
\newcommand{\fdcj}{f^{dcj}}
\newcommand{\fdbk}{f^{dbk}}
\newcommand{\fbgk}{f^{bgk}}
\newcommand{\fbem}{f^{bem}}
\newcommand{\fagl}{f^{agl}}
\newcommand{\fhen}{f^{hen}}
\newcommand{\fdah}{f^{dah}}
\newcommand{\fmdv}{f^{mdv}}
\newcommand{\faej}{f^{aej}}
\newcommand{\fbdu}{f^{bdu}}
\newcommand{\fbcm}{f^{bcm}}
\newcommand{\fman}{f^{man}}
\newcommand{\facp}{f^{acp}}
\newcommand{\fpbq}{f^{pbq}}
\newcommand{\feah}{f^{eah}}
\newcommand{\fgah}{f^{gah}}
\newcommand{\fdjm}{f^{djm}}
\newcommand{\fbeg}{f^{beg}}
\newcommand{\fahj}{f^{ahj}}
\newcommand{\fdch}{f^{dch}}
\newcommand{\fbag}{f^{bag}}
\newcommand{\fbdg}{f^{bdg}}
\newcommand{\fabe}{f^{abe}}
\newcommand{\fbgh}{f^{bgh}}
\newcommand{\fahk}{f^{ahk}}
\newcommand{\facn}{f^{acn}}
\newcommand{\fbdm}{f^{bdm}}
\newcommand{\fmcu}{f^{mcu}}
\newcommand{\fauv}{f^{auv}}
\def\bra#1{%
  \left\langle\smash{#1}{\vphantom1}\right|}
\def\ket#1{%
  \left|\smash{#1}{\vphantom1}\right\rangle}
\def\eq#1{Eq.~(\ref{#1})}
\newcommand{\secn}[1]{Section~\ref{#1}}
\newcommand{\be}{\begin{equation}}
\newcommand{\ee}{\end{equation}}
\newcommand{\beq}{\begin{eqnarray}}
\newcommand{\eeq}{\end{eqnarray}}
\newcommand{\bea}{\begin{eqnarray}}
\newcommand{\eea}{\end{eqnarray}}

\newcommand{\e}{\epsilon}

\newcommand{\as}{\alpha_s}

\title{Multiparton webs beyond three loops} 

\author[a]{Neelima Agarwal,}
\author[b]{Abhinava Danish,}
\author[c]{Lorenzo Magnea,}
\author[b]{Sourav Pal,}
\author[b]{Anurag Tripathi}

\affiliation[a]{Department of Physics, Chaitanya Bharathi Institute of Technology, \\ Gandipet, Hyderabad, Telangana State 500075, India}
\affiliation[b]{Department of Physics, Indian Institute of Technology Hyderabad, \\ Kandi, Sangareddy, Telangana State 502285, India}
\affiliation[c]{Dipartimento di Fisica and Arnold-Regge Center, Universit\`a di Torino, 
\\ and INFN, Sezione di Torino, Via Pietro Giuria 1, I-10125 Torino, Italy}

\emailAdd{neelimaagarwal$\_$physics@cbit.ac.in}
\emailAdd{abhinavadan@gmail.com}
\emailAdd{lorenzo.magnea@unito.it}
\emailAdd{spalexam@gmail.com}
\emailAdd{tripathi@iith.ac.in}

\abstract{Correlators of Wilson-line operators are fundamental ingredients for 
the study of the infrared properties of non-abelian gauge theories. In perturbation 
theory, they are known to exponentiate, and their logarithm can be organised 
in terms of collections of Feynman diagrams called {\it webs}. We study the 
classification of webs to high perturbative orders, proposing a set of tools to 
generate them recursively: in particular, we introduce the concept of {\it Cweb},
or {\it correlator web}, which is a set of skeleton diagrams built with connected 
gluon correlators, instead of individual Feynman diagrams. As an application, 
we enumerate all Cwebs entering the soft anomalous dimension matrix for multi-parton 
scattering amplitudes at four loops, and we compute the {\it mixing matrices} 
for all Cwebs connecting four or five Wilson lines at that loop order, verifying that 
they obey sum rules that were derived or conjectured in the literature. Our results 
provide the colour building blocks for the calculation of the soft anomalous dimension 
matrix at four-loop order.}

\begin{document}
\maketitle
  

\section{Introduction}
\label{Intro}

Studies of the structure of infrared (IR) singularities that appear in scattering 
amplitudes in gauge field theories have a long and rich history, and have 
led to remarkable all-order insights into the organisation of the perturbative 
expansion~\cite{Bloch:1937pw,Sudakov:1954sw,Yennie:1961ad,Kinoshita:1962ur,
Lee:1964is,Grammer:1973db,Mueller:1979ih,Collins:1980ih,Sen:1981sd,
Sen:1982bt,Korchemsky:1987wg,Korchemsky:1988hd,Magnea:1990zb,
Dixon:2008gr,Gardi:2009qi,Becher:2009qa}. In the computation of loop 
corrections to scattering amplitudes, IR singularities arise when a virtual 
particle flowing in a loop becomes soft or collinear to one of the external 
particles. Upon constructing, from amplitudes, a well-defined physical 
observable, these singularities are cancelled by the contribution of real 
emission diagrams, which  must be integrated over the phase space for 
undetected real radiation. The singularities, however, often leave their imprints 
on the perturbative expansion, in the form of potentially large logarithms of 
kinematic variables, which may need to be resummed in order to obtain precise 
predictions. Such resummations\footnote{For an introduction to the relevant 
techniques, see, for example,~\cite{Sterman:1995fz}.} are made possible by the 
universal nature of infrared radiation, which results in factorised expressions for 
scattering amplitudes, where soft and collinear effects are organised by universal 
functions, depending only on the quantum numbers of the external particles 
involved in the scattering, but not on the specific nature of the hard process 
being considered~\cite{Sterman:1995fz,Dixon:2008gr,Feige:2014wja}. These 
soft and collinear functions, in turn, can be expressed as matrix element of 
field operators and Wilson lines, which are the object of the present study. 
We note that such matrix elements play a ubiquitous role, not only for the 
factorisation of scattering amplitudes, but also in many effective theories based 
on QCD~\cite{Manohar:2000dt,Brambilla:2004jw, Becher:2014oda}; furthermore, 
we emphasise that a detailed knowledge of infrared singularities is also important 
for collider phenomenology at finite orders: indeed, the cancellation of singularities 
between squared matrix elements with different numbers of external particles is 
difficult to implement at high orders for complicated collider observables, which 
must be evaluated numerically. In practice, the cancellation must be performed 
analytically, and the construction of general and efficient subtraction procedures 
beyond next-to-leading order (NLO) is an ongoing effort  (see, for example, 
\cite{GehrmannDeRidder:2005cm,Somogyi:2005xz,Catani:2007vq,
Czakon:2010td,Boughezal:2015dva,Sborlini:2016hat,Caola:2017dug,
Herzog:2018ily,Magnea:2018hab,Magnea:2018ebr}).

The methods developed in this paper concern the evaluation of Wilson-line
correlators of the general form
\beq
  {\cal S}_n \left( \gamma_i \right) \, \equiv \, \bra{0} \prod_{k = 1}^n
  \Phi_{r_k} \left(  \gamma_k \right) \ket{0} \, ,
\label{genWLC}
\eeq
where $\Phi_r ( \gamma )$  is a Wilson-line operator evaluated on a 
(smooth) space-time contour $\gamma$, defined by
\beq
  \Phi_r \left(  \gamma \right) \, \equiv \, P \exp \left[ {\rm i} g \!
  \int_\gamma d x \cdot {\bf A}_r (x) \right] \, ,
\label{genWL}
\eeq
where ${\bf A}^\mu_r (x) = A^\mu_a (x) \, {\bf T}_r^a$ is a non-abelian gauge field, 
with ${\bf T}_r^a$ the generators of the representation $r$ of the gauge group. 
The smooth contours $\gamma_k$ can be closed (in which case the correlator
is gauge-invariant), or open: in this case, the correlator is a colour tensor
with open colour indices in representations $r_k$ attached to the ends of each 
Wilson line. While most of our considerations will apply in general to any 
correlator of the form~(\ref{genWLC}), we will be especially interested in the
soft colour operator associated with multi-particle scattering amplitudes in 
gauge theories, which encodes all their soft singularities. This soft operator is of
the form of \eq{genWLC}, with the contours $\gamma_k$ given by semi-infinite 
straight lines extending from the origin along directions $\beta_k$, corresponding
to the four-velocities of the particles participating in the scattering. In this case
we write more explicitly
\beq
  {\cal S}_n \Big( \beta_i \cdot \beta_j, \as (\mu^2), \e \Big) \, \equiv \, 
  \bra{0} \prod_{k = 1}^n \Phi_{\beta_k} \left( \infty, 0 \right) \ket{0} , \quad
  \Phi_\beta \left( \infty, 0 \right) \, \equiv \, P \exp \left[ {\rm i} g \!
  \int_0^\infty d \lambda \, \beta \cdot {\bf A} (\lambda \beta) \right] ,
\label{softWLC}
\eeq
where $\alpha_s = g^2/(4 \pi)$, for simplicity we did not display the representations
to which the Wilson lines belong, and we introduced the dimensional regularisation 
parameter $\e$, setting the space-time dimension to $d = 4 - 2 \e$.

Soft operators of the form (\ref{softWLC}) are highly singular, being affected
by ultraviolet, soft, and, in case $\beta_i^2 = 0$, collinear divergences: as a 
consequence, special care is required to evaluate them~\cite{Mitov:2010xw,
Henn:2013wfa,Gardi:2013saa,Falcioni:2014pka}. In the massless case, this can 
be done by introducing auxiliary regulators: for collinear divergences, one may 
set $\beta_i^2 \neq 0$, for soft divergences one may for example introduce a 
smooth exponential long-distance cutoff on gluon interactions, as discussed 
in Refs.~\cite{Gardi:2013saa,Falcioni:2014pka}, while retaining dimensional 
regularisation for ultraviolet singularities. The bare correlator can then be 
evaluated and renormalised, yielding the desired answer.

Wilson-line correlators of the form of \eq{genWLC} have the following  
basic properties.
\begin{itemize}
\item After renormalisation, $n$-line correlators  obey renormalisation group 
equations which lead, in  dimensional regularisation, to exact exponentiation 
in terms of a soft anomalous dimension matrix $\Gamma_n$. For straight-line 
correlators, of the form  of \eq{softWLC}, one may write
\beq
  \mathcal{S}_n \Big( \beta_i \cdot \beta_j, \as (\mu^2), \e \Big) \, = \, 
  \mathcal{P} \exp \left[ - \frac{1}{2} \int_{0}^{\mu^2} \frac{d \lambda^2}  
  {\lambda^2} \, {\bf \Gamma}_n \Big( \beta_i \cdot \beta_j, \alpha_s (\lambda^2), 
  \e \Big) \right]  \, .
\label{softmatr}
\eeq
It is important to note that, in the massless case, the soft anomalous dimension
${\bf \Gamma}_n$ is affected by collinear singularities, which must be properly 
organised in terms of appropriate jet functions~\cite{Dixon:2008gr}; collinear-finite 
contributions can be computed in the massless case by considering Wilson
lines slightly tilted off the light cone, and taking the light-cone limit in the 
intermediate  stages of the calculation~\cite{Almelid:2015jia}. The soft  
anomalous dimension matrix ${\bf \Gamma}_n$ is a central quantity for the 
study of perturbative non-abelian gauge theories, and has been the focus 
of much theoretical work. It was computed at one loop in~\cite{Kidonakis-1998} 
(see also~\cite{Korchemskaya:1994qp}); at two loops in the massless 
case in~\cite{Aybat:2006wq,Aybat:2006mz}, and in  the massive case 
in~\cite{Mitov:2009sv,Ferroglia:2009ep,Ferroglia:2009ii,Kidonakis:2009ev,
Chien:2011wz}; finally, at three loops in the massless case in~\cite{Almelid:2015jia,
Almelid:2017qju}.
\item General Wilson-line correlators of the form of \eq{genWLC} obey a
non-trivial form of {\it diagrammatic exponentiation}, so that one can write
\beq
  {\cal S}_n \left( \gamma_i \right) \, = \, \exp \Big[ {\cal W}_n \left( \gamma_i \right) 
  \Big]  \, ,
\label{diaxp}
\eeq
where the logarithm of  the correlator, ${\cal W}_n \left( \gamma_i \right)$,
can be directly computed in terms of a subset of the  Feynman diagrams
contributing to ${\cal S}_n (\gamma_i)$. For non-abelian gauge theories, this
was first pointed out in Refs.~\cite{Sterman-1981,Gatheral,Frenkel-1984}, for 
the case of two straight, semi-infinite Wilson lines. For general  configurations,
it was proven in Refs.~\cite{Mitov:2010rp,Gardi:2010rn}. Feynman diagrams
contributing to ${\cal W}_n \left( \gamma_i \right)$  are  collectively called 
{\it webs}. For an abelian theory, webs are connected diagrams; for a 
non-abelian theory, if only two Wilson lines are present, webs are two-eikonal
irreducible diagrams, {\it i.e.} diagrams that do  not become disconnected upon 
cutting only the two Wilson lines; for general, multi-line correlators, webs are 
sets of diagrams that differ among themselves by the ordering of their gluon
attachments to the Wilson lines. The properties of webs will be further discussed
in \secn{Webs}, and a useful generalisation of the concept of web will be 
proposed in \secn{Cwebs}. Clearly, by means of webs, one can directly compute
the soft  anomalous dimension matrix ${\bf \Gamma}_n$.
\item For the case when the contours $\gamma_i$ are semi-infinite, or infinite
straight lines, all loop corrections to the bare correlator, in  the absence of auxiliary 
IR regulators, vanish in dimensional regularisation, since they are given by 
scale-less integrals. Bare correlators are then exactly given by the unit matrix 
in the tensor product of the colour representations of the $n$ Wilson lines. 
The renormalised correlator is therefore a `pure counterterm': in order to 
compute it, one must first construct an IR-regulated version of the bare 
correlator, whose  loop corrections will not vanish, but will in general be 
regulator-dependent; one proceeds then to renormalise the regulated 
correlator, extracting the relevant UV counterterms; the set of these 
counterterms constitutes the renormalized version of the original correlator.
It is important to note that, for general multi-line correlators, multiplicative
renormalisability and  exponentiation combine non-trivially, due to the
non abelian nature of webs, and the calculation of renormalised correlators
includes commutators of counterterms and bare webs, as required~\cite{Gardi:2011yz}.
\item When the countours $\gamma_i$ are light-like, straight, semi-infinite 
Wilson lines, scale invariance imposes strong constraints on the functional
dependence of the soft anomalous dimension matrix ${\bf \Gamma}_n$. Up to 
two loops, ${\bf \Gamma}_n$ can only involve dipole correlations between Wilson
lines \cite{Gardi:2009qi,Becher:2009cu,Becher:2009qa,Gardi-Magnea}; beyond 
two loops, only quadrupole correlations can arise, which must depend on
scale-invariant conformal cross ratios of the form $\rho_{ijkl} \equiv (\beta_i 
\cdot  \beta_j  \beta_k \cdot  \beta_l)/(\beta_i \cdot  \beta_k  \beta_j \cdot  \beta_l)$:
the first such correlations arise at three loops, with at least four Wilson lines,
and were computed in Ref.~\cite{Almelid:2015jia}; further correlations may 
arise only in association with higher-order Casimir operators, which may 
start contributing only at four loops, as discussed in \secn{Colstru}.
\end{itemize}
In this paper, we study the properties of the logarithms of Wilson-line correlators,
${\cal W}_n (\gamma_i)$, with emphasis on their colour structures, extending
earlier studies  to higher orders in perturbation theory. After reviewing existing  
results on diagrammatic exponentiation in \secn{Webs}, in \secn{Cwebs} we 
propose a natural extension of the concept of web, which we believe will prove 
useful for classification purposes and high-order studies. Subsequently, in 
\secn{Repli}, we discuss how our definition leads to a simple recursive method 
to generate webs at $(p+1)$ loops from those arising at $p$ loops, and we discuss
how this recursion can be implemented  in a code based on the `replica trick'
used in Ref.~\cite{Gardi:2010rn}. In \secn{Fourwe}, we apply our method to 
construct the {\it web mixing matrices} at four loops for all webs connecting
four or five Wilson lines, verifying that these matrices satisfy the expected
properties, including conjectures that were proposed at lower orders; furthermore, 
we briefly discuss the colour structures that arise at four loops, verifying the 
compatibility of our results with the discussion in Refs.~\cite{Ahrens:2012qz,
Becher:2019avh}. We conclude, in \secn{Conclu}, with an outlook on possible 
future developments, while an Appendix lists in detail the web mixing matrices 
for all the webs  discussed in the main text.


\section{Diagrammatic exponentiation for multiple Wilson-line correlators}
\label{Webs}

\begin{figure}[H]
	\centering
	\vspace{-3mm}
	\subfloat{\includegraphics[height=4cm,width=4cm]{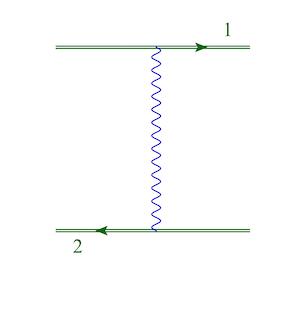} }
	\qquad \qquad \qquad
	\subfloat{\includegraphics[height=4cm,width=4cm]{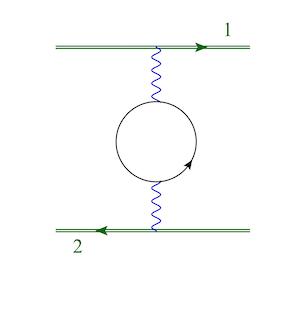} }
	\vspace{-4mm}
	\caption{Two low-order examples of webs with two Wilson lines in an abelian 
	gauge theory.}
	\label{fig:abelian-webs}
\end{figure}

\begin{figure}[H]
	\centering
	\vspace{-3mm}
	\subfloat{\includegraphics[height=4cm,width=4cm]{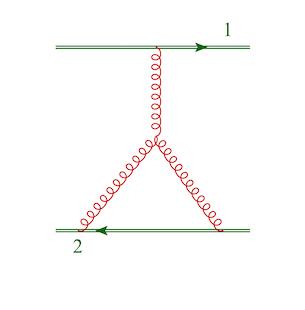} }
	\qquad
	\subfloat{\includegraphics[height=4cm,width=4cm]{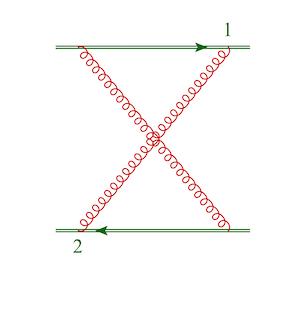} }
	\qquad
	\subfloat{\includegraphics[height=4cm,width=4cm]{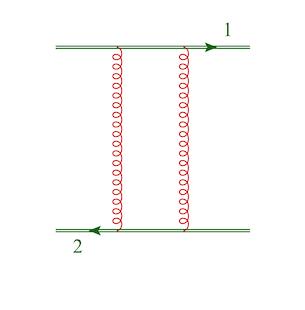} }
	\vspace{-5mm}
	\caption{The first two diagrams contribute to ${\cal W}_2$, while the third 
	diagram, which becomes disconnected by cutting the two Wilson lines, 
	does not.}
	\label{fig:non-abelian-webs}
\end{figure}

\begin{figure}[H]
	\centering
	\vspace{-3mm}
	\subfloat{\includegraphics[height=4cm,width=4cm]{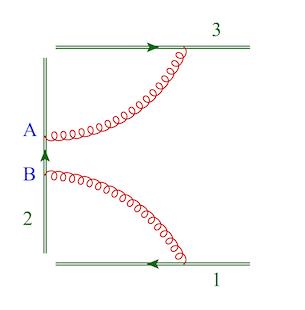} }
	\qquad \qquad \qquad
	\subfloat{\includegraphics[height=4cm,width=4cm]{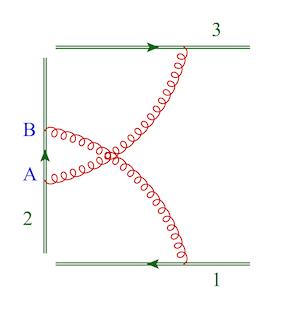} }
	\vspace{-3mm}
	\caption{A simple two-loop, three-line web involving two Feynman diagrams. Note
	that Wilson lines are oriented, and labelled by integers in green; multiple gluon 
	attachments to a given line (whose permutations generate the web) are labelled 
	by capital letters in blue.}
	\label{threelegweb}
\end{figure}

Diagrammatic exponentiation in the eikonal approximation was first observed
in QED~\cite{Grammer:1973db}, where ${\cal W}_n (\gamma_i)$ is given  by
the sum of all connected photon subdiagrams (obtained by removing the 
Wilson lines form the original diagrams): two examples are shown in Figure
\ref{fig:abelian-webs}. For a non-abelian theory, the presence of non-commuting  
colour factors associated with gluon attachments to the Wilson lines invalidates
the simple QED exponentiation: it remains nonetheless true that the logarithms
of Wilson-line correlators have a direct diagrammatic interpretation. For the case 
of two Wilson lines~\cite{Sterman-1981,Gatheral,Frenkel-1984}, ${\cal W}_2$ 
is constructed from the set of diagrams which remain connected after cutting  
the two Wilson lines: these two-eikonal-irreducible diagrams are called {\it webs} 
(see Fig.~\ref{fig:non-abelian-webs} for examples). It is important to note that, 
even in the simple case of two Wilson lines in a colour-singlet configuration, 
webs appear in ${\cal W}_2$ with modified colour factors: more precisely, 
the only colour factors appearing in ${\cal W}_2$ are those that correspond 
to connected gluon subdiagrams, such  as the first  one portrayed in 
Fig.~\ref{fig:non-abelian-webs}. This provides an interesting generalisation 
of the abelian exponentiation, and points to further extensions to the 
multi-Wilson-line case.
\noindent The problem of constructing the colour operator ${\cal W}_n (\gamma_i)$  
for the general case of $n$ Wilson lines was solved in the remarkable series of 
papers~\cite{Mitov:2010rp,Gardi:2010rn,Gardi:2011wa,Gardi:2011yz,Dukes:2013wa,
Gardi:2013ita,Dukes:2013gea,Dukes:2016ger}, while an interesting alternative 
approach was developed in Refs.~\cite{Vladimirov:2014wga,Vladimirov:2015fea}.  
Here we briefly summarise the main results, with special attention to the colour 
structures, which will be the main focus of our paper.
\noindent Let $D$ be a Feynman diagram contributing to the correlator ${\cal  S}_n$. 
Each such diagram can be written as the product of a kinematic factor ${\cal  K}(D)$, 
depending on  the  four velocities $\beta_i$ (or more generally on the contours  
$\gamma_i$ in the case of non-straight Wilson lines) and a colour factor $C(D)$. 
In full generality, the logarithm of the correlator, ${\cal W}_n$, can be written as 
linear combination of the same diagrams, with modified colour factors. We write
\beq
  {\cal W}_n (\gamma_i)  \, = \, \sum_D {\cal K} (D)  \, \widetilde{C}(D) \, ,
\label{eq:webeq}
\eeq
where $\widetilde{C}(D)$ is referred to as \textit{Exponentiated Color Factor} 
(ECF) for diagram $D$. The crucial point in \eq{eq:webeq} is of course that
a large  number of diagrams have vanishing ECFs, and therefore do not
contribute to the logarithm of the correlator: for example, for $n = 2$, one 
can show that all two-eikonal reducible diagrams have $\widetilde{C}(D) 
= 0$. In general, ECFs are linear combinations of the ordinary color factors
of sets of  diagrams that differ only by the order of their gluon attachments
to the Wilson lines. This naturally leads to the general  definition of a {\it web}
not as a single diagram, but as a set of diagrams that can be obtained from 
any representative element  by permuting the gluon attachments to the Wilson 
lines:  a simple example of a two-loop, three-line web involving two diagrams is
presented in Fig.~\ref{threelegweb}.

\noindent As a consequence of these considerations, the sum in \eq{eq:webeq} is
naturally organised as a sum over webs, and for each diagram belonging
to a given web $w$ the ECF is expressed as 
\beq
	\widetilde{C} (D)  \, = \, \sum_{D' \in w} R_w (D, D') \, C(D') \, ,
\label{eq:ecf}
\eeq
where the sum extends to the diagrams belonging to web $w$, and $R_w 
(D, D')$ is called {\it web mixing matrix}. Combining \eq{eq:webeq} and 
\eq{eq:ecf} one can express the Wilson-line correlator as
\beq
  {\cal S}_n \, = \, \exp \left[ \sum_{D,D'} {\cal K} (D) R (D, D') C(D')  \right] \, ,
\label{Snwebs}
\eeq
where the sum extends to all diagrams appearing in the correlator, order by 
order in perturbation theory, and the matrix $R$ is block-diagonal, with blocks
corresponding to individual webs. In turn, each web $w$ can be written as
\beq
  w \, = \, \sum_{D \in  w} {\cal  K} (D) \, \widetilde{C} (D) \, = \,
  \sum_{D,D' \in  w} {\cal K} (D) \, R_w (D, D') \, C (D')  \, .
\label{eq:webredef}
\eeq
In this language, ${\cal W}_n = \sum w_n$, where the sum extends to all
webs arising in the presence of $n$ Wilson lines, order by order in perturbation 
theory.

Clearly, web mixing matrices are crucial quantities for the purpose of computing
Wilson-line correlators, and therefore the soft anomalous dimension matrix.  
Their properties were extensively studied in Refs.~\cite{Mitov:2010rp,Gardi:2010rn,
Gardi:2011wa,Gardi:2011yz,Dukes:2013wa,Gardi:2013ita,Dukes:2013gea}, and
can be summarised as follows.
\begin{itemize}
\item Web mixing matrices are idempotent, {\it i.e.} $\forall w, \,\, R_w^2 =  R_w$, 
as a consequence of which their eigenvalues can only be either 1 or 0.
\item Operating on the vector of color factors for the diagrams of web $w$, the  
web mixing matrix $R_w$ acts as a projection operator, selecting a subset of 
the possible colour factors, in number equal to its rank.
\item To all orders in perturbation theory, it can be shown~\cite{Gardi:2013ita}  
that  the set of colour factors surviving the projection is the set of colour factors 
of connected gluon diagrams: this is the general form of the non-abelian exponentiation
theorem.
\item Acting on the vector of kinematic factors for the diagrams of web $w$, in
the presence of an infrared regulator, the web mixing matrix  $R_w$ projects onto 
kinematic factors that do not contain ultraviolet sub-divergences. This is crucial, 
in order to be  able to isolate UV counterterms with no dependence on the infrared 
regulator.
\item The elements of web mixing matrices obey the row sum rule $\sum_{D'} 
R_w (D, D') = 0$, implying that at least one of  the eigenvalues of  $R_w$ must 
vanish.
\end{itemize} 
In addition to these properties, which were proved in Refs.~\cite{Gardi:2011wa,
Gardi:2011yz,Gardi:2013ita}, in the case of semi-infinite Wilson lines radiating 
out of a common origin, the matrix elements of web mixing matrices are
also conjectured to obey a column sum rule which can be stated as follows.
Given a diagram $D$, consider the set of its connected subdiagrams (after  
removing the Wilson lines), $\{ D_{\rm c}^i \subset D  \}$; we say that a 
connected subdiagram $D_{\rm c}^i$ can be shrunk to the common origin 
of the Wilson lines if all the vertices connecting the subdiagram to the Wilson 
lines can be moved to the origin without encountering vertices associated
with other subdiagrams. 
 For a given diagram $D$, we define the {\it column weight} of diagram 
 $D$, $s(D)$, as the number of different ways in which the connected subdiagrams 
 $D_{\rm c}^i$ can be {\it sequentially} shrunk, so that all gluon attachments to 
 Wilson lines in $D$ are moved to the common origin. This means, in practice, 
 that if all gluon attachments are entangled, so that no subdiagram can be shrunk 
 without shrinking the whole diagram, then $s(D) = 0$. On the other hand, if, for 
 example, a single subdiagram $D_{\rm c}^i$ can be shrunk without affecting the 
 others, this provides a non-trivial sequence for the the shrinking of the whole 
 diagram, so that $s(D) = 1$: this is the situation for the two diagrams portrayed 
 in Fig.~\ref{threelegweb}. With this definition, it is conjectured that \cite{Gardi:2011yz}.
\begin{itemize}
\item The elements of web mixing matrices obey the column sum rule
$\, \sum_{D} \, s(D) \, R_w (D, D')  = 0$.
\end{itemize} 
In what follows, row and column sum rules will be illustrated in a number of 
examples, and the conjectured column sum rule will be verified to hold for
all four-loop webs connecting four or five Wilson lines.

We conclude this section by noting that the characterisation of web mixing 
matrices as projectors can be made explicit by introducing, for each web,
a diagonalising matrix $Y_w$, such that
\beq
  Y_w R_w Y_w^{-1} \, = \, {\rm diag}  \left( \lambda_1, ... , \lambda_{p_w} \right)  
  \, = \, {\bf 1}_{r_w} \oplus {\bf  0}_{p_w - r_w} \,  ,
\label{Yw}
\eeq
where $p_w$ is the number of diagrams for web $w$, and thus the dimension
of the matrix $R_w$, while $r_w < p_w$ is the rank  of $R_w$.  Without loss of 
generality, we have ordered the eigenvalues of $R_w$, labelled by $\lambda_i$, 
so that null eigenvalues appear in the last positions. With these conventions, 
\eq{eq:webredef} can be rewritten in matrix notation as
\beq 
\label{eq:Web}
  w & = &  \left( {\cal K}^T Y_w^{-1} \right) Y_w R_w Y_w^{-1}  
  \left( Y_w C \right) \nonumber \\ 
  & = &  \sum_{h =  1}^{r_w} \left(  {\cal K}^T Y_w^{-1}  \right)_h 
  \left( Y_w C \right)_h  \, ,
\eeq
where ${\cal K}$ is the vector of kinematic factors and $C$ is the vector of 
colour factors for web $w$. It is clear that only $r_w$ ECFs will contribute
to web $w$:  the non-abelian exponentiation theorem tells us that they will
be colour factors which, by the Feynman rules, would be associated to 
connected gluon subdiagrams.


\section{From gluon webs to correlator webs}
\label{Cwebs}

As  discussed in \secn{Webs}, webs for multi-parton scattering amplitudes are
defined as sets of diagrams connecting gluons to Wilson lines, containing diagrams 
which are related to one another by permutations of gluon attachments to the
Wilson lines. In this section, we present a definition for a closely related structure,
where fixed-order Feynman diagrams are replaced by `skeleton' diagrams, which
are built out of connected gluon correlators, instead of gluon propagators and 
vertices. The basic reason to introduce these structures is that they strongly
simplify the counting and organisation of contributions to the logarithm of
Wilson-line correlators, especially at high orders, where radiative corrections
to gluon subdiagrams become important and proliferate; furthermore, as we 
will see, using connected correlators does not affect the definition and structure
of web mixing matrices, which are derived exclusively from the ordering of
gluon attachments to the Wilson lines, and are not affected by gluon interactions
away from the Wilson lines.

With this in mind, we define a {\it correlator web}, or {\it Cweb} as a set of 
skeleton diagrams, built out of connected gluon correlators attached to Wilson 
lines, and closed under permutations of the gluon attachments to each Wilson 
line.

Clearly, the main difference between webs and Cwebs is the fact that Cwebs
are not fixed-order quantities, but admit their own perturbative expansion in powers 
of $g$. In  a manner similar to webs, it is not easy to find a non-ambiguous notation 
to uniquely identify Cwebs: indeed, to some extent, they reflect the full complexity
of the perturbative expansion. Below, we will use the notation $W_n (k_1, 
\ldots  , k_n)$ for a Cweb involving $n$ Wilson lines\footnote{Recall that
in \secn{Webs} we used lower-case $w$ for ordinary webs, while we reserve
upper-case $W$ for Cwebs.}, with  $k_i$ gluon attachments to the $i$-th Wilson 
line. It is clear that beyond the lowest orders several different $n$-line Cwebs 
will share the same number of attachments to the Wilson lines: one can then 
refine the notation, using $W_n^{(c_2, \ldots , c_p)} (k_1, \ldots  , k_n)$ for a 
Cweb with the prescribed attachments, constructed out of $c_m$ $m$-point 
connected gluon correlators:  one should keep in mind, however, that, at sufficiently 
high-orders, also this notation becomes ambiguous. We note also that there is 
an obvious degeneracy in the counting of Cwebs, since Cwebs that differ only 
by a permutation of their Wilson lines are structurally identical, and it is trivial 
to include them in the calculation of the full correlator, simply by summing over 
Wilson-line labels. As a consequence, for classification purposes, we will break 
this degeneracy by picking a specific Wilson-line ordering, choosing $k_1 \leq
k_2 \leq \ldots \leq k_n$.

\begin{figure}[H]
	\centering
	\vspace{-3mm}
	\subfloat[$W_1^{(1)} (2)$]{\includegraphics[height=3.7cm,width=4cm]{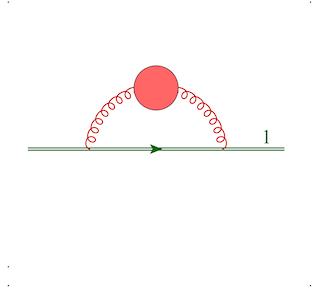} }
	\qquad \qquad \qquad
	\subfloat[$W_2^{(1)} (1,1)$]{\includegraphics[height=3.7cm,width=4cm]{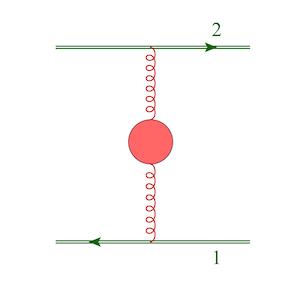} }
	\vspace{1mm}
	\caption{The only two Cwebs whose perturbative expansion starts at 
	${\cal O}(g^2)$.}
	\label{cwebg2}
\end{figure}

Taking into account the fact that the perturbative expansion for an $m$-point 
connected gluon correlator starts at ${\cal O} (g^{m - 2})$, we can write the 
perturbative expansion for a Cweb, with a prescribed correlator content and 
number of attachments, as
\beq
  W_n^{(c_2, \ldots , c_p)} (k_1, \ldots  , k_n)  \, = \, 
  g^{\, \sum_{i = 1}^n k_i \, + \,  \sum_{r = 2}^p c_r (r - 2)} \, \sum_{j = 0}^\infty \,
  W_{n, \, j}^{(c_2, \ldots , c_p)} (k_1, \ldots  , k_n) \, g^{2 j} \, ,
\label{pertCweb}
\eeq
which defines the perturbative coefficients of the Cweb, $W_{n, \, j}$.
Based on \eq{pertCweb}, it is natural to classify Cwebs based on the perturbative
order where they receive their lowest-order contribution, which is given by the
power of $g$ in the prefactor of \eq{pertCweb}; one may then easily design a 
recursive procedure to construct all Cwebs order by order. We begin by noticing 
that only two Cwebs have lowest-order contributions at order $g^2$: a self-energy
insertion with a two-point connected gluon correlator attached to a single Wilson 
line, which we denote by  $W_1^{(1)} (2)$, and the configuration with a  two-gluon
correlator joining two Wilson lines, which we denote by $W_2^{(1)} (1,1)$. They  
are depicted in Fig.~\ref{cwebg2}. 
\noindent In the massless case (with Wilson lines on the light cone), 
the self-energy Cweb vanishes identically, since, by the eikonal Feynman 
rules, it is proportional to the square of the Wilson-line four-velocity vector, 
$\beta^2$, so one is left with a single non-vanishing ${\cal  O}(g^2)$ Cweb. 
Starting from this  initial condition, one may systematically construct Cwebs 
starting at higher perturbative orders: keeping in mind that one must imagine 
having an unlimited supply of yet-uncoupled Wilson lines, one may proceed 
by performing the following moves.
\begin{itemize}
\item Add a two-gluon connected correlator  connecting any two Wilson lines
(including Wilson lines that  had no attachments at lower orders).
\item Connect an existing $m$-point correlator to any Wilson line (again, 
including Wilson lines with no attachments at lower orders), turning  it into
an  $(m+1)$-point correlator.
\item Connect an existing $m$-point correlator to an existing $n$-point 
correlator, resulting in an $(n+m)$-point correlator.
\end{itemize}
Executing all moves is clearly redundant, since the same Cweb is generated  
more than once through different sequences of moves: upon performing all
moves, one must therefore remove multiply-counted Cwebs. This procedure
can be considerably streamlined by taking into account known properties of
webs, which naturally generalise to Cwebs. Specifically,
\begin{itemize}
\item webs (and thus Cwebs) that are given by the product of two or more
disconnected lower-order webs (so that there are subsets of Wilson lines
not joined by any gluons) do not contribute to the logarithm of the
correlator, ${\cal W}_n$;
\item as mentioned above, in a massless theory all self-energy webs 
(Cwebs), where all gluon lines attach to the same Wilson line, vanish as
a consequence of the eikonal Feynman rules. Thus, any Cweb containing
a connected gluon correlator attaching to a single Wilson line will vanish.
\end{itemize}
It turns out that both these rules can be applied to trim the recursive 
procedure: more precisely, moves that lead to a disconnected Cweb, or
(in  the massless theory) to a self-energy Cweb can be immediately 
discarded. This is less than obvious, since a disconnected Cweb can 
become connected at the next recursive step, and similarly a
self-energy Cweb can become connected to other Wilson lines upon
adding more gluons. It is however easy to convince oneself that all
non-vanishing Cwebs that are reached by the recursion through 
intermediate stages involving either self-energy  or disconnected 
Cwebs, are also reached through sequences of steps involving only
non-vanishing Cwebs. The recursion can therefore be stopped whenever
a vanishing contribution of these two kinds is reached.
\begin{figure}[H]
	\centering
	\vspace{-3mm}
	\subfloat[$W_2^{(0,1)} (1,2)$]{\includegraphics[height=3.2cm,width=3.2cm]{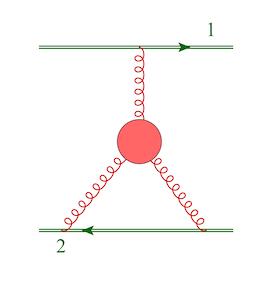} } 
	\quad
	\subfloat[$W_2^{(2)} (2,2)$]{\includegraphics[height=3.2cm,width=3.2cm]{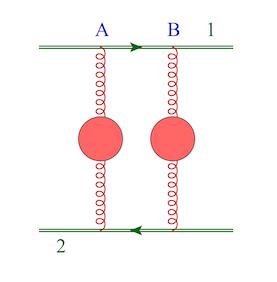} }
	\quad
	\subfloat[$W_3^{(0,1)} (1,1,1)$]{\includegraphics[height=3.2cm,width=3.2cm]{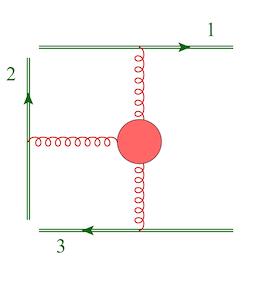} }
	\quad
	\subfloat[$W_3^{(2)} (1,2,1)$]{\includegraphics[height=3.2cm,width=3.2cm]{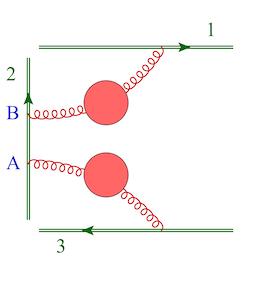} }
	\caption{Representative skeleton diagrams for the four non-vanishing Cwebs 
	in a massless theory whose perturbative expansion starts at ${\cal O}(g^4)$.
	Cwebs (a) and (c) comprise a single skeleton diagram, while Cwebs (b) and 
	(d) comprise two skeleton diagrams, obtained by permuting the labelled 
	attachments.}
	\label{Cwebg4}
\end{figure}   
\noindent Using these recursion criteria, it is easy to enumerate inequivalent 
Cwebs at low orders. In the massless theory, we find four inequivalent Cwebs
starting at ${\cal O}(g^4)$, which we label $W_2^{(0,1)} (1,2)$, $W_2^{(2)} 
(2,2)$,  $W_3^{(0,1)} (1,1,1)$ and $W_3^{(2)} (1,1,2)$: they  are displayed in 
Fig.~\ref{Cwebg4} (up to permutations of the  Wilson lines). Similarly, at ${\cal O} 
(g^6)$ we find fourteen new Cwebs, depicted in Figures 6, 7 and 8. Out of these, 
four Cwebs connect two Wilson lines, and their labels are $W_2^{(0,0,1)} (1,3)$,
$W_2^{(0,0,1)} (2,2)$, $W_2^{(1,1)} (2,3)$ and $W_2^{(3)} (3,3)$; six Cwebs 
connect three Wilson lines: $W_3^{(0,0,1)} (1,1,2)$, $W_3^{(1,1)} (1,1,3)$, 
$W_{3, \, {\rm I}}^{(1,1)} (1,2,2)$, $W_{3, \, {\rm II}}^{(1,1)} (1,2,2)$, $W_3^{(3)} 
(1,2,3)$ and $W_3^{(3)} (2,2,2)$. Notice that here we find the first occurrence 
of two Cwebs with the same correlator content and attachments: they are 
distinguished by different attachments of the three-gluon correlator to the 
Wilson lines, and we tag them by different roman numeral indices. Finally, 
still at ${\cal O} (g^6)$ we find four Cwebs connecting four Wilson lines:
$W_4^{(0,0,1)} (1,1,1,1)$, $W_4^{(1,1)} (1,1,1,2)$, $W_4^{(3)} (1,1,2,2)$ 
and $W_4^{(3)} (1,1,1,3)$.   
\begin{figure}[H]
	\centering
	\vspace{-3mm}
	\subfloat[$W_2^{(0,0,1)} (1,3)$]{\includegraphics[height=3.2cm,width=3.2cm]{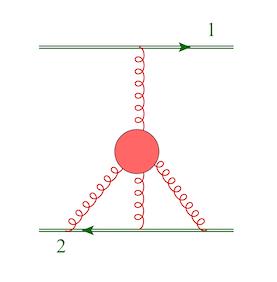} }
	\quad
	\subfloat[$W_2^{(0,0,1)} (2,2)$]{\includegraphics[height=3.2cm,width=3.2cm]{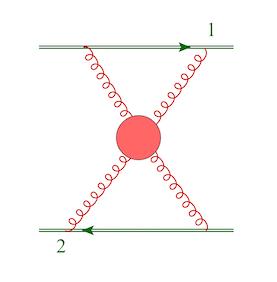} }
	\quad
	\subfloat[$W_2^{(1,1)} (2,3)$]{\includegraphics[height=3.2cm,width=3.2cm]{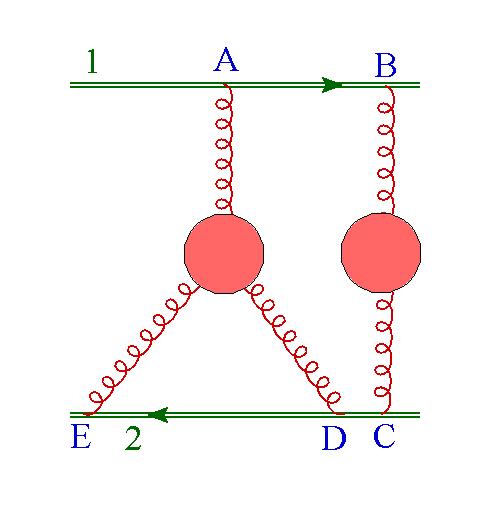} }
	\quad
	\subfloat[$W_2^{(3)} (3,3)$]{\includegraphics[height=3.2cm,width=3.2cm]{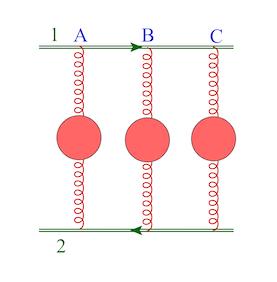} }
	\caption{Representative skeleton diagrams for the four non-vanishing Cwebs 
	in a massless theory connecting two Wilson lines, and whose perturbative 
	expansion starts at ${\cal O}(g^6)$. Cwebs (a) and (b) have a single skeleton 
	diagram, while Cwebs (c) and (d) have six.}
	\label{Cwebg6_2}
\end{figure}   
\begin{figure}[H]
	\centering
	\vspace{-3mm}
	\subfloat[$W_3^{(0,0,1)} (1,1,2)$]{\includegraphics[height=2.8cm,width=3.2cm]{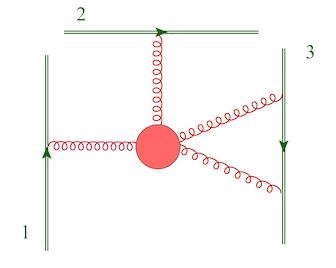} }
	\quad
	\subfloat[$W_3^{(1,1)} (1,1,3)$]{\includegraphics[height=2.8cm,width=3.2cm]{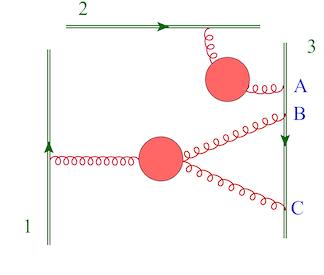} }
	\quad
	\subfloat[$W_{3, \, {\rm I}}^{(1,1)} (2,1,2)$]{\includegraphics[height=2.8cm,width=3.2cm]{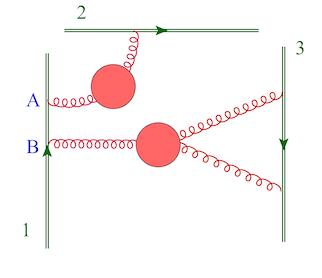} }
        \quad
	\subfloat[$W_{3, \, {\rm II}}^{(1,1)} (2,1,2)$]{\includegraphics[height=2.8cm,width=3.2cm]{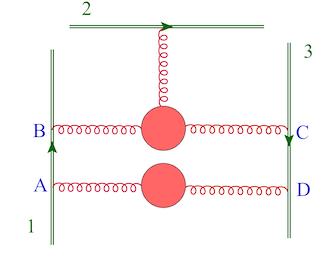} }
	\quad
	\subfloat[$W_3^{(3)} (3,2,1)$]{\includegraphics[height=2.8cm,width=3.2cm]{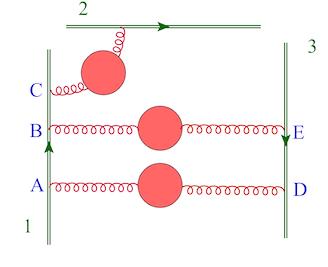} }
	\quad
	\subfloat[$W_3^{(3)} (2,2,2)$]{\includegraphics[height=2.8cm,width=3.2cm]{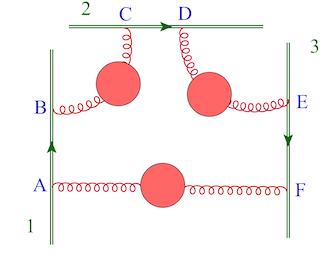} }
	\caption{Representative skeleton diagrams for the six non-vanishing Cwebs 
	in a massless theory connecting three Wilson lines, and whose perturbative 
	expansion starts at ${\cal O}(g^6)$. Their respective numbers of skeleton 
	diagrams are $\{1,3,2,4,6,8\}$.}
	\label{Cwebg6_3}
\end{figure}   
\begin{figure}[H]
	\centering
	\vspace{-3mm}
	\subfloat[$W_4^{(0,0,1)} (1,1,1,1)$]{\includegraphics[height=3.2cm,width=3.2cm]{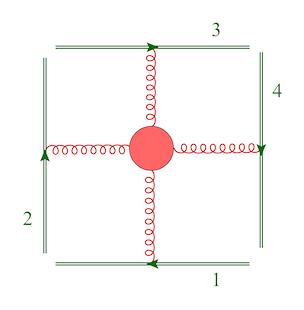} }
	\quad
	\subfloat[$W_4^{(1,1)} (1,2,1,1)$]{\includegraphics[height=3.2cm,width=3.2cm]{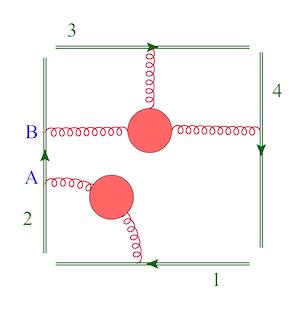} }
	\quad
	\subfloat[$W_4^{(3)} (1,2,1,2)$]{\includegraphics[height=3.2cm,width=3.2cm]{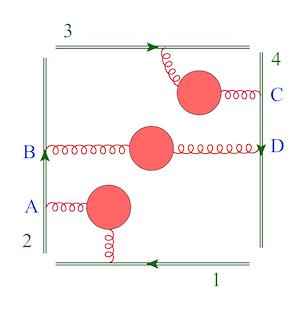} }
	\quad
	\subfloat[$W_4^{(3)} (1,3,1,1)$]{\includegraphics[height=3.2cm,width=3.2cm]{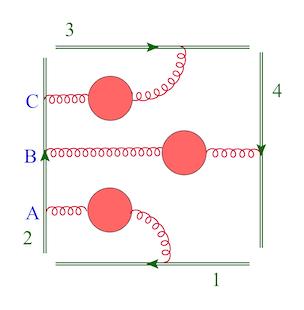} }
	\caption{Representative skeleton diagrams for the four non-vanishing Cwebs 
	in a massless theory connecting four Wilson lines, and whose perturbative 
	expansion starts at ${\cal O}(g^6)$. Their respective numbers of skeleton 
	diagrams are $\{1,2,4,6\}$.}
	\label{Cwebg6_4}
\end{figure}   
At ${\cal  O}(g^8)$, we find a total of 60 new Cwebs: 8 connecting two lines,
22 connecting three lines, 21 connecting 4 lines and 9 connecting five lines. 
Four-loop Cwebs connecting four and five lines will be discussed in more 
detail in \secn{Fourwe}, and are listed in the Appendix.

We conclude this section by discussing briefly the symmetry properties of Cwebs
and the counting of skeleton diagrams entering each Cweb. First, we note that
Cwebs are built out of connected gluon correlators, each of which is constrained
by Bose symmetry and gauge invariance. In simple cases, this poses strong 
limitations on the colour structures entering the Cweb: for example, colour 
conservation forces the gluon two-point function to be diagonal in colour, and
thus proportional to the unit matrix in the adjoint representation of the gauge 
group; thus, to any order, a two-point function joining Wilson lines $i$ and 
$j$ will always generate the dipole structure ${\bf T}_i \cdot {\bf T}_j  \equiv 
{\bf T}_i^a \, {\bf T}_{j, \, a}$. Similarly the three-point (off-shell) correlator is conjectured 
to be proportional to the structure constants $f_{a b c}$ to all orders in perturbation 
theory, a conjecture which has been verified to high orders (see for example 
Refs.~\cite{vanRitbergen:1998pn,Chetyrkin:2017bjc}), and which implies the 
complete antisymmetry of the corresponding kinematic factor. For correlators 
with $n>3$ gluons, several colour structures are possible (including building  
blocks for higher-order Casimir operators) and decoding the  constraints 
imposed by colour conservation and Bose symmetry becomes more intricate: 
these constraints are however crucial for the construction of the full soft 
anomalous dimension matrix (see for example~\cite{Almelid:2015jia,Almelid:2017qju}), 
and they are efficiently summarised to all orders in the language of Cwebs.

Concerning the counting of skeleton diagrams contributing to a given Cweb,
which gives the dimension of the corresponding  web mixing matrix, we note 
that if a connected correlator is attached to a given Wilson line with $p$ gluons, 
the permutations of those gluons should not be taken into account in the 
enumeration of contributing diagrams, since Bose symmetry is embedded
in the structure of the correlator and all $p$ gluons are equivalent. The 
dimension of the web mixing matrix should therefore be computed not by
counting permutations of gluons on each Wilson line, but rather by counting
shuffles of the subsets of gluons emerging from each correlator. Thus, for
example, if a Wilson line has a total of five attachments, three of them 
emerging from one correlator, and the remaining two from another one,
that line will contribute to the dimension of the web mixing matrix a factor 
of 10 (the number of possible shuffles of two sets of three and two cards),
rather than a factor of 120 (the number of permutations of five cards). We
note also that, in the language of Cwebs, the bulk of the growth of the 
number of diagrams at high orders in perturbation theory is hidden in
the internal structure of the contributing gluon correlators:  thus, for 
example, in our language, the four Feynman diagrams contributing
to $W_4^{(0,0,1)}(1,1,1,1)$ at ${\cal O}(g^6)$ are understood as four 
contributions  to that same Cweb at that order, rather than four distinct 
webs; similarly $W_5^{(0,0,0,1)} (1,1,1,1,1)$ receives contributions from
25 Feynman diagrams at ${\cal O}(g^8)$.

We now move on  to the discussion of the calculation of web mixing matrices
with the method of replicas~\cite{Gardi:2010rn}, before discussing the  
four-loop case in greater detail.


\section{A replica-trick algorithm to generate Cweb mixing matrices}
\label{Repli} 

As in other combinatorial problems involving exponentiation, such as the 
construction of effective actions, Wilson-line correlators can be studied  efficiently 
by means of algorithms constructed with the replica method~\cite{MezaPariVira}.
Here we will briefly discuss the application of the replica method to infrared
exponentiation, following Refs.~\cite{Laenen:2008gt,Gardi:2010rn}, and then
outline our algorithm for the construction of web mixing matrices.

In order to introduce the replica method, consider the path integral expression
for the Wilson-line correlator in \eq{genWLC}
\beq
  {\cal S}_n \left( \gamma_i \right) \, = \, \int {\cal D} A_\mu^a  \,\,
  {\rm e}^{{\rm i} S \left(  A_\mu^a \right)} \,  \prod_{k = 1}^n
  \Phi \left(  \gamma_k \right) \, = \, \exp \Big[ {\cal W}_n (\gamma_i) \Big] \, ,
\label{genWNCpath}
\eeq
where $S$ is the classical action\footnote{In the presence of matter fields, we can 
imagine having integrated them out and included their effect in $S$, since they do
not play a role in the following argument: only gluons couple to Wilson lines, and 
matter fields appear only in loops.}.  In order to compute ${\cal W}_n (\gamma_i)$,
we may imagine building a {\it replicated theory}, replacing the single gluon field 
$A_\mu^a$ with $N_r$ identical copies $A_\mu^{a, \, i}$  ($i = 1, \ldots, N_r$), which 
do not interact with each other. Under this replacement, one has $S \left(  A_\mu^a 
\right)  \rightarrow \sum_{i = 1}^{N_r} S \left(  A_\mu^{a, \, i} \right)$; if, furthermore, 
we replace each Wilson line in \eq{genWNCpath} by the product of $N_r$ Wilson 
lines, each belonging to one replica of the theory, one readily realises that the
replicated correlator is given by
\beq
  {\cal S}_n^{\, {\rm repl.}} \left( \gamma_i \right) \, = \,   \Big[ 
  {\cal S}_n \left( \gamma_i \right) \Big]^{N_r} \, = \, \exp \Big[ N_r \,
  {\cal W}_n (\gamma_i) \Big] \, =  \, {\bf 1} + N_r \, {\cal W}_n (\gamma_i) 
  + {\cal O} (N_r^2) \, .
\label{exprepl}
\eeq
As a consequence, in order to compute ${\cal W}_n (\gamma_i)$ order by order
in perturbation theory, one may compute the replicated correlator, and then extract
from the result the term of order $N_r$. 

Importantly, while gluon fields belonging to different replicas do not interact, they 
all belong to the same gauge group: therefore, the colour matrices associated with 
their attachments to the Wilson lines do not commute, and their ordering is relevant.
On the other hand, in a Cweb, each connected gluon correlator can be assigned a 
unique replica number, since there are no interaction vertices connecting different
replicas.  It is clear then that the contributions of different skeleton diagrams to the 
replicated correlator in \eq{exprepl} will be simply related to those of the same 
diagrams in the unreplicated theory, by means of combinatorial factors, counting 
the multiplicities associated with the presence of different replicas.  The computation
of ${\cal W}_n (\gamma_i)$ in terms of the skeleton diagrams contributing to 
${\cal S}_n (\gamma_i)$ is thus reduced to the computation of these combinatorial 
factors. The necessary steps, listed below, were identified in Refs.~\cite{Laenen:2008gt,
Gardi:2010rn}, and can be naturally adapted to the language of Cwebs.

\begin{itemize}
\item Given a Cweb, assign a replica number $i$ ($1 \leq i \leq N_r$) to each 
connected gluon correlator present in the web. Note that if only one connected 
correlator is present, only one replica can contribute to any diagram. It is then 
easy to show that all diagrams of the Cweb contribute to ${\cal W}_n (\gamma_i)$ 
with unit weight: in other words, there is no mixing matrix.
\item Define a {\it replica ordering operator} $R$ which acts by ordering the color 
generators on each Wilson line according to their replica numbers. Note that the
Wilson lines are naturally oriented, so this operation is unambiguous. If we denote 
by ${\bf T}_k^{(i)}$ the group generator associated with the emission of a gluon in 
replica $i$ from the  $k$-th Wilson line, 
the replica-ordering operator acts on a product ${\bf T}_k^{(i)} {\bf T}_k^{(j)}$ by 
relabelling the generators, exchanging their replica numbers $i$ and $j$, if $i > j$,
while leaving them unchanged if $i \leq j$.
\item In order to compute the colour factor of a skeleton diagram in the replicated  
theory, it is now sufficient to note that any non-trivial action by the replica ordering
operator $R$ effectively replaces the selected skeleton diagram with another one 
drawn from the same Cweb. The colour factor in the replicated theory is thus a 
linear combination of the colour factors of all skeleton diagrams in the Cweb, 
with multiplicities given by the number of possible replica orderings of the gluon
attachments on every Wilson line.
\item Algorithmically, for a Cweb $W_n^{(c_2, \ldots, c_p)} (k_1, \ldots, k_n)$, built
out of $m = \sum_{r = 2}^p c_r$ connected correlators, one needs to determine
two relevant numbers: the number of possible hierarchies between the $m$ 
replica numbers of the correlators, $h(m)$, and, for every fixed hierarchy $h$, 
the multiplicity with which that hierarchy can occur in the presence of $N_r$ 
replicas  $M_{N_r}(h)$. The determination of $h(m)$ is made non-trivial by the 
fact that the case of equal replica numbers must be treated separately: the 
sequence $h(m)$ is however well known~\cite{IntSeq}, and given by the so-called 
Fubini numbers\footnote{The Fubini numbers admit a generating function, and
they can be defined by
\beq
   \frac{1}{2 - \exp (x)} - 1 \, \equiv \, \sum_{m=1}^\infty h(m) \, \frac{x^m}{m!} \,\, .
\label{genfu}
\eeq}. 
In the first instances, for $m =  \{1,2,3,4,5,6\}$ one finds $h(m) = \{1,3,13,75,541,4683\}$. 
The multiplicity of a given hierarchy, on the other hand, is easily seen to be given by
\beq
  M_{N_r}(h) \, = \, \frac{N_r!}{\big( N_r - n_r(h) \big)! \,\, n_r(h)!}  \, ,
\label{multhi}
\eeq
where $n_r(h)$ is the number of distinct replicas present for hierarchy $h$. 
To give a concrete example, for $m = 5$, labelling the replica numbers of 
the 5 correlators with $i_k$, $(k = 1, \ldots, 5)$, and picking  the hierarchy
$i_1 = i_2 <  i_3 = i_4 < i_5$, we have $n_r(h) = 3$, and thus $M_{N_r} (h)
= N_r (N_r - 1) (N_r - 2)/6$.
\item Finally, the exponentiated color factors for a skeleton diagram $D$ is 
given by
\beq
  C_{N_r}^{\, {\rm repl.}}  (D) \, = \, \sum_h M_{N_r} (h) \, R \big[ C(D) \big| h 
  \big]  \, ,
\label{expocolf}
\eeq
where $R \big[ C(D) \big|  h  \big]$ is the color factor of the skeleton diagram 
obtained from $D$ through  the action of the replica-ordering operator $R$ for
the case of hierarchy $h$. The Cweb mixing matrix is built out of the coefficients
$M_{N_r}(h)$, which are polynomials in $N_r$, picking  terms that are linear in
$N_r$: many examples will be described in \secn{Fourwe} and in the Appendix.
Note that in the presence of $m$-point correlators, with $m \geq 4$, each such 
correlator can contribute different `internal' colour factors, for example different 
permutations of products of structure constants. Since, however, the information 
on the mixing matrix is encoded in the coefficients $M_{N_r}(h)$, the different 
colour factors arising from the internal structure of the correlators can be treated 
one by one, without affecting the mixing of the diagrams.
\item We note that, given a Cweb $W_n^{(c_2, \ldots, c_p)} (k_1, \ldots, k_n)$,
the dimension $d_W$ of its mixing matrix $R_W$ can be expressed as 
follows. Let $s_W (k)$ be the number of shuffles that can be performed with 
the  attachments to the $k$-th Wilson line, given the arrangement of connected
correlators for the Cweb under consideration. One would na\"ively conclude
that the dimension of $R_W$ is the product of the factors $s_W(k)$ over all
Wilson lines. This however fails to take into account an important degeneracy,
which already appears in the simple two-line Cweb $W^{(2)}_2 (2,2)$ at two
loops: counting shuffles separately on each line yields $d_W = 4$, which is 
wrong, because the two shuffles on the second Wilson line can be obtained 
from the shuffles on the first line by exchanging the two gluons, which is
manifestly a symmetry of the Cweb. In order to take into account this 
degeneracy, one must divide by the number of available permutations 
of subsets of $m$-point correlators that have the property of being attached 
to the same sets of $m$ Wilson lines.
\end{itemize} 

\begin{figure}[H]
	\centering
	\subfloat[]{\includegraphics[height=2.5cm,width=4.5cm]{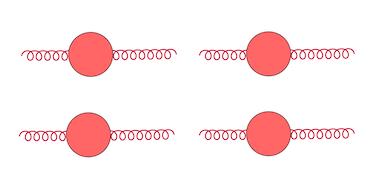} } 
	\qquad
	\qquad
	\qquad
	\subfloat[]{\includegraphics[height=2.5cm,width=4.5cm]{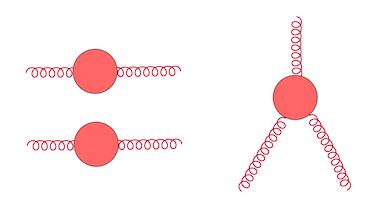} } \newline \\
	\subfloat[]{\includegraphics[height=2.5cm,width=4.5cm]{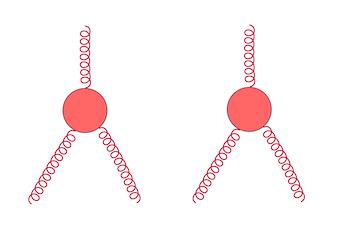} }
	\qquad
        \subfloat[]{\includegraphics[height=2.5cm,width=4.5cm]{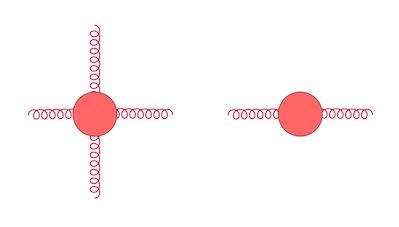} }
        \qquad
        \subfloat[]{\includegraphics[height=2.5cm,width=3.3cm]{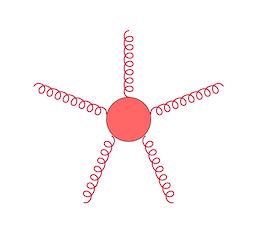} }\newline \\
	\caption{Combinations of connected correlators that can form Cwebs at four loops.}
\label{fig:connected}
\end{figure}

\noindent In order to compute Cweb mixing matrices at four loops, extending the 
results Refs.~\cite{Gardi:2010rn,Gardi:2011yz,Gardi:2013ita}, we have developed 
an in-house Mathematica code which we describe very briefly below. 
\begin{itemize}
\item The first step is to generate all the Cwebs that appear at four loops 
(${\cal{O}} (g^8)$), in particular those involving four and five Wilson lines. 
To do so, we note that at four loops (${\cal{O}} (g^8)$), all possible Cwebs 
can be obtained by combining the connected correlators shown in 
Fig.~\ref{fig:connected}, and attaching them to $2 \leq n \leq 5$ Wilson 
lines. One may begin by attaching four two-point correlators in all possible ways 
to the Wilson lines. Next one proceeds to Cwebs generated by attaching two 
two-point correlators and one three-point correlator, in all possible ways, and 
similary with the other types, obtained by using the other building blocks in
Fig.~\ref{fig:connected}. We note that the five-point correlator at this order
can only give a trivial Cweb, since it contains only a single skeleton diagram.
\item The code assigns a distinct replica variable to each of the correlators 
present in a given Cweb: for example, four two-gluon correlators will be
assigned replica indices $i, j, k$ and $l$. Then the correlators are sequentially
attached to the Wilson lines in lexicographic order, beginning with the one with
index $i$, attached between Wilson line 1 and 2, proceeding  to the one with 
index $j$, attached in all possible ways, so as to generate a set of 'partial 
skeleton diagrams'. This procedure continues till all correlators are exhausted. 
Clearly,  this procedure will generate several Cwebs that are identical (as far as 
color is concerned), since they are related to each other by a mere renaming 
of the Wilson lines: duplicates are discarded. At this stage, we have only one
diagram per Cweb.  	
\item The code then takes one of the above diagrams and generates all the 
other diagrams of the corresponding Cweb by permuting (or more precisely 
shuffling) the gluon attachments on each of the Wilson lines.
\item The next step is to generate all the multiplicities associated to the possible 
hierachies, corresponding to the entries of the Table 1 of Ref.~\cite{Gardi:2010rn}.
A subroutine creates all possible different hierarchies $h$ for each diagram; 
the code then proceeds to obtain all other colums of the table, and finally 
leads to the exponentiated colour factors $\tilde C(D)$, from which we obtain 
the mixing matrix $R$. Finally, $R$ is diagonalised and the diagonalising 
matrix $Y$ is recorded.
\item The code automatically discards self-energy Cwebs, where
a connected correlator is attached only to a single Wilson line, but, in principle,
keeps disconnected Cwebs, so that the vanishing of the corresponding $R$
matrix works as a test.
\end{itemize}

\noindent We note that the run time of the code increases steeply with the increase 
in the number of connected correlators, which causes a rapid increase in the number 
of available replica hierarchies. The code has been checked by reproducing all
lower-order results available in the literature, and by verifying the two known properties 
of mixing matrices: their idempotence, and the row sum rule, which hold 
true for all the $R$ matrices that we have computed. Furthermore, one can verify 
that all different Cwebs at ${\cal O}(g^8)$ have been constructed, by applying the
recursive construction described in \secn{Cwebs}. Finally, as shown explicitly in 
\secn{Fourwe} and in the Appendix, all $R$ matrices we have computed satisfy 
the conjectured column sum rule discussed in \secn{Webs}.

\section{A selection of four-loop Cwebs}
\label{Fourwe}

In this section we present in some detail the calculation of two four-loop Cwebs,
involving respectively four and five Wilson lines. This will allow us to introduce 
some notation which will be useful to simplify the full listing of results for similar 
webs, which is presented in the Appendix. For each Cweb, we will present
the mixing matrix $R$, the diagonalising matrix $Y$, and the exponentiated 
colour factors.


\subsection {A four-loop, four-line Cweb}
\label{foufou}

\begin{figure}[H]
\captionsetup[subfloat]{labelformat=empty}
\centering
		\subfloat[][${\bf C}_1$]{\includegraphics[height=4cm,width=4cm]{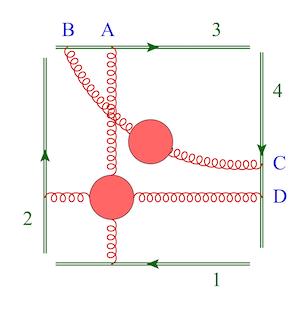} }
		\qquad \qquad
		\subfloat[][${\bf C}_2$]{\includegraphics[height=4cm,width=4cm]{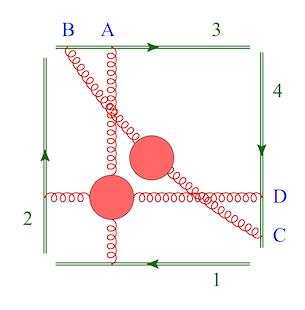} }
                \qquad \qquad
		\subfloat[][${\bf C}_3$]{\includegraphics[height=4cm,width=4cm]{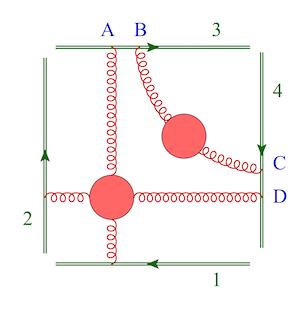} }
		\qquad \qquad
		\subfloat[][${\bf C}_4$]{\includegraphics[height=4cm,width=4cm]{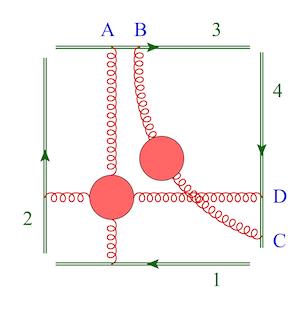} }
\caption{The four skeleton diagrams of the Cweb $W_{4, \, {\rm I}}^{(1,0,1)} (1,1,2,2)$.} 
\label{fig:Webone}
\end{figure}

As an example, we have selected the Cweb $W_{4, \, {\rm I}}^{(1,0,1)} (1,1,2,2)$, 
which comprises four skeleton diagrams depicted in Fig.~\ref{fig:Webone}. We 
note that in this case the Cweb label includes a roman numeral, to distinguish 
it from a second Cweb with identical correlator and attachment content, $W_{4, 
\, {\rm II}}^{(1,0,1)} (1,1,2,2)$, where however the four-gluon correlator has two
attachments to the same Wilson line. That Cweb, discussed in the Appendix,
involves only two skeleton diagrams, and therefore has a $2\times2$ mixing 
matrix. In this case, it is evident by looking at any one of the diagrams that
there are four shuffles to be considered, so that the mixing matrix will have 
dimension four. In order to write down an explicit expression for the matrices 
$R$ and $Y$, it is necessary to introduce an ordering among the diagrams: in 
this case, the ordering is displayed in Fig.~\ref{fig:Webone}, but in general it can
be identified by labelling the gluon attachments to be shuffled, and giving the
sequence of the shuffles to be considered. As an example, in Fig.~\ref{fig:Webone},
we have labelled the attachments on Wilson line $3$ by $A$ and $B$, and 
those on Wilson line $4$ by $C$ and $D$: the shuffles associated with the
four skeleton diagrams can be labelled by giving the sequences of the 
attachments, in the orderings defined by the orientation of the Wilson lines.
In this case ${\bf C}_1 = \{\{BA\},\{CD\}\}$, ${\bf C}_2 = \{\{BA\},\{DC\}\}$, 
${\bf C}_3 = \{\{AB\},\{CD\}\}$, and ${\bf C}_4 = \{\{AB\},\{DC\}\}$. Having
chosen the ordering of diagrams, it is straightforward to apply the algorithm
and obtain the exponentiated colour factors. 
\noindent We find that the mixing matrix $R$ and the diagonalising matrix $Y$  
are given by
\begin{align}
\begin{split}
&
  R \, = \, \displaystyle{\left(
  \begin{array}{cccc}
  \frac{1}{2} & \, 0 & \, 0 & \, -\frac{1}{2} \\
  - \frac{1}{2} & \, 1 & \, 0 & \, -\frac{1}{2} \\
  - \frac{1}{2} & \, 0 & \, 1 & \, -\frac{1}{2} \\
  - \frac{1}{2} & \, 0 & \, 0 & \, \frac{1}{2} \\
  \end{array}
  \right) } \, , \qquad \qquad
  Y \, = \, \left(
  \begin{array}{cccc}
  - 1 & \, 0 & \, 0 & \, 1 \\
  - 1 & \, 0 & \, 1 & \, 0 \\
  - 1 & \, 1 & \, 0 & \, 0 \\
  1 & \, 0 & \, 0 & \, 1 \\
  \end{array}
  \right) \, , \end{split}
\label{eq:web1}
\end{align} 
The expected properties of the mixing matrix are easily verified: $R$ is idempotent,
{\it i.e.} $R^2 = R$, the matrix elements in each row sum to zero, and furthermore
the column sum rule is obeyed. Indeed, in this case the vector built out of the
indices $s(D)$ for the four diagrams in Fig.~\ref{fig:Webone} is given by
$S \equiv \{ s({\bf C}_1), s({\bf C}_2), s({\bf C}_3), s({\bf C}_4) \} = \{1, 0, 0, 1\}$:
in the first and last diagrams one can move the gluon attachments of one of the 
two correlators to the origin without affecting the other correlator, which is not
possible for the second and the third diagram. One then readily verifies that
the column vector $S \cdot R$ is a null vector. Finally, we observe (upon
diagonalisation by means of the matrix $Y$) that the mixing matrix has rank
$r = 3$: as a consequence, there are three independent exponentiated colour 
factors, which can be chosen as the combinations $\left( Y {\bf C} \right)_i$,
where $i = \{1,2,3\}$. In order to display explicit expressions for the colour 
factors, we have to make a choice for the internal colour structure of the 
four-point correlator, which appears here at tree level, and thus is built out 
of three possible products of pairs of structure constants. As an example, we 
display here the colour factors arising  from the combination $f^{a_1 a_2 x} 
f^{a_3 a_4 x}$, where the gluons emerging from the four point correlator and 
attaching to line $i$ have colour index $a_i$, with $i = 1,2,3,4$. The other two 
possible colour factors for this Cweb can be obtained by simple permutations.
In the case we examine, the emerging colour combinations are
\beq
  (Y {\bf C})_1 & = &  i \fabg \fcdg \fedh \ta 1 \tb 2 \te 3 \tc 3 \thh 4 - 
  i \fabg \fcdg \fcej \ta 1 \tb 2 \tj 3 \td 4 \te 4 \, , \nonumber \\ [3mm]
  (Y {\bf C})_2 & = & - i \fabg \fcdg \fcej \ta 1 \tb 2 \tj 3 \td 4 \te 4 \, ,
  \label{ecfwebone} \\ [3mm] 
  (Y {\bf C})_3 & = & i \fabg \fcdg \fedh \ta 1 \tb 2 \tc 3 \te 3 \thh 4 \, . \nonumber
\eeq
We observe that these exponentiated color factors correspond to fully 
connected Feynman diagrams, so that the non-abelian exponentiation 
theorem of Ref.~\cite{Gardi:2013ita} is, as expected, satisfied.


\subsection {A four-loop, five-line Cweb}
\label{foufi}

\begin{figure}[H]
	\captionsetup[subfloat]{labelformat=empty}
	\centering
	\subfloat[][]{\includegraphics[height=4cm,width=4cm]{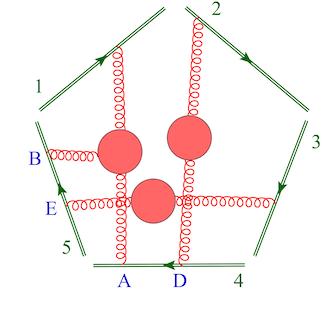} }
	\vspace{-4mm}
	\caption{A sample skeleton diagram contributing to $W_{5, \, {\rm I}}^{(2,1)} 
	(1,1,1,2,2)$.} 
	\label{fivesample}
\end{figure}

As an explicit example of a four-loop, five-line Cweb, we select the one labelled
by $W_{5, \, {\rm I}}^{(2,1)} (1,1,1,2,2)$, one of two Cwebs with this particular
set of attachments and correlators. A sample skeleton diagram contributing
this Cweb is shown in Fig.~\ref{fivesample}: it is immediate to note that there
are four possible shuffles of the labels, so that the mixing matrix will have  
dimension four.
\noindent  The four skeleton diagrams comprising the Cweb can be labelled with 
the shuffles of the gluon attachments to the Wilson lines: we choose the sequence 
${\bf  C}_1 = \{\{DA\},\{EB\}\}$, ${\bf  C}_2 = \{\{DA\},\{BE\}\}$, ${\bf  C}_3 = 
\{\{AD\},\{EB\}\}$, and ${\bf  C}_4 = \{\{AD\},\{BE\}\}$, where, as usual, the letters 
are ordered along the oriented Wilson lines following the arrows: with this choice,
Fig.~\ref{fivesample} portrays diagram ${\bf  C}_1$. By inspection, one realizes
that for diagram ${\bf C}_1$ it is possible to shrink to the origins of the Wilson 
lines, independently of each other, both the two-gluon correlator joining lines 3 
and 5, and the other two-gluon correlator, joining lines 2 and 4, without affecting 
the three-gluon correlator. The  value of the parameter $s$ for this diagram is 
therefore $s({\bf  C}_1) = 2$. By a similar reasoning, one concludes that the other 
three diagrams have $s({\bf  C}_i) = 1$, for $i = 2, 3$, while $s({\bf  C}_4) = 2$. 
The calculation of the mixing matrix leads to
\begin{align}
  \begin{split}
  &
  R \, = \, \displaystyle{ \left(
	\begin{array}{cccc}
	\frac{1}{6} & -\frac{1}{6} & -\frac{1}{6} & \frac{1}{6} \\
      - \frac{1}{3} & \frac{1}{3} & \frac{1}{3} & -\frac{1}{3} \\
      - \frac{1}{3} & \frac{1}{3} & \frac{1}{3} & -\frac{1}{3} \\
	\frac{1}{6} & -\frac{1}{6} & -\frac{1}{6} & \frac{1}{6} \\
	\end{array}
	\right) } \, ,  \qquad \qquad
  Y \, = \, \left(
	\begin{array}{cccc}
	1 & -1 & -1 & 1 \\
	-1 & 0 & 0 & 1 \\
	2 & 0 & 1 & 0 \\
	2 & 1 & 0 & 0 \\
	\end{array}
	\right) \, .
  \end{split}
\end{align} 
One observes that both the row and the column sum rules are satisfied. Furthermore,
diagonalising the mixing matrix, one finds that it has rank $r = 1$, so that there is
only one exponentiated colour factor. It is
\beq
  (Y {\bf C})_1 \, = \,   - i \fabc \fadh \fbeg \thh 1 \tg 2 \tc 3 \td 4 \te 5 \, .
\label{ecffive}
\eeq
In the Appendix, we will similarly treat all the other four-loop Cwebs connecting 
four and five Wilson lines.


\subsection{A note on colour structures at four loops}
\label{Colstru}

Four-loop colour structures are especially interesting for the study of gauge-theory
scattering amplitudes, since at this order for the first time the soft anomalous 
dimension matrix can receive contributions from quartic Casimir operators of the
gauge algebra. The relevance of higher-order Casimir invariants was first noted 
in this context in~\cite{Gardi:2009qi}, but it was in fact known from the early 
days of QCD~\cite{Cvitanovic:2008zz}. The presence of quartic Casimir 
contributions at four loops in the cusp anomalous dimension is one of only 
two possible sources for the violation of the dipole structure of the soft anomalous 
dimension matrix for massless theories, which holds up to two loops. Remarkably, 
the cusp anomalous dimension for QCD was recently computed analytically at four 
loops~\cite{Henn:2019swt,vonManteuffel:2020vjv}, following the precise numerical 
predictions of Refs.~\cite{Moch:2017uml,Moch:2018wjh}, and the presence of a 
non-vanishing contribution from quartic Casimir operators was confirmed. Studying 
the implications of such contributions for the multiparticle soft anomalous dimension 
matrix is therefore now an open and very interesting problem, also in light of the results 
of Refs.~\cite{Catani:2019nqv,Dixon:2019lnw}. Of course, the ultimate understanding
of the role played by these contributions, and their implications for collinear  
factorisation, will depend upon a full calculation of the kinematic factors for the 
relevant Cwebs: indeed, we note that collinear factorisation remains compatible 
with the three-loop expression of the soft anomalous dimension matrix only thanks
to a remarkable connection between matrices with different number of partons,
and after enforcing the  constraints of colour conservation~\cite{Almelid:2015jia}.
\begin{figure}[H]
	\captionsetup[subfloat]{labelformat=empty}
	\centering
	\subfloat[][]{\includegraphics[height=4cm,width=8cm]{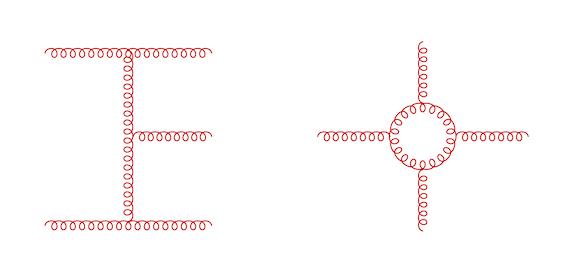} }
	\vspace{-8mm}
	\caption{Gluon diagrams representing the structure of the exponentiated colour 
	factors appearing at four loops.} 
	\label{fig:gen-connect}
\end{figure}
\noindent We may, in any case, make a few remarks in this issue, based purely on 
the analysis of the colour structure of four-loops Cwebs. First, we observe that all 
the exponentiated colour structures that arise in four-loop Cwebs connecting
four and five Wilson lines (listed in the Appendix) correspond to the connected
gluon diagrams depicted in Fig.~\ref{fig:gen-connect}, with the open ends of
the gluon lines attaching to generic permutations of Wilson lines. The gluon-loop
structure on the right of Fig.~\ref{fig:gen-connect} is of course built out of products 
of structure constants $f_{abc}$, however, upon symmetrization, could in principle 
yield a quartic Casimir contribution. We note, however, that, for all Cwebs whose 
perturbative expansion starts at ${\cal O} (g^8)$, the gluon-loop structure cancels
in the exponentiated colour factors $(Y {\bf C})_i$, before their recombination 
with kinematic factors. 
It would appear that the only possible source of quartic Casimir contributions 
 from four- and five-leg Cwebs at four loops is in the one-loop correction 
 to the ${\cal O} (g^6)$ Cweb $W_4^{(0,0,1)} (1,1,1,1)$, depicted in 
 Fig.~\ref{Cwebg6_4}(a). Indeed, at one loop, the four-gluon correlator featuring  
 in that  Cweb (which must be Bose symmetric), can develop a symmetric colour 
 structure, yielding a contribution of the form
 \beq
   W_4^{(0,0,1)} (1,1,1,1) \Big|_{\rm quartic} \, = \, {\cal K}^{ijkl} \, 
   {\bf T}_i^a {\bf T}_j^b {\bf T}_k^c {\bf T}_l^d \, d^{\, ({\rm r})}_{abcd} \, ,
 \label{quartC}
 \eeq
 where ${\cal K}$ is a kinematic factor, and $d^{\, ({\rm r})}_{abcd}$ is the completely 
 symmetrized trace of four generators of the gauge algebra in representation $r$.
 We note, however, that the colour structures directly arising from the diagrams are 
 not all independent, and, upon enforcing colour conservation, which implies the 
 operator constraint $\sum_i {\bf  T_i} = 0$, they must be reduced to a basis, 
 following for example the analysis of Ref.~\cite{Gardi:2013ita}. Since antisymmetric
 combination of generators on the same Wilson line can be eliminated using the
 gauge algebra, this step leads to the appearance of symmetric combination of 
 generators, which can again yield contributions of quartic Casimir type.
These results and considerations are in agreement with the analysis of 
Refs.~\cite{Ahrens:2012qz,Becher:2019avh}, with the effective vertex analysis in 
Ref.~\cite{Gardi:2013ita}, and with the generating functional approach of
Ref.~\cite{Vladimirov:2017ksc}, where the role played by symmetric colour 
structures in the exponentiated soft function is emphasised.


\section{Summary and Outlook}
\label{Conclu}

The study of the infrared structure of perturbative gauge amplitudes is an active
developing field, with implications for both theoretical studies, concerning the
mathematical properties of gauge theories, and for high-energy phenomenology
at colliders. These studies are very advanced, and we now have a well-developed
understanding of infrared factorisation and exponentiation, with the soft anomalous 
dimension matrix fully known at three loops for massless particles~\cite{Almelid:2015jia},
the cusp anomalous dimension computed to four loops~\cite{Henn:2019swt,
vonManteuffel:2020vjv}, and several high-order radiative amplitudes available
(see for example~\cite{Catani:2019nqv,Dixon:2019lnw}). The current frontier
is the calculation of the soft anomalous dimension matrix for multiparticle scattering
at three loops including massive particles, and the exploration of the four-loop 
domain; in general, all these studies bring to the fore the relevance, for gauge 
theory calculations, of correlators involving Wilson lines, possibly together 
with gauge and matter fields, which provide leading-power approximations to 
scattering amplitudes and cross sections in soft and collinear limits.

In the present paper, we have developed a set of tools for the analysis of soft 
anomalous dimensions at high orders, and we have studied the exponentiated 
colour structures arising at the four-loop level in multiparticle amplitudes. We 
have introduced the concept of a {\it correlator web}, or Cweb, which, we believe,
will be useful for the classification and study of exponentiated correlators
at high orders: Cwebs include their own radiative correction, are easily generated
and enumerated, since their number grows only moderately as a function of the
perturbative order, and may help to clarify and implement symmetry properties
of gluon correlators and the consequences of colour conservation.

We have enumerated all Cwebs for a massless theory up to four loops, and 
we have computed their mixing matrices and exponentiated colour factors
for all cases involving four and five Wilson lines. We observe that all exponentiated 
colour factors correspond to completely connected gluon diagrams, as expected
from the non-abelian exponentiation theorem~\cite{Gardi:2013ita}. Furthermore,
we note that structures compatible with the presence of quartic Casimir 
contributions are present, as expected, but they cancel in Cwebs arising 
at ${\cal O} (g^8)$, while they survive in radiative corrections to Cwebs
arising at lower orders. Finally, we verify the properties of mixing matrices
that were proved or conjectured in earlier studies~\cite{Gardi:2010rn,
Gardi:2011wa,Gardi:2011yz,Dukes:2013wa,Gardi:2013ita,Dukes:2013gea}.

The combinatorial complexity of the calculation of web mixing matrices grows 
rapidly with the perturbative order, most notably because of the proliferation
of hierarchies that arise in the application of the replica method, which are
counted by the Fubini numbers. We have developed a code which can handle
this complexity at four loops with minimal computing resources, but we believe
that going to yet higher orders is likely to require significant optimisation, the 
development of new tools, or the deployment of considerably larger computing  
power. That not withstanding, our results in this paper provide a number of 
needed ingredients for the calculation of infrared divergences at four loops, and  
we believe that the tools that we have developed will be useful to further our 
understanding at higher orders as well.


\section*{Acknowledgments}

\noindent  We thank Einan Gardi and Niamh Maher for useful discussions concerning
the choice of colour bases and the role of colour conservation. AT and LM would like to thank MHRD Govt. of India for the GIAN grant 
(171008M01), ``The Infrared Structure of Perturbative Gauge Theories'' and for the
SPARC grant (P578) ``Perturbative QCD for Precision Physics at the LHC'', which
were crucial to the completion  of the present research. AT and SP would also like 
to thank the University of Turin and INFN Turin for warm hospitality during the course 
of this work. AD would like to thank CSIR, Govt. of India for a JRF fellowship. SP 
would also like to thank MHRD Govt. of India for an SRF fellowship. LM would like
to thank IIT Hyderabad for their warm hospitality, and acknowledges the contribution
of the Italian Ministry of University and Research (MIUR), grant PRIN 2017LNEEZ.


\appendix


\section{Appendix}
\label{App}

In this appendix we give results for all the webs that appear at 4 loops in the 
scattering amplitude, that can connect 4 or 5 Wilson lines. Throughout the list, 
$R$, $Y$ and $D$ denote the mixing matrix, the diagonalizing matrix and 
the diagonal matrix respectively. $D$ is represented as $D = \D{r}$, where 
$r$ is the rank of the mixing matrix $R$. We display only one skeleton 
diagram per web, and we explicitly give the order of the shuffles that generate 
the other diagrams, which is tied to the order the columns of the mixing matrix 
in the chosen basis. Finally, we give the expressions for the exponentiated 
colour  factors, which, in all cases, correspond to fully  connected gluon diagrams, 
as expected. For Cwebs involving four-point correlators, the colour factors that
we present correspond to one of three possible permutations of structure 
constants arising from the internal structure of the correlator, as in \secn{foufou}.
We omit from the list the Cwebs that are composed of a single skeleton diagram, 
such as $W_5^{(0,0,0,1)} (1,1,1,1,1)$, whose mixing matrix is just a number, 
$R = 1$.

\vspace{2mm}


\subsection{Cwebs connecting four Wilson lines}

\vspace{2mm}


\begin{itemize}


\item[{\bf 1}.]  $\textbf{W}_{4, \, \text{I}}^{(1,0,1)}(1,1,2,2)$

This Cweb has four diagrams, one of which is displayed below. The table gives
the chosen order of the four shuffles of the gluon attachments, and the corresponding
$S$ factors.
\begin{minipage}{0.5\textwidth}
	\begin{figure}[H]
	        \vspace{-2mm}
		\includegraphics[height=4cm,width=4cm]{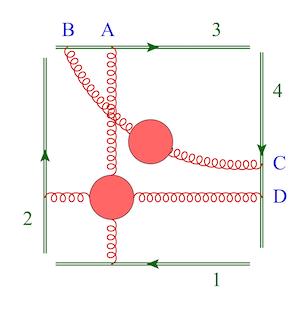}
	\end{figure}
\end{minipage} 
\hspace{-2cm}
\begin{minipage}{0.46\textwidth}
\begin{tabular}{ | c | c | c |}
	\hline
	\textbf{Diagrams} & \textbf{Sequences} & \textbf{S-factors} \\ \hline
	$C_1$ & $\lbrace\lbrace BA \rbrace,\lbrace CD \rbrace \rbrace$ & 1 \\ \hline
	$C_2$ & $\lbrace\lbrace BA \rbrace,\lbrace DC \rbrace \rbrace$ & 0\\ \hline
	$C_3$ & $\lbrace\lbrace AB \rbrace,\lbrace CD \rbrace \rbrace$ & 0\\ \hline
	$C_4$ & $\lbrace\lbrace AB \rbrace,\lbrace DC \rbrace \rbrace$ & 1\\ \hline
	\end{tabular}
\label{tab:abcd1}	
\end{minipage}

\noindent The $R$, $Y$ and $D$ matrices are given by
\begin{align}
\begin{split}
&
R=\displaystyle{\left(
\begin{array}{cccc}
\frac{1}{2} & 0 & 0 & -\frac{1}{2} \\
-\frac{1}{2} & 1 & 0 & -\frac{1}{2} \\
-\frac{1}{2} & 0 & 1 & -\frac{1}{2} \\
-\frac{1}{2} & 0 & 0 & \frac{1}{2} \\
\end{array}
\right) }\,, \qquad \qquad
Y=\left(
\begin{array}{cccc}
-1 & 0 & 0 & 1 \\
-1 & 0 & 1 & 0 \\
-1 & 1 & 0 & 0 \\
1 & 0 & 0 & 1 \\
\end{array}
\right)
\,,\qquad \qquad
D = \D{3} \, .
\end{split}
\label{eq:web1app}
\end{align}
Finally, the exponentiated colour factors are
\begin{eqnarray}
(YC)_1 &=&  i \fabg \fcdg \fedh \ta 1 \tb 2 \te 3 \tc 3 \thh 4 - i \fabg \fcdg \fcej \ta 1 
\tb 2 \tj 3 \td 4 \te 4 \, , \nonumber \\ \nonumber \\
(YC)_2 &=& -i \fabg \fcdg \fcej \ta 1 \tb 2 \tj 3 \td 4 \te 4 \, ,\nonumber \\ \nonumber \\ 
(YC)_3 &=& i \fabg \fcdg \fedh \ta 1 \tb 2 \te 3 \tc 3 \thh 4 -\fabg \fcdg \fcej \fedh 
\ta 1 \tb 2 \tj 3 \thh 4 \, .
\end{eqnarray} 

\vspace{2mm}


\item[{\bf 2}.] $\textbf{W}_{4,\text {II}}^{(1,0,1)}(1,1,2,2)$

This Cweb has two diagrams, one of which is displayed below. The table gives
the chosen order of the two shuffles of the gluon attachments, and the corresponding
$S$ factors.
\begin{minipage}{0.5\textwidth}
		\begin{figure}[H]
		\vspace{-2mm}
			\includegraphics[height=4cm,width=4cm]{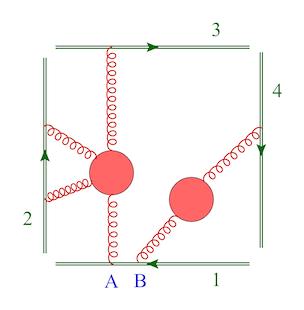}
		\end{figure}
\end{minipage} 
\hspace{-2cm}
\begin{minipage}{0.45\textwidth}
		\begin{tabular}{ | c | c | c |}
			\hline
			\textbf{Diagrams} & \textbf{Sequences} & \textbf{S-factors} \\ \hline
			$C_1$ & $\lbrace\lbrace BA\rbrace\rbrace$ & 1 \\ \hline
		$C_2$ & $\lbrace\lbrace AB\rbrace\rbrace$ & 1\\ \hline
		\end{tabular}
		\label{tab:abcd2}	
\end{minipage}

\noindent The $R$, $Y$ and $D$ matrices are given by
\begin{align}
\begin{split}
	&
	R=\left(
	\begin{array}{cc}
	\frac{1}{2} & -\frac{1}{2} \\
	-\frac{1}{2} & \frac{1}{2} \\
	\end{array}
	\right)\,, \qquad
	Y=\left(
	\begin{array}{cc}
	-1 & 1 \\
	1 & 1 \\
	\end{array}
	\right)
	\,,\qquad
	D= \D{1} \, .
	\end{split}
	\end{align}
Finally, the only exponentiated colour factor is	
\begin{eqnarray}
(YC)_1 &=& -i \fabg \fcdg \faeh \thh 1 \tb 2 \tc 2 \td 3 \te 4  \, .
\end{eqnarray} 

\vspace{2mm}


\item[{\bf 3}.] $\textbf{W}_4^{(1,0,1)}(1,1,1,3)$ \\

\vspace{-3mm}
This Cweb has three diagrams, one of which is displayed below. The table gives
the chosen order of the three shuffles of the gluon attachments, and the corresponding
$S$ factors.
\begin{minipage}{0.5\textwidth}
		\begin{figure}[H]
		         \vspace{-2mm}
			\includegraphics[height=4cm,width=4cm]{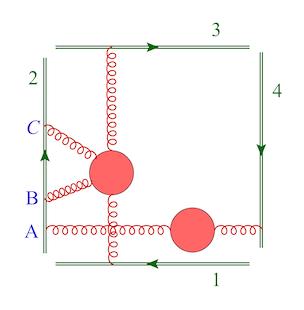}
		\end{figure}
	\end{minipage} 
	\hspace{-2cm}
	\begin{minipage}{0.45\textwidth}
		\begin{tabular}{ | c | c | c |}
			\hline
			\textbf{Diagrams} & \textbf{Sequences} & \textbf{S-factors} \\ \hline
			$C_1$ & $\lbrace\lbrace ABC\rbrace\rbrace$ & 1 \\ \hline
			$C_2$ & $\lbrace\lbrace BAC\rbrace\rbrace$ & 0\\ \hline
			$C_3$ & $\lbrace\lbrace BCA\rbrace\rbrace$ & 1 \\
			\hline
		\end{tabular}
		\label{tab:abcd3}	
	\end{minipage} 

\noindent The $R$, $Y$ and $D$ matrices are given by
\begin{align}
\begin{split}
&
R=\left(
\begin{array}{ccc}
\frac{1}{2} & 0 & -\frac{1}{2} \\
-\frac{1}{2} & 1 & -\frac{1}{2} \\
-\frac{1}{2} & 0 & \frac{1}{2} \\
\end{array}
\right)
\,,\qquad
Y=\left(
\begin{array}{ccc}
-1 & 0 & 1 \\
-1 & 1 & 0 \\
1 & 0 & 1 \\
\end{array}
\right)
\,,\qquad
D=\D{2} \, .
\end{split}
\end{align}
Finally, the exponentiated colour factors are	
\begin{eqnarray}
(YC)_1 &=&  -i \fabg \fcdg \fcej \ta 1 \tb 2 \tj 2 \td 3 \te 4 +i \fabg \fcdg \febh 
\ta 1 \thh 2 \tc 2 \td 3 \te 4 \, ,\nonumber \\ \nonumber  \\
(YC)_2 &=& -i \fabg \fcdg \fcej \ta 1 \tb 2 \tj 2 \td 3  \te 4  \, .
\end{eqnarray}

\vspace{1cm}


\item[{\bf 4}.] $\textbf W_{4, \rm I}^{(0,2)}(1,1,2,2)$ \\
	
\vspace{-3mm}
This Cweb has two diagrams, one of which is displayed below. The table gives
the chosen order of the two shuffles of the gluon attachments, and the corresponding
$S$ factors.
\begin{minipage}{0.5\textwidth}
		\begin{figure}[H]
		 \vspace{-2mm}
			\includegraphics[height=4cm,width=4cm]{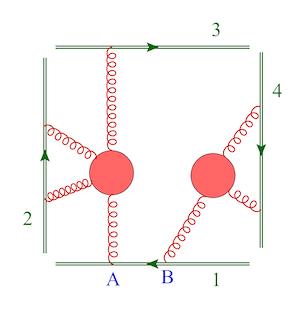}
		\end{figure}
	\end{minipage} 
	\hspace{-2cm}
	\begin{minipage}{0.45\textwidth}
		\begin{tabular}{ | c | c | c |}
			\hline
			\textbf{Diagrams} & \textbf{Sequences} & \textbf{S-factors} \\ \hline
			$C_1$ & $\lbrace\lbrace BA\rbrace\rbrace$ & 1 \\ \hline
			$C_2$ & $\lbrace\lbrace AB\rbrace\rbrace$ & 1\\ \hline
		\end{tabular}
		\label{tab:abcd4}	
	\end{minipage} 

\noindent The $R$, $Y$ and $D$ matrices are given by
\begin{align}
\begin{split}
&
R=\left(
\begin{array}{cc}
\frac{1}{2} & -\frac{1}{2} \\
-\frac{1}{2} & \frac{1}{2} \\
\end{array}
\right)\,, \qquad
Y=\left(
\begin{array}{cc}
-1 & 1 \\
1 & 1 \\
\end{array}
\right)
\,,\qquad
D= \D{1} \, .
\label{bro1}
\end{split}
\end{align}
Finally, the only exponentiated colour factor is	
\begin{eqnarray}
(YC)_1 &=& i \fabc \fdag \fdef \tg 1 \tb 2 \tc 3 \te 4 \tf 4  \, .
\label{bro2}
\end{eqnarray} 

\vspace{2mm}


\item[{\bf 5}.] $\text W_4^{(0,2)}(1,1,1,3)$ \\

\vspace{-3mm}
This is a three-diagram Cweb, represented by

\begin{minipage}{0.5\textwidth}
	\begin{figure}[H]
		\vspace{-2mm}
		\includegraphics[height=4cm,width=4cm]{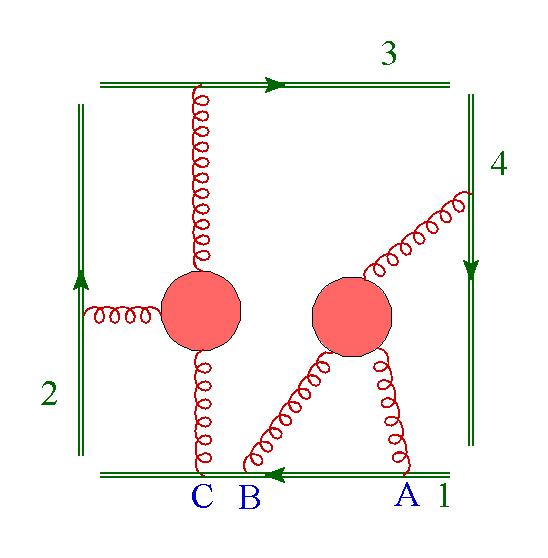}
	\end{figure}
\end{minipage} 
\hspace{-2cm}
\begin{minipage}{0.45\textwidth}
	\begin{tabular}{ | c | c | c |}
		\hline
		\textbf{Diagrams} & \textbf{Sequences} & \textbf{S-factors} \\ \hline
		$C_1$ & $\lbrace\lbrace ABC\rbrace\rbrace$ & 1 \\ \hline
		$C_2$ & $\lbrace\lbrace ACB\rbrace\rbrace$ & 0\\ \hline
			$C_3$ & $\lbrace\lbrace CAB\rbrace\rbrace$ & 1\\ \hline
	\end{tabular}
	\label{tab:abcd5}	
\end{minipage} 

\noindent The $R$, $Y$ and $D$ matrices are given by
\begin{align}
\begin{split}
&
R=\left(
\begin{array}{ccc}
\frac{1}{2} & 0 & -\frac{1}{2} \\
-\frac{1}{2} & 1 & -\frac{1}{2} \\
-\frac{1}{2} & 0 & \frac{1}{2} \\
\end{array}
\right)\,, \qquad
Y=\left(
\begin{array}{ccc}
-1 & 0 & 1 \\
-1 & 1 & 0 \\
1 & 0 & 1 \\
\end{array}
\right)
\,,\qquad
D= \D{2} \, .
\end{split}
\end{align}
Finally, the exponentiated colour factors are	
\begin{eqnarray}
(YC)_1 &=&  i \fabc \fdef \ffaj \te 1 \tj 1 \tb 2 \tc 3 \td 4-i \fabc \faeg \fdef 
\tg 1 \tf 1 \tb 2 \tc 3 \td 4  \, ,\nonumber \\ \nonumber \\
(YC)_2 &=& -i \fabc \faeg \fdef \tg 1 \tf 1 \tb 2 \tc 3 \td 4  \, .
\end{eqnarray}  

\pagebreak


\item[{\bf 6}.] $\text W_4^{(2,1)}(1,1,1,4)$ \\

\vspace{-3mm}
This is a more complicated Cweb, containing twelve diagrams. We find

\vspace{4mm}
\begin{minipage}{0.5\textwidth}
		\begin{figure}[H]
		        \vspace{-6mm}
			\includegraphics[height=4cm,width=4cm]{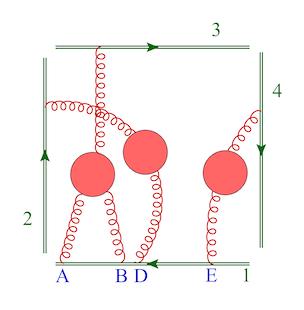}
		\end{figure}
	\end{minipage} 
	\hspace{-1cm}
	\begin{minipage}{0.45\textwidth}
		\begin{tabular}{ | c | c | c |}
			\hline
			\textbf{Diagrams} & \textbf{Sequences} & \textbf{S-factors} \\ \hline
			$C_1$ & $\lbrace\lbrace EDBA\rbrace\rbrace$ & 1 \\ \hline
			$C_2$ & $\lbrace\lbrace DEBA\rbrace\rbrace$ & 1\\ \hline
			$C_3$ & $\lbrace\lbrace EBDA\rbrace\rbrace$ & 0\\ \hline
			$C_4$ & $\lbrace\lbrace BEDA\rbrace\rbrace$ & 0\\ \hline
			$C_5$ & $\lbrace\lbrace DBEA\rbrace\rbrace$ & 0\\ \hline
			$C_6$ & $\lbrace\lbrace BDEA\rbrace\rbrace$ & 0\\ \hline
			$C_7$ & $\lbrace\lbrace EBAD\rbrace\rbrace$ & 1\\ \hline
			$C_8$ & $\lbrace\lbrace BEAD\rbrace\rbrace$ & 0\\ \hline
			$C_9$ & $\lbrace\lbrace BAED\rbrace\rbrace$ & 1\\ \hline
			$C_{10}$ & $\lbrace\lbrace DBAE\rbrace\rbrace$ & 1\\ \hline
			$C_{11}$ & $\lbrace\lbrace BDAE\rbrace\rbrace$ & 0\\ \hline
			$C_{12}$ & $\lbrace\lbrace BADE\rbrace\rbrace$ & 1\\ \hline
		\end{tabular}
		\label{tab:abcd6}	
	\end{minipage}

\noindent The $R$, $Y$ and $D$ matrices are given by
\begin{align}
\begin{split}
&
R=\frac{1}{6} \left(
\begin{array}{cccccccccccc}
2 & -1 & 0 & 0 & 0 & 0 & -1 & 0 & -1 & -1 & 0 & 2 \\
-1 & 2 & 0 & 0 & 0 & 0 & -1 & 0 & 2 & -1 & 0 & -1 \\
-1 & -1 & 3 & 0 & 0 & 0 & -1 & 0 & -1 & 2 & -3 & 2 \\
-1 & 2 & -3 & 6 & -3 & 0 & 2 & -3 & -1 & 2 & -3 & 2 \\
-1 & -1 & 0 & 0 & 3 & 0 & 2 & -3 & 2 & -1 & 0 & -1 \\
2 & -1 & -3 & 0 & -3 & 6 & 2 & -3 & 2 & 2 & -3 & -1 \\
-1 & -1 & 0 & 0 & 0 & 0 & 2 & 0 & -1 & 2 & 0 & -1 \\
-1 & 2 & 0 & 0 & -3 & 0 & -1 & 3 & -1 & 2 & 0 & -1 \\
-1 & 2 & 0 & 0 & 0 & 0 & -1 & 0 & 2 & -1 & 0 & -1 \\
-1 & -1 & 0 & 0 & 0 & 0 & 2 & 0 & -1 & 2 & 0 & -1 \\
2 & -1 & -3 & 0 & 0 & 0 & 2 & 0 & -1 & -1 & 3 & -1 \\
2 & -1 & 0 & 0 & 0 & 0 & -1 & 0 & -1 & -1 & 0 & 2 \\
\end{array}
\right)\,, \qquad \\  \\ & \hspace{8mm}
Y=\left(
\begin{array}{cccccccccccc}
1 & -1 & 0 & 0 & 0 & 0 & 0 & 0 & -1 & 0 & 0 & 1 \\
1 & -1 & -1 & 0 & 0 & 0 & 1 & 0 & -1 & 0 & 1 & 0 \\
0 & -1 & 0 & 0 & 0 & 0 & 1 & 0 & -1 & 1 & 0 & 0 \\
0 & 0 & 1 & -1 & 0 & 0 & -1 & 1 & 0 & 0 & 0 & 0 \\
1 & -1 & 0 & -1 & 0 & 1 & 0 & 0 & 0 & 0 & 0 & 0 \\
0 & -1 & 1 & -1 & 1 & 0 & 0 & 0 & 0 & 0 & 0 & 0 \\
-1 & 0 & 0 & 0 & 0 & 0 & 0 & 0 & 0 & 0 & 0 & 1 \\
0 & 1 & 1 & 0 & 0 & 0 & 0 & 0 & 0 & 0 & 1 & 0 \\
1 & 1 & 0 & 0 & 0 & 0 & 0 & 0 & 0 & 1 & 0 & 0 \\
0 & -1 & 0 & 0 & 0 & 0 & 0 & 0 & 1 & 0 & 0 & 0 \\
1 & 0 & 0 & 0 & 1 & 0 & 0 & 1 & 0 & 0 & 0 & 0 \\
1 & 1 & 0 & 0 & 0 & 0 & 1 & 0 & 0 & 0 & 0 & 0 \\
\end{array}
\right)
\,,\qquad
D= \D{6} \, .
\end{split}
\end{align}
Therefore, there are six exponentiated colour structures, given by
\begin{eqnarray}
(YC)_1 &=& -i\fabc \fbgk \fdeg \ta 1 \tkk 1 \td 2 \tc 3 \te 4-i\fabc \fagl \fdeg \tl 1 \tb 1 \td 2 \tc 3 \te 4 \, ,\nonumber \\ \nonumber \\
(YC)_2 &=& -i\fabc \fbgk \fdeg \ta 1 \tkk 1 \td 2 \tc 3 \te 4 -i\fabc \fbem \fdah \thh 1 \tm 1 \td 2 \tc 3 \te 4 \nonumber \\&& -i\fabc \fdah \fhen \tn 1 \tb 1 \td 2 \tc 3 \te 4 -i\fabc \fagl \fdeg \tl 1 \tb 1 \td 2 \tc 3 \te 4  \, ,\nonumber \\ \nonumber \\
(YC)_3 &=& i\fabc \fbem \fmdv \ta 1 \tv 1 \td 2 \tc 3 \te 4 + i\fabc \faej \fbdu \tj 1 \tu 1 \td 2 \tc 3 \te 4 \nonumber \\&& -i\fabc \fbem  \fdah \thh 1 \tm 1 \td 2 \tc 3 \te 4 -i\fabc \fdah \fhen \tn 1 \tb 1 \td 2 \tc 3 \te 4 \nonumber \\&& -i\fabc \fagl \fdeg \tl 1 \tb 1 \td 2 \tc 3 \te 4  \, ,  \nonumber \\ \nonumber \\
(YC)_4 &=& i\fabc \fbem \fdah \thh 1 \tm 1 \td 2 \tc 3 \te 4  \, ,\nonumber \\ \nonumber \\
(YC)_5 &=& -i \fabc \fbgk \fdeg \ta 1 \tkk 1 \td 2 \tc 3 \te 4 \, ,\nonumber \\ \nonumber \\
(YC)_6 &=& - i\fabc \fbem \fmdv \ta 1 \tv 1 \td 2 \tc 3 \te 4 \, .
\end{eqnarray}  

\vspace{2mm}


\item[{\bf 7}.] $\text W_{4,\, {\rm I}}^{(2,1)}(1,1,2,3)$ \\

\vspace{-3mm}
This is a six-diagram Cweb, the first of three with same correlator and attachment content, 
for which  we find

\vspace{-2mm}
\begin{minipage}{0.5\textwidth}
		\begin{figure}[H]
		\vspace{-2mm}
		\includegraphics[height=4cm,width=4cm]{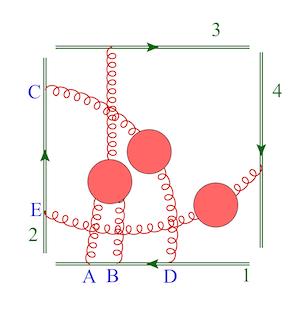}
		\end{figure}
	\end{minipage} 
	\hspace{-2cm}
	\begin{minipage}{0.48\textwidth}
		\begin{tabular}{ | c | c | c |}
			\hline
			\textbf{Diagrams} & \textbf{Sequences} & \textbf{S-factors} \\ \hline
			$C_1$ & $\lbrace \lbrace EC\rbrace,\lbrace DBA\rbrace\rbrace$ & 1 \\ \hline
			$C_2$ & $\lbrace \lbrace CE\rbrace,\lbrace DBA\rbrace\rbrace$ & 2 \\ \hline
			$C_3$ & $\lbrace \lbrace EC\rbrace,\lbrace BDA\rbrace\rbrace$ & 0 \\ \hline
			$C_4$ & $\lbrace \lbrace CE\rbrace,\lbrace BDA\rbrace\rbrace$ & 0 \\ \hline
			$C_5$ & $\lbrace \lbrace EC\rbrace,\lbrace BAD\rbrace\rbrace$ & 2 \\ \hline
			$C_6$ & $\lbrace \lbrace CE\rbrace,\lbrace BAD\rbrace\rbrace$ & 1 \\ \hline
		\end{tabular}
		\label{tab:abcd7}	
	\end{minipage} 

\noindent The $R$, $Y$ and $D$ matrices are given by
\begin{align}
\begin{split}
&
R=\left(
\begin{array}{cccccc}
\frac{1}{3} & -\frac{1}{3} & 0 & 0 & -\frac{1}{3} & \frac{1}{3} \\
-\frac{1}{6} & \frac{1}{6} & 0 & 0 & \frac{1}{6} & -\frac{1}{6} \\
-\frac{1}{6} & \frac{1}{6} & \frac{1}{2} & -\frac{1}{2} & -\frac{1}{3} & \frac{1}{3} \\
\frac{1}{3} & -\frac{1}{3} & -\frac{1}{2} & \frac{1}{2} & \frac{1}{6} & -\frac{1}{6} \\
-\frac{1}{6} & \frac{1}{6} & 0 & 0 & \frac{1}{6} & -\frac{1}{6} \\
\frac{1}{3} & -\frac{1}{3} & 0 & 0 & -\frac{1}{3} & \frac{1}{3} \\
\end{array}
\right) , \quad
Y=\left(
\begin{array}{cccccc}
1 & -1 & 0 & 0 & -1 & 1 \\
1 & -1 & -1 & 1 & 0 & 0 \\
-1 & 0 & 0 & 0 & 0 & 1 \\
\frac{1}{2} & 0 & 0 & 0 & 1 & 0 \\
-\frac{1}{2} & 0 & 1 & 1 & 0 & 0 \\
\frac{1}{2} & 1 & 0 & 0 & 0 & 0 \\
\end{array}
\right) , \quad 
D = \D{2} \, .
\end{split}
\end{align}
This yields two colour structures,
\begin{eqnarray}
(YC)_1 &=&  -i\fabc \fbdh \fdej \ta 1 \thh 1 \tj 2 \tc 3 \te 4 +i \fabc \fdag \fdej 
\tg 1 \tb 1 \tj 2 \tc 3 \te 4 \, ,\nonumber \\ \nonumber \\
(YC)_2 &=& - i\fabg \fbdh \fdej \ta 1 \thh 1 \tj 2 \tc 3 \te 4 \, .
\end{eqnarray} 

\vspace{2mm}


\item[{\bf 8}.] $\text W_{4, \, {\rm II}}^{(2,1)}(1,1,2,3)$ \\

\vspace{-3mm}
Another six-diagram Cweb, the second of three with same correlator and attachment 
content, for which  we find

\vspace{-2mm}
\begin{minipage}{0.5\textwidth}
	\begin{figure}[H]
		\vspace{-2mm}
		\includegraphics[height=4cm,width=4cm]{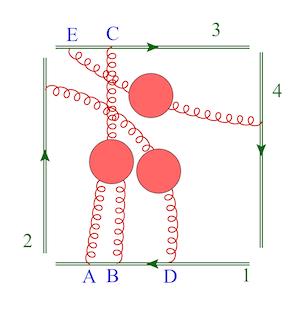}
	\end{figure}
\end{minipage} 
\hspace{-2cm}
\begin{minipage}{0.47\textwidth}
	\begin{tabular}{ | c | c | c |}
		\hline
		\textbf{Diagrams} & \textbf{Sequences} & \textbf{S-factors} \\ \hline
		$C_1$ & $\lbrace \lbrace EC\rbrace,\lbrace DBA\rbrace\rbrace$ & 2 \\ \hline
		$C_2$ & $\lbrace \lbrace CE\rbrace,\lbrace DBA\rbrace\rbrace$ & 1 \\ \hline
		$C_3$ & $\lbrace \lbrace EC\rbrace,\lbrace BDA\rbrace\rbrace$ & 0 \\ \hline
		$C_4$ & $\lbrace \lbrace CE\rbrace,\lbrace BDA\rbrace\rbrace$ & 0 \\ \hline
		$C_5$ & $\lbrace \lbrace EC\rbrace,\lbrace BAD\rbrace\rbrace$ & 1 \\ \hline
		$C_6$ & $\lbrace \lbrace CE\rbrace,\lbrace BAD\rbrace\rbrace$ & 2 \\ \hline
	\end{tabular}
	\label{tab:abcd8}	
\end{minipage} 

\noindent The $R$, $Y$ and $D$ matrices are given by
\begin{align}
\begin{split}
&
R=\left(
\begin{array}{cccccc}
\frac{1}{6} & -\frac{1}{6} & 0 & 0 & -\frac{1}{6} & \frac{1}{6} \\
-\frac{1}{3} & \frac{1}{3} & 0 & 0 & \frac{1}{3} & -\frac{1}{3} \\
-\frac{1}{3} & \frac{1}{3} & \frac{1}{2} & -\frac{1}{2} & -\frac{1}{6} & \frac{1}{6} \\
\frac{1}{6} & -\frac{1}{6} & -\frac{1}{2} & \frac{1}{2} & \frac{1}{3} & -\frac{1}{3} \\
-\frac{1}{3} & \frac{1}{3} & 0 & 0 & \frac{1}{3} & -\frac{1}{3} \\
\frac{1}{6} & -\frac{1}{6} & 0 & 0 & -\frac{1}{6} & \frac{1}{6} \\
\end{array}
\right) , \quad
Y=\left(
\begin{array}{cccccc}
1 & -1 & 0 & 0 & -1 & 1 \\
1 & -1 & -1 & 1 & 0 & 0 \\
-1 & 0 & 0 & 0 & 0 & 1 \\
2 & 0 & 0 & 0 & 1 & 0 \\
1 & 0 & 1 & 1 & 0 & 0 \\
2 & 1 & 0 & 0 & 0 & 0 \\
\end{array}
\right) , \quad 
D= \D{2} \, .
\end{split}
\end{align}
The exponentiated colour factor are
\begin{eqnarray}
(YC)_1 &=& -i \fabc \fbdh \fcej \ta 1 \thh 1 \td 2 \tj 3 \te 4 + i \fabc \fcej \fdag 
\tg 1 \tb 1 \td 2 \tj 3 \te 4 \, , \nonumber \\ \nonumber \\
(YC)_2 &=&  -i \fabc \fbdh \fcej \ta 1 \thh 1 \td 2 \tj 3 \te 4 \, .
\end{eqnarray} 

\vspace{2mm}


\item[{\bf 9}.] $\text W_{4, \, {\rm III}}^{(2,1)} (1,1,2,3)$ \\

\vspace{-3mm}
Yet another six-diagram Cweb, the third of three with same correlator and attachment 
content. We find

\vspace{-2mm}
\begin{minipage}{0.5\textwidth}
	\begin{figure}[H]
		\vspace{-2mm}
		\includegraphics[height=4cm,width=4cm]{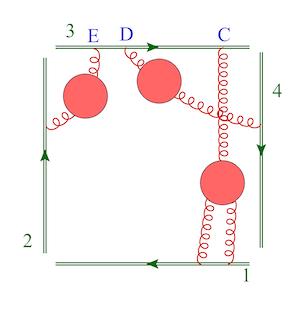}
	\end{figure}
\end{minipage} 
\hspace{-2cm}
\begin{minipage}{0.45\textwidth}
	\begin{tabular}{ | c | c | c |}
		\hline
		\textbf{Diagrams} & \textbf{Sequences} & \textbf{S-factors} \\ \hline
		$C_1$ & $\lbrace \lbrace EDC\rbrace\rbrace$ & 1 \\ \hline
		$C_2$ & $\lbrace \lbrace DEC\rbrace\rbrace$ & 1 \\ \hline
		$C_3$ & $\lbrace \lbrace ECD\rbrace\rbrace$ & 1 \\ \hline
		$C_4$ & $\lbrace \lbrace CED\rbrace\rbrace$ & 1 \\ \hline
		$C_5$ & $\lbrace \lbrace DCE\rbrace\rbrace$ & 1 \\ \hline
		$C_6$ & $\lbrace \lbrace CDE\rbrace\rbrace$ & 1 \\ \hline
	\end{tabular}
	\label{tab:abcd9}	
\end{minipage}

\noindent The $R$, $Y$ and $D$ matrices are given by
\begin{align}
\begin{split}
&
R=\left(
\begin{array}{cccccc}
\frac{1}{3} & -\frac{1}{6} & -\frac{1}{6} & -\frac{1}{6} & -\frac{1}{6} & \frac{1}{3} \\
-\frac{1}{6} & \frac{1}{3} & -\frac{1}{6} & \frac{1}{3} & -\frac{1}{6} & -\frac{1}{6} \\
-\frac{1}{6} & -\frac{1}{6} & \frac{1}{3} & -\frac{1}{6} & \frac{1}{3} & -\frac{1}{6} \\
-\frac{1}{6} & \frac{1}{3} & -\frac{1}{6} & \frac{1}{3} & -\frac{1}{6} & -\frac{1}{6} \\
-\frac{1}{6} & -\frac{1}{6} & \frac{1}{3} & -\frac{1}{6} & \frac{1}{3} & -\frac{1}{6} \\
\frac{1}{3} & -\frac{1}{6} & -\frac{1}{6} & -\frac{1}{6} & -\frac{1}{6} & \frac{1}{3} \\
\end{array}
\right) , \quad
Y=\left(
\begin{array}{cccccc}
1 & -1 & 0 & -1 & 0 & 1 \\
0 & -1 & 1 & -1 & 1 & 0 \\
-1 & 0 & 0 & 0 & 0 & 1 \\
1 & 1 & 0 & 0 & 1 & 0 \\
0 & -1 & 0 & 1 & 0 & 0 \\
1 & 1 & 1 & 0 & 0 & 0 \\
\end{array}
\right) , \quad 
D=\D{2} \, ,
\end{split}
\end{align}
which leads to the colour factors
\begin{eqnarray}
(YC)_1 &=& -i\fabc \fcgh \fdeg \ta 1 \tb 1 \te 2 \thh 3 \td 4 \, , \nonumber \\ 
\nonumber \\
(YC)_2 &=&  - i\fabc \fcej \fdjk \ta 1 \tb 1 \te 2 \tkk 3 \td 4 \, .
\end{eqnarray} 

\vspace{2mm}


\item[{\bf 10}.] $\text W_{4 ,\, {\rm I}}^{(2,1)}(1,2,2,2)$ \\

\vspace{-2mm}
This is the first of five Cwebs  with the same correlator and attachment content. 
Its four diagrams are

\vspace{-2mm}
\begin{minipage}{0.5\textwidth}
	\begin{figure}[H]
		\vspace{-2mm}
		\includegraphics[height=4cm,width=4cm]{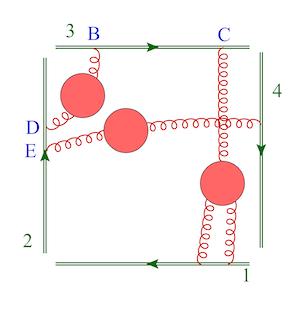}
	\end{figure}
\end{minipage} 
\hspace{-2cm}
\begin{minipage}{0.46\textwidth}
	\begin{tabular}{ | c | c | c |}
		\hline
		\textbf{Diagrams} & \textbf{Sequences} & \textbf{S-factors} \\ \hline
		$C_1$ & $\lbrace \lbrace ED\rbrace,\lbrace BC \rbrace\rbrace$ & 1 \\ 
		\hline
			$C_2$ & $\lbrace \lbrace ED\rbrace,\lbrace CB \rbrace\rbrace$ & 2 \\ 
		\hline
			$C_3$ & $\lbrace \lbrace DE\rbrace,\lbrace BC \rbrace\rbrace$ & 2 \\ 
		\hline
			$C_4$ & $\lbrace \lbrace DE\rbrace,\lbrace CB \rbrace\rbrace$ & 1 \\ 
		\hline
	 \end{tabular}
	\label{tab:abcd10}	
\end{minipage}

\noindent The $R$, $Y$ and $D$ matrices are given by
\begin{align}
\begin{split}
&
R=\left(
\begin{array}{cccc}
\frac{1}{3} & -\frac{1}{3} & -\frac{1}{3} & \frac{1}{3} \\
-\frac{1}{6} & \frac{1}{6} & \frac{1}{6} & -\frac{1}{6} \\
-\frac{1}{6} & \frac{1}{6} & \frac{1}{6} & -\frac{1}{6} \\
\frac{1}{3} & -\frac{1}{3} & -\frac{1}{3} & \frac{1}{3} \\
\end{array}
\right)\,, \qquad 
Y=\left(
\begin{array}{cccc}
1 & -1 & -1 & 1 \\
-1 & 0 & 0 & 1 \\
\frac{1}{2} & 0 & 1 & 0 \\
\frac{1}{2} & 1 & 0 & 0 \\
\end{array}
\right)
\,,\qquad
D= \D{1}
\end{split}
\end{align}
There is therefore only one colour structure,
\begin{eqnarray}
(YC)_1 &=& -i\fabc f^{cdh} \fdeg \ta 1 \tb 1 \tg 2 \thh 3 \te 4 \, .
\end{eqnarray} 

\vspace{2mm}


\item[{\bf 11}.] $\text W_{4 ,\, {\rm II}}^{(2,1)}(1,2,2,2)$ \\

\vspace{-2mm}
This is the second of five Cwebs  with the same correlator and attachment content. 
It has eight diagrams, which can be organised as follows.

\vspace{2mm}
\begin{minipage}{0.5\textwidth}
	\begin{figure}[H]
		\vspace{-2mm}
		\includegraphics[height=4cm,width=4cm]{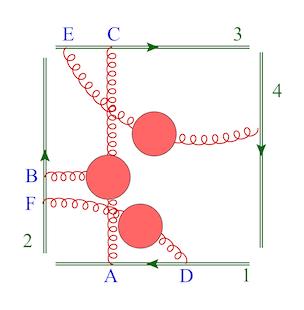}
	\end{figure}
\end{minipage} 
\hspace{-2cm}
\begin{minipage}{0.53\textwidth}
	\begin{tabular}{ | c | c | c |}
		\hline
		\textbf{Diagrams} & \textbf{Sequences} & \textbf{S-factors} \\ \hline
		$C_1$ & $\lbrace \lbrace DA\rbrace,\lbrace FB \rbrace, \lbrace EC\rbrace \rbrace$ & 2 \\ 
		\hline
		$C_2$ & $\lbrace \lbrace DA\rbrace,\lbrace FB \rbrace, \lbrace CE\rbrace \rbrace$ & 1 \\ 
		\hline
		$C_3$ & $\lbrace \lbrace DA\rbrace,\lbrace BF \rbrace, \lbrace EC\rbrace \rbrace$ & 0 \\ 
		\hline
		$C_4$ & $\lbrace \lbrace DA\rbrace,\lbrace BF \rbrace, \lbrace CE\rbrace \rbrace$ & 0 \\ 
		\hline
		$C_5$ & $\lbrace \lbrace AD\rbrace,\lbrace FB \rbrace, \lbrace EC\rbrace \rbrace$ & 0 \\ 
		\hline
		$C_6$ & $\lbrace \lbrace AD\rbrace,\lbrace FB \rbrace, \lbrace CE\rbrace \rbrace$ & 0 \\ 
		\hline
		$C_7$ & $\lbrace \lbrace AD\rbrace,\lbrace BF \rbrace, \lbrace EC\rbrace \rbrace$ & 1 \\ 
		\hline
		$C_8$ & $\lbrace \lbrace AD\rbrace,\lbrace BF \rbrace, \lbrace CE\rbrace \rbrace$ & 2 \\ 
		\hline
	\end{tabular}
	\label{tab:abcd11}	
\end{minipage} \\ 

\noindent The $R$, $Y$ and $D$ matrices are given by
\begin{align}
\begin{split}
&
R=\left(
\begin{array}{cccccccc}
\frac{1}{6} & -\frac{1}{6} & 0 & 0 & 0 & 0 & -\frac{1}{6} & \frac{1}{6} \\
-\frac{1}{3} & \frac{1}{3} & 0 & 0 & 0 & 0 & \frac{1}{3} & -\frac{1}{3} \\
-\frac{1}{3} & \frac{1}{3} & \frac{1}{2} & -\frac{1}{2} & 0 & 0 & -\frac{1}{6} & \frac{1}{6} \\
\frac{1}{6} & -\frac{1}{6} & -\frac{1}{2} & \frac{1}{2} & 0 & 0 & \frac{1}{3} & -\frac{1}{3} \\
-\frac{1}{3} & \frac{1}{3} & 0 & 0 & \frac{1}{2} & -\frac{1}{2} & -\frac{1}{6} & \frac{1}{6} \\
\frac{1}{6} & -\frac{1}{6} & 0 & 0 & -\frac{1}{2} & \frac{1}{2} & \frac{1}{3} & -\frac{1}{3} \\
-\frac{1}{3} & \frac{1}{3} & 0 & 0 & 0 & 0 & \frac{1}{3} & -\frac{1}{3} \\
\frac{1}{6} & -\frac{1}{6} & 0 & 0 & 0 & 0 & -\frac{1}{6} & \frac{1}{6} \\
\end{array}
\right) , \qquad
Y=\left(
\begin{array}{cccccccc}
1 & -1 & 0 & 0 & 0 & 0 & -1 & 1 \\
1 & -1 & 0 & 0 & -1 & 1 & 0 & 0 \\
1 & -1 & -1 & 1 & 0 & 0 & 0 & 0 \\
-1 & 0 & 0 & 0 & 0 & 0 & 0 & 1 \\
2 & 0 & 0 & 0 & 0 & 0 & 1 & 0 \\
1 & 0 & 0 & 0 & 1 & 1 & 0 & 0 \\
1 & 0 & 1 & 1 & 0 & 0 & 0 & 0 \\
2 & 1 & 0 & 0 & 0 & 0 & 0 & 0 \\
\end{array}
\right) , \qquad \\ \\ &
D= \D{3} \, .
\end{split}
\end{align}
The three colour factors are
\begin{eqnarray}
(YC)_1 &=& i \fabc \fceg \fdbj \td 1 \ta 1 \tj 2 \tg 3 \te 4-i \fabc \fadh \fceg 
\thh 1 \tb 2 \td 2 \tg 3 \te 4  \, , \nonumber \\ \nonumber \\
(YC)_2 &=&-i \fabc \fadh \fceg \thh 1 \tb 2 \td 2 \tg 3 \te 4 \, , \nonumber \\ \nonumber \\
(YC)_3 &=& i \fabc \fceg \fdbj \td 1 \ta 1 \tj 2 \tg 3 \te 4 -\fabc \fadh \fceg 
\fdbj \thh 1 \tj 2 \tg 3 \te 4 \, .
\end{eqnarray} 

\vspace{2mm}


\item[{\bf 12}.] $\text W_{4, \, \rm III}^{(2,1)}(1,2,2,2)$ \\

\vspace{-2mm}
The third of five Cwebs  with the same correlator and attachment content, also
has eight diagrams.

\begin{minipage}{0.5\textwidth}
	\begin{figure}[H]
		\vspace{-2mm}
		\includegraphics[height=4cm,width=4cm]{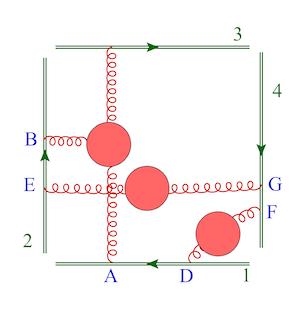}
	\end{figure}
\end{minipage} 
\hspace{-2cm}
\begin{minipage}{0.53\textwidth}
	\begin{tabular}{ | c | c | c |}
		\hline
		\textbf{Diagrams} & \textbf{Sequences} & \textbf{S-factors} \\ \hline
		$C_1$ & $\lbrace \lbrace DA\rbrace,\lbrace EB \rbrace, \lbrace GF\rbrace \rbrace$ & 1 \\ 
		\hline
		$C_2$ & $\lbrace \lbrace DA\rbrace,\lbrace EB \rbrace, \lbrace FG\rbrace \rbrace$ & 1 \\ 
		\hline
		$C_3$ & $\lbrace \lbrace DA\rbrace,\lbrace BE \rbrace, \lbrace GF\rbrace \rbrace$ & 0 \\ 
		\hline
		$C_4$ & $\lbrace \lbrace DA\rbrace,\lbrace BE \rbrace, \lbrace FG\rbrace \rbrace$ & 1 \\ 
		\hline
		$C_5$ & $\lbrace \lbrace AD\rbrace,\lbrace EB \rbrace, \lbrace GF\rbrace \rbrace$ & 1 \\ 
		\hline
		$C_6$ & $\lbrace \lbrace AD\rbrace,\lbrace EB \rbrace, \lbrace FG\rbrace \rbrace$ & 0 \\ 
		\hline
		$C_7$ & $\lbrace \lbrace AD\rbrace,\lbrace BE \rbrace, \lbrace GF\rbrace \rbrace$ & 1 \\ 
		\hline
		$C_8$ & $\lbrace \lbrace AD\rbrace,\lbrace BE \rbrace, \lbrace FG\rbrace \rbrace$ & 1 \\ 
		\hline
	\end{tabular}
	\label{tab:abcd12}	
\end{minipage}

\vspace{2mm}
\noindent The $R$, $Y$ and $D$ matrices are given by
\begin{align}
\begin{split}
&
R=\left(
\begin{array}{cccccccc}
\frac{1}{3} & -\frac{1}{6} & 0 & -\frac{1}{6} & -\frac{1}{6} & 0 & -\frac{1}{6} & \frac{1}{3} \\
-\frac{1}{6} & \frac{1}{3} & 0 & -\frac{1}{6} & -\frac{1}{6} & 0 & \frac{1}{3} & -\frac{1}{6} \\
-\frac{2}{3} & \frac{1}{3} & 1 & -\frac{2}{3} & \frac{1}{3} & 0 & -\frac{2}{3} & \frac{1}{3} \\
-\frac{1}{6} & -\frac{1}{6} & 0 & \frac{1}{3} & \frac{1}{3} & 0 & -\frac{1}{6} & -\frac{1}{6} \\
-\frac{1}{6} & -\frac{1}{6} & 0 & \frac{1}{3} & \frac{1}{3} & 0 & -\frac{1}{6} & -\frac{1}{6} \\
\frac{1}{3} & -\frac{2}{3} & 0 & \frac{1}{3} & -\frac{2}{3} & 1 & \frac{1}{3} & -\frac{2}{3} \\
-\frac{1}{6} & \frac{1}{3} & 0 & -\frac{1}{6} & -\frac{1}{6} & 0 & \frac{1}{3} & -\frac{1}{6} \\
\frac{1}{3} & -\frac{1}{6} & 0 & -\frac{1}{6} & -\frac{1}{6} & 0 & -\frac{1}{6} & \frac{1}{3} \\
\end{array}
\right) , \qquad
Y=\left(
\begin{array}{cccccccc}
2 & -1 & -1 & 0 & -1 & 0 & 0 & 1 \\
1 & 0 & -1 & 0 & -1 & 0 & 1 & 0 \\
1 & -1 & 0 & 0 & -1 & 1 & 0 & 0 \\
1 & -1 & -1 & 1 & 0 & 0 & 0 & 0 \\
-1 & 0 & 0 & 0 & 0 & 0 & 0 & 1 \\
0 & -1 & 0 & 0 & 0 & 0 & 1 & 0 \\
1 & 1 & 0 & 0 & 1 & 0 & 0 & 0 \\
1 & 1 & 0 & 1 & 0 & 0 & 0 & 0 \\
\end{array}
\right) , \qquad \\ \\ &
D=\D{4} \, .
\end{split}
\end{align}
The four colour structures are 
\begin{eqnarray}
(YC)_1 &=& -i\fabc \febh \fedk  \ta 1 \td 1 \thh 2 \tc 3 \tkk 4-i\fabc \fdag \fedk 
\tg 1 \tb 2 \te 2 \tc 3 \tkk 4 \nonumber \\ && - i \fabc \fdag \febh 
\tg 1 \thh 2 \tc 3 \te 4 \td 4 \, ,\nonumber \\ \nonumber \\
(YC)_2 &=&-i\fabc \fdag \febh \tg 1 \thh 2 \tc 3 \td 4 \te 4 \, ,  \nonumber \\ \nonumber \\
(YC)_3 &=&-i\fabc \fdag \fedk \tg 1 \tb 2 \te 2 \tc 3 \tkk 4  \, ,  \nonumber \\ \nonumber \\
(YC)_4 &=&-i\fabc \febh \fedk  \ta 1 \td 1 \thh 2 \tc 3 \tkk 4 \, .
\end{eqnarray} 

\vspace{2mm}


\item[{\bf 13}.] $\text W_{4 , \, {\rm IV}}^{(2,1)}(1,1,2,3)$\\

\vspace{-2mm}
The fourth Cweb of this set has twelve diagrams.

\vspace{3mm}
\begin{minipage}{0.5\textwidth}
	\begin{figure}[H]
		\vspace{-2mm}
		\includegraphics[height=4cm,width=4cm]{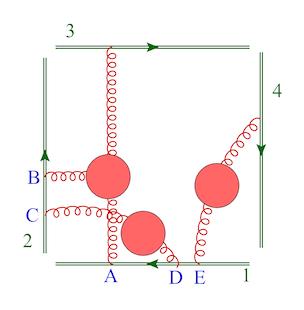}
	\end{figure}
\end{minipage} 
\hspace{-2cm}
\begin{minipage}{0.48\textwidth}
	\begin{tabular}{ | c | c | c |}
		\hline
		\textbf{Diagrams} & \textbf{Sequences} & \textbf{S-factors} \\ \hline
		$C_1$ & $\lbrace \lbrace CB\rbrace,\lbrace EDA \rbrace\rbrace$ & 1 \\ 
		\hline
		$C_2$ & $\lbrace \lbrace BC\rbrace,\lbrace EDA \rbrace\rbrace$ & 0 \\ 
		\hline
		$C_3$ & $\lbrace \lbrace CB\rbrace,\lbrace DEA \rbrace\rbrace$ & 1 \\ 
		\hline
		$C_4$ & $\lbrace \lbrace BC\rbrace,\lbrace DEA \rbrace\rbrace$ & 0 \\ 
		\hline
		$C_5$ & $\lbrace \lbrace CB\rbrace,\lbrace EAD \rbrace\rbrace$ & 0 \\ 
		\hline
		$C_6$ & $\lbrace \lbrace BC\rbrace,\lbrace EAD \rbrace\rbrace$ & 1 \\ 
		\hline
		$C_7$ & $\lbrace \lbrace CB\rbrace,\lbrace AED \rbrace\rbrace$ & 0 \\ 
		\hline
		$C_8$ & $\lbrace \lbrace BC\rbrace,\lbrace AED \rbrace\rbrace$ & 1 \\ 
		\hline
		$C_9$ & $\lbrace \lbrace CB\rbrace,\lbrace DAE \rbrace\rbrace$ & 1 \\ 
		\hline
		$C_{10}$ & $\lbrace \lbrace BC\rbrace,\lbrace DAE \rbrace\rbrace$ & 0 \\ 
		\hline
		$C_{11}$ & $\lbrace \lbrace CB\rbrace,\lbrace ADE \rbrace\rbrace$ & 0 \\
		\hline
		$C_{12}$ & $\lbrace \lbrace BC\rbrace,\lbrace ADE \rbrace\rbrace$ & 1 \\ 
		\hline 
	\end{tabular}
	\label{tab:abcd13}	
\end{minipage}

\vspace{2mm}
\noindent The $R$, $Y$ and $D$ matrices are given by
\begin{align}
\begin{split}
&
R=\frac{1}{6} \left(
\begin{array}{cccccccccccc}
2 & 0 & -1 & 0 & 0 & -1 & 0 & -1 & -1 & 0 & 0 & 2 \\
-1 & 3 & -1 & 0 & 0 & -1 & 0 & -1 & 2 & -3 & 0 & 2 \\
-1 & 0 & 2 & 0 & 0 & -1 & 0 & 2 & -1 & 0 & 0 & -1 \\
2 & -3 & -4 & 6 & 0 & -1 & 0 & 2 & 2 & -3 & 0 & -1 \\
-1 & 0 & -1 & 0 & 3 & -1 & 0 & -1 & 2 & 0 & -3 & 2 \\
-1 & 0 & -1 & 0 & 0 & 2 & 0 & -1 & 2 & 0 & 0 & -1 \\
-1 & 0 & 2 & 0 & -3 & 2 & 6 & -4 & -1 & 0 & -3 & 2 \\
-1 & 0 & 2 & 0 & 0 & -1 & 0 & 2 & -1 & 0 & 0 & -1 \\
-1 & 0 & -1 & 0 & 0 & 2 & 0 & -1 & 2 & 0 & 0 & -1 \\
2 & -3 & -1 & 0 & 0 & 2 & 0 & -1 & -1 & 3 & 0 & -1 \\
2 & 0 & -1 & 0 & -3 & 2 & 0 & -1 & -1 & 0 & 3 & -1 \\
2 & 0 & -1 & 0 & 0 & -1 & 0 & -1 & -1 & 0 & 0 & 2 \\
\end{array}
\right) \, , 
\end{split}
\end{align}
\begin{align}
\begin{split}
\hspace{1cm} Y=\left(
\begin{array}{cccccccccccc}
1 & 0 & -1 & 0 & 1 & -1 & -1 & 0 & 0 & 0 & 0 & 1 \\
1 & 0 & -1 & 0 & 0 & 0 & -1 & 0 & 0 & 0 & 1 & 0 \\
1 & -1 & -1 & 0 & 1 & 0 & -1 & 0 & 0 & 1 & 0 & 0 \\
0 & 0 & -1 & 0 & 1 & 0 & -1 & 0 & 1 & 0 & 0 & 0 \\
0 & 0 & 0 & 0 & 1 & -1 & -1 & 1 & 0 & 0 & 0 & 0 \\
1 & -1 & -1 & 1 & 0 & 0 & 0 & 0 & 0 & 0 & 0 & 0 \\
-1 & 0 & 0 & 0 & 0 & 0 & 0 & 0 & 0 & 0 & 0 & 1 \\
0 & 0 & 1 & 0 & 1 & 0 & 0 & 0 & 0 & 0 & 1 & 0 \\
0 & 1 & 1 & 0 & 0 & 0 & 0 & 0 & 0 & 1 & 0 & 0 \\
1 & 0 & 1 & 0 & 0 & 0 & 0 & 0 & 1 & 0 & 0 & 0 \\
0 & 0 & -1 & 0 & 0 & 0 & 0 & 1 & 0 & 0 & 0 & 0 \\
1 & 0 & 1 & 0 & 0 & 1 & 0 & 0 & 0 & 0 & 0 & 0 \\
\end{array}
\right)
\,,\qquad
D=\D{6}  \, .
\nonumber
\end{split}
\end{align}
There are thus six colour structures
\begin{eqnarray}
(YC)_1 &=& -i\fabc \fbdk \fdeg \ta 1 \tg 1 \tkk 2 \tc 3 \te 4 - i\fabc \faeh \fbdk 
\td 1 \thh 1 \tkk 2 \tc 3 \te 4 \nonumber \\ && +i\fabc \fdeg \fgal 
\tl 1 \td 2 \tb 2 \tc 3 \te 4  \, , \nonumber \\ \nonumber \\
(YC)_2 &=& i\fabc \fdeg \fgal  \tl 1 \tb 2 \td 2 \tc 3 \te 4 \, , \nonumber \\ \nonumber \\
(YC)_3 &=& -i\fabc \fbdk \fdeg \ta 1 \tg 1 \tkk 2 \tc 3 \te 4 -i\fabc \faeh \fbdk 
\td 1 \thh 1 \tkk 2 \tc 3 \te 4 \nonumber \\ && + i\fabc \faeh \fhdm 
\tm 1 \td 2 \tb 2 \tc 3 \te 4 \, ,  \nonumber \\ \nonumber \\
(YC)_4 &=& i\fabc \faeh \fhdm \tm 1 \tb 2 \td 2 \tc 3 \te 4 \, , \nonumber \\ \nonumber \\
(YC)_5 &=& -i\fabc \faeh \fbdk \td 1 \thh 1 \tkk 2 \tc 3 \te 4 \, ,\nonumber \\ \nonumber \\
(YC)_6 &=&  - i\fabc \fbdk \fdeg \ta 1 \tg 1 \tkk 2 \tc 3 \te 4 \, .
\end{eqnarray} 

\vspace{2mm}


\item[{\bf 14}.] $\text  W_{4 ,\, {\rm V}}^{(2,1)}(1,1,2,3)$ \\

\vspace{-2mm}
The last Cweb of this set has six diagrams.

\begin{minipage}{0.5\textwidth}
		\begin{figure}[H]
		\vspace{-2mm}
		\includegraphics[height=4cm,width=4cm]{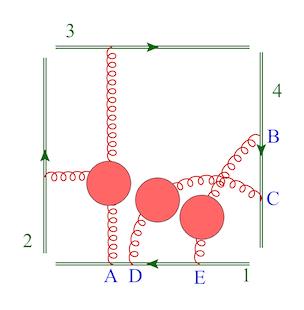}
		\end{figure}
	\end{minipage} 
	\hspace{-2cm}
	\begin{minipage}{0.48\textwidth}
		\begin{tabular}{ | c | c | c |}
			\hline
			\textbf{Diagrams} & \textbf{Sequences} & \textbf{S-factors} \\ \hline
			$C_1$ & $\lbrace \lbrace BC\rbrace,\lbrace EDA \rbrace\rbrace$ & 1 \\ 
			\hline
			$C_2$ & $\lbrace \lbrace CB\rbrace,\lbrace EDA \rbrace\rbrace$ & 0 \\ 
			\hline
			$C_3$ & $\lbrace \lbrace BC\rbrace,\lbrace EAD \rbrace\rbrace$ & 1 \\ 
			\hline
			$C_4$ & $\lbrace \lbrace CB\rbrace,\lbrace EAD \rbrace\rbrace$ & 0 \\ 
			\hline
			$C_5$ & $\lbrace \lbrace BC\rbrace,\lbrace AED \rbrace\rbrace$ & 1 \\ 
			\hline
			$C_6$ & $\lbrace \lbrace CB\rbrace,\lbrace AED \rbrace\rbrace$ & 0 \\ 
			\hline
		\end{tabular}
		\label{tab:abcd14}	
	\end{minipage}

\noindent The $R$, $Y$ and $D$ matrices are given by
\begin{align}
\begin{split}
&
R=\left(
\begin{array}{cccccc}
\frac{1}{6} & 0 & -\frac{1}{3} & 0 & \frac{1}{6} & 0 \\
-\frac{1}{3} & \frac{1}{2} & -\frac{1}{3} & 0 & \frac{2}{3} & -\frac{1}{2} \\
-\frac{1}{3} & 0 & \frac{2}{3} & 0 & -\frac{1}{3} & 0 \\
\frac{1}{6} & -\frac{1}{2} & -\frac{1}{3} & 1 & \frac{1}{6} & -\frac{1}{2} \\
\frac{1}{6} & 0 & -\frac{1}{3} & 0 & \frac{1}{6} & 0 \\
\frac{2}{3} & -\frac{1}{2} & -\frac{1}{3} & 0 & -\frac{1}{3} & \frac{1}{2} \\
\end{array}
\right), \quad
Y=\left(
\begin{array}{cccccc}
2 & -1 & -2 & 0 & 0 & 1 \\
1 & 0 & -2 & 0 & 1 & 0 \\
1 & -1 & -1 & 1 & 0 & 0 \\
-2 & 1 & 0 & 0 & 0 & 1 \\
-1 & 0 & 0 & 0 & 1 & 0 \\
2 & 0 & 1 & 0 & 0 & 0 \\
\end{array}
\right) , \quad 
D= \D{3} 
\end{split}
\end{align}
The three colour structures are
\begin{eqnarray}
(YC)_1 &=& -i \fabc f^{geh} f^{adg} \thh 1 \tb 2 \tc 3 \td 4 \te 4 \, ,\nonumber \\ \nonumber \\
(YC)_2 &=&- i \fabc \fdeg \faeh \td 1 \thh 1 \tb 2 \tc 3 \tg 4  \, ,\nonumber \\ \nonumber \\
(YC)_3 &=& -i \fabc f^{adg} f^{deh} \tg 1 \te 1 \tb 2 \tc 3 \thh 4 \, .
\end{eqnarray} 

\vspace{2mm}


\item[{\bf 15}.] $\text W_4^{(4)}(2,2,2,2)$ \\

\vspace{-2mm}
This Cweb, comprising sixteen diagrams, generalises a memebr of an infinite 
series of highly symmetrical `Multiple Gluon Exchange Webs', studied in 
Ref.~\cite{Falcioni:2014pka}. Explicit calculations confirm the structure of the 
mixing matrix  predicted there. We find

\vspace{5mm}
\begin{minipage}{0.5\textwidth}
		\begin{figure}[H]
		\vspace{-2mm}
		\includegraphics[height=4cm,width=4cm]{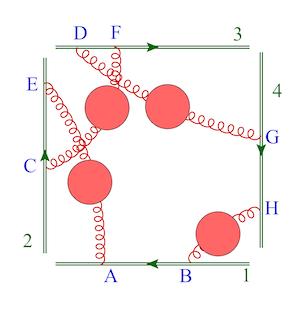}
		\end{figure}
	\end{minipage} 
	\hspace{-3.0cm}
	\begin{minipage}{0.60\textwidth}
		\begin{tabular}{ | c | c | c |}
			\hline
			\textbf{Diagrams} & \textbf{Sequences} & \textbf{S-factors} \\ \hline
			$C_1$ & $\lbrace \lbrace BA\rbrace,\lbrace CE \rbrace, \lbrace DF\rbrace, \lbrace GH\rbrace \rbrace$ & 2 \\
			\hline
			$C_2$ & $\lbrace \lbrace BA\rbrace,\lbrace CE \rbrace, \lbrace DF\rbrace, \lbrace HG\rbrace \rbrace$ & 1 \\ 
			\hline
			$C_3$ & $\lbrace \lbrace BA\rbrace,\lbrace CE \rbrace, \lbrace FD\rbrace, \lbrace GH\rbrace \rbrace$ & 1 \\
			\hline
			$C_4$ & $\lbrace \lbrace BA\rbrace,\lbrace CE \rbrace, \lbrace FD\rbrace, \lbrace HG\rbrace \rbrace$ & 4 \\ 
			\hline 
			$C_5$ & $\lbrace \lbrace BA\rbrace,\lbrace EC \rbrace, \lbrace DF\rbrace, \lbrace GH\rbrace \rbrace$ & 1 \\
			\hline
			$C_6$ & $\lbrace \lbrace BA\rbrace,\lbrace EC \rbrace, \lbrace DF\rbrace, \lbrace HG\rbrace \rbrace$ & 2 \\ 
			\hline
			$C_7$ & $\lbrace \lbrace BA\rbrace,\lbrace EC \rbrace, \lbrace FD\rbrace, \lbrace GH\rbrace \rbrace$ & 0 \\ 
			\hline 
			$C_8$ & $\lbrace \lbrace BA\rbrace,\lbrace EC \rbrace, \lbrace FD\rbrace, \lbrace HG\rbrace \rbrace$ & 1 \\ 
			\hline
			$C_9$ & $\lbrace \lbrace AB\rbrace,\lbrace CE \rbrace, \lbrace DF\rbrace, \lbrace GH\rbrace \rbrace$ & 1 \\ 
			\hline
			$C_{10}$ & $\lbrace \lbrace AB\rbrace,\lbrace CE \rbrace, \lbrace DF\rbrace, \lbrace HG\rbrace \rbrace$ & 0 \\ 
			\hline
			$C_{11}$ & $\lbrace \lbrace AB\rbrace,\lbrace CE \rbrace, \lbrace FD\rbrace, \lbrace GH\rbrace \rbrace$ & 2 \\ 
			\hline
			$C_{12}$ & $\lbrace \lbrace AB\rbrace,\lbrace CE \rbrace, \lbrace FD\rbrace, \lbrace HG\rbrace \rbrace$ & 1 \\ 
			\hline
			$C_{13}$ & $\lbrace \lbrace AB\rbrace,\lbrace EC \rbrace, \lbrace DF\rbrace, \lbrace GH\rbrace \rbrace$ & 4 \\
			\hline 
			$C_{14}$ & $\lbrace \lbrace AB\rbrace,\lbrace EC \rbrace, \lbrace DF\rbrace, \lbrace HG\rbrace \rbrace$ & 1 \\ 
			\hline
			$C_{15}$ & $\lbrace \lbrace AB\rbrace,\lbrace EC \rbrace, \lbrace FD\rbrace, \lbrace GH\rbrace \rbrace$ & 1 \\
			\hline 
			$C_{16}$ & $\lbrace \lbrace AB\rbrace,\lbrace EC \rbrace, \lbrace FD\rbrace, \lbrace HG\rbrace \rbrace$ & 2 \\
			\hline
			\end{tabular}
		\label{tab:abcd15}	
	\end{minipage}

\noindent The $R$, $Y$ and $D$ matrices are given by
\begin{align}
\begin{split}
&
R=\frac{1}{12} \left(
\begin{array}{cccccccccccccccc}
2 & -1 & -1 & 0 & -1 & 0 & 0 & 1 & -1 & 0 & 0 & 1 & 0 & 1 & 1 & -2 \\
-2 & 3 & 1 & -2 & 1 & -2 & 0 & 1 & -1 & 0 & 2 & -1 & 2 & -1 & -3 & 2 \\
-2 & 1 & 3 & -2 & -1 & 2 & 0 & -1 & 1 & 0 & -2 & 1 & 2 & -3 & -1 & 2 \\
0 & -1 & -1 & 2 & 1 & 0 & 0 & -1 & 1 & 0 & 0 & -1 & -2 & 1 & 1 & 0 \\
-2 & 1 & -1 & 2 & 3 & -2 & 0 & -1 & 1 & 0 & 2 & -3 & -2 & 1 & -1 & 2 \\
0 & -1 & 1 & 0 & -1 & 2 & 0 & -1 & 1 & 0 & -2 & 1 & 0 & -1 & 1 & 0 \\
6 & -3 & -9 & 6 & -9 & 6 & 12 & -9 & -3 & 0 & 6 & -3 & 6 & -3 & -9 & 6 \\
2 & 1 & -1 & -2 & -1 & -2 & 0 & 3 & -3 & 0 & 2 & 1 & 2 & 1 & -1 & -2 \\
-2 & -1 & 1 & 2 & 1 & 2 & 0 & -3 & 3 & 0 & -2 & -1 & -2 & -1 & 1 & 2 \\
6 & -9 & -3 & 6 & -3 & 6 & 0 & -3 & -9 & 12 & 6 & -9 & 6 & -9 & -3 & 6 \\
0 & 1 & -1 & 0 & 1 & -2 & 0 & 1 & -1 & 0 & 2 & -1 & 0 & 1 & -1 & 0 \\
2 & -1 & 1 & -2 & -3 & 2 & 0 & 1 & -1 & 0 & -2 & 3 & 2 & -1 & 1 & -2 \\
0 & 1 & 1 & -2 & -1 & 0 & 0 & 1 & -1 & 0 & 0 & 1 & 2 & -1 & -1 & 0 \\
2 & -1 & -3 & 2 & 1 & -2 & 0 & 1 & -1 & 0 & 2 & -1 & -2 & 3 & 1 & -2 \\
2 & -3 & -1 & 2 & -1 & 2 & 0 & -1 & 1 & 0 & -2 & 1 & -2 & 1 & 3 & -2 \\
-2 & 1 & 1 & 0 & 1 & 0 & 0 & -1 & 1 & 0 & 0 & -1 & 0 & -1 & -1 & 2 \\
\end{array}
\right)  \, , \qquad \\ \\ &
\hspace{8mm} 
Y=\left(
\begin{array}{cccccccccccccccc}
-3 & 2 & 2 & -1 & 2 & -1 & -1 & 0 & 2 & -1 & -1 & 0 & -1 & 0 & 0 & 1 \\
-1 & 0 & 1 & 0 & 1 & 0 & -1 & 0 & 1 & 0 & -1 & 0 & -1 & 0 & 1 & 0 \\
-1 & 1 & 0 & 0 & 1 & -1 & 0 & 0 & 1 & -1 & 0 & 0 & -1 & 1 & 0 & 0 \\
-1 & 1 & 1 & -1 & 0 & 0 & 0 & 0 & 1 & -1 & -1 & 1 & 0 & 0 & 0 & 0 \\
-1 & 1 & 1 & -1 & 1 & -1 & -1 & 1 & 0 & 0 & 0 & 0 & 0 & 0 & 0 & 0 \\
1 & 0 & 0 & 0 & 0 & 0 & 0 & 0 & 0 & 0 & 0 & 0 & 0 & 0 & 0 & 1 \\
0 & 1 & 0 & 0 & 0 & 0 & 0 & 0 & 0 & 0 & 0 & 0 & 0 & 0 & 1 & 0 \\
0 & 0 & 1 & 0 & 0 & 0 & 0 & 0 & 0 & 0 & 0 & 0 & 0 & 1 & 0 & 0 \\
-1 & -\frac{1}{2} & -\frac{1}{2} & 0 & 0 & 0 & 0 & 0 & 0 & 0 & 0 & 0 & 1 & 0 & 0 & 0 \\
-2 & 0 & -1 & 0 & 0 & 0 & 0 & 0 & 0 & 0 & 0 & 1 & 0 & 0 & 0 & 0 \\
0 & -\frac{1}{2} & \frac{1}{2} & 0 & 0 & 0 & 0 & 0 & 0 & 0 & 1 & 0 & 0 & 0 & 0 & 0 \\
2 & 1 & 0 & 0 & 0 & 0 & 0 & 0 & 1 & 0 & 0 & 0 & 0 & 0 & 0 & 0 \\
-2 & -1 & 0 & 0 & 0 & 0 & 0 & 1 & 0 & 0 & 0 & 0 & 0 & 0 & 0 & 0 \\
0 & \frac{1}{2} & -\frac{1}{2} & 0 & 0 & 1 & 0 & 0 & 0 & 0 & 0 & 0 & 0 & 0 & 0 & 0 \\
2 & 0 & 1 & 0 & 1 & 0 & 0 & 0 & 0 & 0 & 0 & 0 & 0 & 0 & 0 & 0 \\
1 & \frac{1}{2} & \frac{1}{2} & 1 & 0 & 0 & 0 & 0 & 0 & 0 & 0 & 0 & 0 & 0 & 0 & 0 \\
\end{array}
\right)
\,,\qquad 
D= \D{5} \, ,
\end{split}
\end{align}
and the five colour structure  are
\begin{eqnarray}
(YC)_1 &=& i \fach \fdbk \fdcj \ta 1 \tb 1 \thh 2 \tj 3 \tkk 4 + i\fabg \fdbk \fdcj 
\tg 1 \tc 2 \ta 2 \tj 3 \tkk 4 \nonumber \\ && - i \fabg \fach \fdbk 
\tg 1 \thh 2 \tc 3 \td 3 \tkk 4 -i \fabg \fach \fdcj 
\tg 1 \thh 2 \tj 3 \tb 4 \td 4 \, ,\nonumber \\&&  \nonumber \\
(YC)_2 &=& -i\fabg \fach \fdcj \tg 1 \thh 2 \tj 3 \tb 4 \td 4 \, ,  \nonumber \\ \nonumber \\
(YC)_3 &=& -i \fabg \fach \fdbk \tg 1 \thh 2 \tc 3 \td 3 \tkk 4  \, , \nonumber \\ \nonumber \\
(YC)_4 &=&  i \fabg \fdbk \fdcj \tg 1 \tc 2 \ta 2 \tj 3 \tkk 4 -\fabg \fach \fdbk \fdcj 
\tg 1 \thh 2 \tj 3 \tkk 4 \, ,\nonumber \\ \nonumber \\
(YC)_5 &=& i \fach \fdbk \fdcj \ta 1 \tb 1 \thh 2 \tj 3 \tkk 4 \, .
\end{eqnarray}

\vspace{2mm}


\item[{\bf 16}.] $\text W_4^{(4)}(1,1,2,4)$ \\

\vspace{-2mm}
We now come  to one of the  largest Cwebs of the  set, with twenty-four diagrams.

\vspace{5mm}
\begin{minipage}{0.5\textwidth}
		\begin{figure}[H]
		\vspace{-2mm}
		\includegraphics[height=4cm,width=4cm]{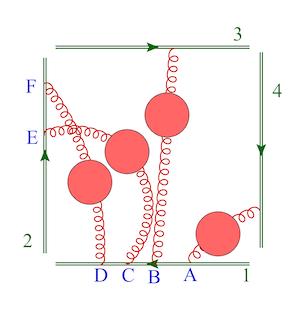}
		\end{figure}
	\end{minipage} 
	\hspace{-3.0cm}
	\begin{minipage}{0.50\textwidth}
		\begin{tabular}{ | c | c | c |}
			\hline
			\textbf{Diagrams} & \textbf{Sequences} & \textbf{S-factors} \\ \hline
			$C_1$ & $\lbrace \lbrace ABCD\rbrace,\lbrace EF \rbrace\rbrace$ & 1 \\ 
			\hline
			$C_2$ & $\lbrace \lbrace ABCD\rbrace,\lbrace FE \rbrace\rbrace$ & 0 \\ 
			\hline
			$C_3$ & $\lbrace \lbrace BACD\rbrace,\lbrace EF \rbrace\rbrace$ & 1 \\ 
			\hline
			$C_4$ & $\lbrace \lbrace BACD\rbrace,\lbrace FE \rbrace\rbrace$ & 0 \\ 
			\hline
			$C_5$ & $\lbrace \lbrace ACBD\rbrace,\lbrace EF \rbrace\rbrace$ & 1 \\ 
			\hline
			$C_6$ & $\lbrace \lbrace ACBD\rbrace,\lbrace FE \rbrace\rbrace$ & 0 \\ 
			\hline
			$C_7$ & $\lbrace \lbrace CABD\rbrace,\lbrace EF \rbrace\rbrace$ & 1 \\ 
			\hline
			$C_8$ & $\lbrace \lbrace CABD\rbrace,\lbrace FE \rbrace\rbrace$ & 0 \\ 
			\hline
			$C_9$ & $\lbrace \lbrace BCAD\rbrace,\lbrace EF \rbrace\rbrace$ & 1 \\ 
			\hline
			$C_{10}$ & $\lbrace \lbrace BCAD\rbrace,\lbrace FE \rbrace\rbrace$ & 0 \\ 
			\hline
			$C_{11}$ & $\lbrace \lbrace CBAD\rbrace,\lbrace EF \rbrace\rbrace$ & 1 \\ 
			\hline
			$C_{12}$ & $\lbrace \lbrace CBAD\rbrace,\lbrace FE \rbrace\rbrace$ & 0 \\ 
			\hline
			$C_{13}$ & $\lbrace \lbrace ACDB\rbrace,\lbrace EF \rbrace\rbrace$ & 1 \\ 
			\hline
			$C_{14}$ & $\lbrace \lbrace ACDB\rbrace,\lbrace FE \rbrace\rbrace$ & 0 \\ 
			\hline
			$C_{15}$ & $\lbrace \lbrace CADB\rbrace,\lbrace EF \rbrace\rbrace$ & 1 \\ 
			\hline
			$C_{16}$ & $\lbrace \lbrace CADB\rbrace,\lbrace FE \rbrace\rbrace$ & 0 \\ 
			\hline
			$C_{17}$ & $\lbrace \lbrace CDAB\rbrace,\lbrace EF \rbrace\rbrace$ & 1 \\ 
			\hline
			$C_{18}$ & $\lbrace \lbrace CDAB\rbrace,\lbrace FE \rbrace\rbrace$ & 0 \\ 
			\hline
			$C_{19}$ & $\lbrace \lbrace BCDA\rbrace,\lbrace EF \rbrace\rbrace$ & 1 \\ 
			\hline
			$C_{20}$ & $\lbrace \lbrace BCDA\rbrace,\lbrace FE \rbrace\rbrace$ & 0 \\ 
			\hline
			$C_{21}$ & $\lbrace \lbrace CBDA\rbrace,\lbrace EF \rbrace\rbrace$ & 1 \\ 
			\hline
			$C_{22}$ & $\lbrace \lbrace CBDA\rbrace,\lbrace FE \rbrace\rbrace$ & 0 \\ 
			\hline
			$C_{23}$ & $\lbrace \lbrace CDBA\rbrace,\lbrace EF \rbrace\rbrace$ & 1 \\ 
			\hline
			$C_{24}$ & $\lbrace \lbrace CDBA\rbrace,\lbrace FE \rbrace\rbrace$ & 0 \\ 
			\hline
	
		\end{tabular}
		\label{tab:abcd16}	
	\end{minipage}

\vspace{1cm}

\noindent The $R$, $Y$ and $D$ matrices are given by

\begin{align}
\hspace{-1cm}
\begin{split}
&
R=\frac{1}{6} \left(
\begin{array}{cccccccccccccccccccccccc}
1 & 0 & 0 & 0 & -1 & 0 & -1 & 0 & -1 & 0 & 1 & 0 & 0 & 0 & 1 & 0 & 0 & 0 & 0 & 0 & 1 & 0 & -1 & 0 \\
-1 & 2 & 1 & -1 & -1 & 0 & -1 & 0 & -1 & 0 & 1 & 0 & 1 & -1 & 1 & 0 & 1 & -1 & 1 & -1 & 1 & 0 & -3 & 2 \\
0 & 0 & 1 & 0 & -1 & 0 & 1 & 0 & -1 & 0 & -1 & 0 & 0 & 0 & 1 & 0 & -1 & 0 & 0 & 0 & 1 & 0 & 0 & 0 \\
1 & -1 & -1 & 2 & -1 & 0 & 1 & 0 & -1 & 0 & -1 & 0 & 1 & -1 & 1 & 0 & -3 & 2 & 1 & -1 & 1 & 0 & 1 & -1 \\
-1 & 0 & 0 & 0 & 2 & 0 & 0 & 0 & 0 & 0 & 0 & 0 & -1 & 0 & 0 & 0 & 0 & 0 & 1 & 0 & -2 & 0 & 1 & 0 \\
0 & -1 & 1 & -1 & -1 & 3 & 0 & 0 & 0 & 0 & 0 & 0 & 0 & -1 & 0 & 0 & 1 & -1 & -1 & 2 & 1 & -3 & -1 & 2 \\
-1 & 0 & 1 & 0 & 0 & 0 & 2 & 0 & 0 & 0 & -2 & 0 & 0 & 0 & 0 & 0 & -1 & 0 & 0 & 0 & 0 & 0 & 1 & 0 \\
0 & -1 & -1 & 2 & 3 & -3 & -4 & 6 & 3 & -3 & -2 & 0 & -2 & 2 & 3 & -3 & 0 & -1 & -2 & 2 & 3 & -3 & -1 & 2 \\
0 & 0 & -1 & 0 & 0 & 0 & 0 & 0 & 2 & 0 & 0 & 0 & 1 & 0 & -2 & 0 & 1 & 0 & -1 & 0 & 0 & 0 & 0 & 0 \\
1 & -1 & 0 & -1 & 0 & 0 & 0 & 0 & -1 & 3 & 0 & 0 & -1 & 2 & 1 & -3 & -1 & 2 & 0 & -1 & 0 & 0 & 1 & -1 \\
1 & 0 & -1 & 0 & 0 & 0 & -2 & 0 & 0 & 0 & 2 & 0 & 0 & 0 & 0 & 0 & 1 & 0 & 0 & 0 & 0 & 0 & -1 & 0 \\
-1 & 2 & 0 & -1 & 3 & -3 & -2 & 0 & 3 & -3 & -4 & 6 & -2 & 2 & 3 & -3 & -1 & 2 & -2 & 2 & 3 & -3 & 0 & -1 \\
0 & 0 & 0 & 0 & -1 & 0 & 1 & 0 & 1 & 0 & -1 & 0 & 1 & 0 & -1 & 0 & 0 & 0 & -1 & 0 & 1 & 0 & 0 & 0 \\
1 & -1 & 1 & -1 & -1 & 0 & 1 & 0 & 1 & 0 & -1 & 0 & -1 & 2 & -1 & 0 & 1 & -1 & -3 & 2 & 1 & 0 & 1 & -1 \\
0 & 0 & 1 & 0 & 0 & 0 & 0 & 0 & -2 & 0 & 0 & 0 & -1 & 0 & 2 & 0 & -1 & 0 & 1 & 0 & 0 & 0 & 0 & 0 \\
1 & -1 & -1 & 2 & 0 & 0 & 0 & 0 & 1 & -3 & 0 & 0 & 0 & -1 & -1 & 3 & 0 & -1 & -1 & 2 & 0 & 0 & 1 & -1 \\
0 & 0 & -1 & 0 & 1 & 0 & -1 & 0 & 1 & 0 & 1 & 0 & 0 & 0 & -1 & 0 & 1 & 0 & 0 & 0 & -1 & 0 & 0 & 0 \\
1 & -1 & -3 & 2 & 1 & 0 & -1 & 0 & 1 & 0 & 1 & 0 & 1 & -1 & -1 & 0 & -1 & 2 & 1 & -1 & -1 & 0 & 1 & -1 \\
0 & 0 & 0 & 0 & 1 & 0 & -1 & 0 & -1 & 0 & 1 & 0 & -1 & 0 & 1 & 0 & 0 & 0 & 1 & 0 & -1 & 0 & 0 & 0 \\
1 & -1 & 1 & -1 & 1 & 0 & -1 & 0 & -1 & 0 & 1 & 0 & -3 & 2 & 1 & 0 & 1 & -1 & -1 & 2 & -1 & 0 & 1 & -1 \\
1 & 0 & 0 & 0 & -2 & 0 & 0 & 0 & 0 & 0 & 0 & 0 & 1 & 0 & 0 & 0 & 0 & 0 & -1 & 0 & 2 & 0 & -1 & 0 \\
-1 & 2 & 1 & -1 & 1 & -3 & 0 & 0 & 0 & 0 & 0 & 0 & -1 & 2 & 0 & 0 & 1 & -1 & 0 & -1 & -1 & 3 & 0 & -1 \\
-1 & 0 & 0 & 0 & 1 & 0 & 1 & 0 & 1 & 0 & -1 & 0 & 0 & 0 & -1 & 0 & 0 & 0 & 0 & 0 & -1 & 0 & 1 & 0 \\
-3 & 2 & 1 & -1 & 1 & 0 & 1 & 0 & 1 & 0 & -1 & 0 & 1 & -1 & -1 & 0 & 1 & -1 & 1 & -1 & -1 & 0 & -1 & 2 \\
\end{array}
\right), \qquad \\ \vspace{5mm} &
\hspace{15mm}
Y=\left(
\begin{array}{cccccccccccccccccccccccc}
 -2 & 1 & 2 & -1 & 0 & 0 & 2 & 0 & 0 & 0 & -2 & 0 & 0 & 0 & 0 & 0 & 0 & -1 & 0 & 0 & 0 & 0 & 0 & 1 \\
 -1 & 0 & 1 & 0 & 0 & 0 & 2 & 0 & 0 & 0 & -2 & 0 & 0 & 0 & 0 & 0 & -1 & 0 & 0 & 0 & 0 & 0 & 1 & 0 \\
 -1 & 1 & 2 & -1 & 0 & -1 & 1 & 0 & -1 & 0 & -1 & 0 & -1 & 1 & 1 & 0 & 0 & -1 & 0 & 0 & 0 & 1 & 0 & 0 \\
 0 & 0 & 1 & 0 & -1 & 0 & 1 & 0 & -1 & 0 & -1 & 0 & 0 & 0 & 1 & 0 & -1 & 0 & 0 & 0 & 1 & 0 & 0 & 0 \\
 0 & 0 & 2 & -1 & 0 & 0 & 0 & 0 & -2 & 0 & 0 & 0 & -2 & 1 & 2 & 0 & 0 & -1 & 0 & 1 & 0 & 0 & 0 & 0 \\
 0 & 0 & 1 & 0 & 0 & 0 & 0 & 0 & -2 & 0 & 0 & 0 & -1 & 0 & 2 & 0 & -1 & 0 & 1 & 0 & 0 & 0 & 0 & 0 \\
 0 & 0 & 0 & 0 & -1 & 1 & 1 & -1 & 0 & 0 & 0 & 0 & 1 & -1 & -1 & 1 & 0 & 0 & 0 & 0 & 0 & 0 & 0 & 0 \\
 -1 & 1 & 1 & -1 & 0 & 0 & 1 & -1 & 0 & 0 & -1 & 1 & 0 & 0 & 0 & 0 & 0 & 0 & 0 & 0 & 0 & 0 & 0 & 0 \\
 0 & 0 & 1 & -1 & -1 & 1 & 1 & -1 & -1 & 1 & 0 & 0 & 0 & 0 & 0 & 0 & 0 & 0 & 0 & 0 & 0 & 0 & 0 & 0 \\
 2 & -1 & 0 & 0 & 0 & 0 & 0 & 0 & 0 & 0 & 0 & 0 & 0 & 0 & 0 & 0 & 0 & 0 & 0 & 0 & 0 & 0 & 0 & 1 \\
 1 & 0 & 0 & 0 & 0 & 0 & 0 & 0 & 0 & 0 & 0 & 0 & 0 & 0 & 0 & 0 & 0 & 0 & 0 & 0 & 0 & 0 & 1 & 0 \\
 0 & 0 & -1 & 1 & 0 & 1 & 0 & 0 & 0 & 0 & 0 & 0 & 0 & 0 & 0 & 0 & 0 & 0 & 0 & 0 & 0 & 1 & 0 & 0 \\
 0 & 0 & 0 & 0 & 1 & 0 & 0 & 0 & 0 & 0 & 0 & 0 & 0 & 0 & 0 & 0 & 0 & 0 & 0 & 0 & 1 & 0 & 0 & 0 \\
 -2 & 1 & -1 & 1 & -1 & 0 & 0 & 0 & 0 & 0 & 0 & 0 & 0 & 0 & 0 & 0 & 0 & 0 & 0 & 1 & 0 & 0 & 0 & 0 \\
 -1 & 0 & 0 & 0 & -1 & 0 & 0 & 0 & 0 & 0 & 0 & 0 & 0 & 0 & 0 & 0 & 0 & 0 & 1 & 0 & 0 & 0 & 0 & 0 \\
 0 & 0 & 2 & -1 & 0 & 0 & 0 & 0 & 0 & 0 & 0 & 0 & 0 & 0 & 0 & 0 & 0 & 1 & 0 & 0 & 0 & 0 & 0 & 0 \\
 0 & 0 & 1 & 0 & 0 & 0 & 0 & 0 & 0 & 0 & 0 & 0 & 0 & 0 & 0 & 0 & 1 & 0 & 0 & 0 & 0 & 0 & 0 & 0 \\
 -1 & 1 & 0 & 0 & 0 & 0 & 0 & 0 & 0 & 1 & 0 & 0 & 0 & 0 & 0 & 1 & 0 & 0 & 0 & 0 & 0 & 0 & 0 & 0 \\
 -1 & 0 & -1 & 0 & -1 & 0 & 0 & 0 & 0 & 0 & 0 & 0 & 0 & 0 & 1 & 0 & 0 & 0 & 0 & 0 & 0 & 0 & 0 & 0 \\
 0 & 1 & -1 & 1 & 1 & 0 & 0 & 0 & 0 & 0 & 0 & 0 & 0 & 1 & 0 & 0 & 0 & 0 & 0 & 0 & 0 & 0 & 0 & 0 \\
 1 & 0 & 0 & 0 & 1 & 0 & 0 & 0 & 0 & 0 & 0 & 0 & 1 & 0 & 0 & 0 & 0 & 0 & 0 & 0 & 0 & 0 & 0 & 0 \\
 -1 & 0 & 1 & 0 & 0 & 0 & 0 & 0 & 0 & 0 & 1 & 0 & 0 & 0 & 0 & 0 & 0 & 0 & 0 & 0 & 0 & 0 & 0 & 0 \\
 1 & 0 & 1 & 0 & 1 & 0 & 0 & 0 & 1 & 0 & 0 & 0 & 0 & 0 & 0 & 0 & 0 & 0 & 0 & 0 & 0 & 0 & 0 & 0 \\
 1 & 0 & -1 & 0 & 0 & 0 & 1 & 0 & 0 & 0 & 0 & 0 & 0 & 0 & 0 & 0 & 0 & 0 & 0 & 0 & 0 & 0 & 0 & 0 \\
\end{array}
\right)\,,\qquad \\ \vspace{5mm} & 
\hspace{5mm} D= \D{9}  \, .
\end{split}
\end{align}
As a consequence,  there are nine  colour structures.
\begin{eqnarray}
(YC)_1 &=& i \fabe \fbgh \fcdg \thh 1 \ta 1 \te 2 \tc 3 \td 4  + 
i \fahk \fbgh \fcdg \tkk 1 \tb 2 \ta 2 \tc 3 \td 4 - \nonumber \\ && 
2 \fabe \fahk \fbgh \fcdg \tkk 1 \te 2 \tc 3 \td 4 \, , \nonumber \\ \nonumber \\
(YC)_2 &=& i\fabe \fbgh \fcdg \thh 1 \ta 1 \te 2 \tc 3 \td 4 + 
i\fahk \fbgh \fcdg \tkk 1 \tb 2 \ta 2 \tc 3 \td 4 \nonumber \\&&
- \fabe \fahk \fbgh \fcdg \tkk 1 \te 2 \tc 3 \td 4 \, , \nonumber \\ \nonumber \\
(YC)_3 &=& i \fabe \fbgh \fcdg \thh 1 \ta 1 \te 2 \tc 3 \td 4 - 
\fabe \fahk \fbgh \fcdg \tkk 1 \te 2 \tc 3 \td 4 \nonumber \\&&
- i \fabe \facn \fbdm \tn 1 \tm 1 \te 2 \tc 3 \td 4 -i \fauv \fbdm \fmcu  
\tv 1 \tb 2 \ta 2 \tc 3 \td 4  \nonumber \\&& +  \fabe \fauv \fbdm \fmcu 
\tv 1 \te 2 \tc 3 \td 4 \, , \nonumber \\ \nonumber \\
(YC)_4 &=&  -i \fauv \fbdm \fmcu \tv 1 \tb 2 \ta 2 \tc 3 \td 4 + 
\fabe \fauv \fbdm \fmcu \tv 1 \te 2 \tc 3 \td 4 \nonumber \\ &&
+ i\fabe f^{adm} \fmcu \tu 1 \tb 1 \te 2 \tc 3 \td 4 \, , \nonumber \\ \nonumber \\
(YC)_5 &=& -i\fabe \fbdm \facn \tn 1 \tm 1 \te 2 \tc 3 \td 4 + 
i\fauv \fbdm f^{cmu} \tv 1 \tb 2 \ta 2 \tc 3 \td 4 \nonumber \\&& 
-2 \fabe \fauv \fbdm f^{cmu} \tv 1 \te 2 \tc 3 \td 4 +
i \fabe \fbdm f^{cmu} \tu 1 \ta 1 \te 2 \tc 3 \td 4 \nonumber \\&& 
+ i f^{adm} f^{bcn} f^{mnq} \tq 1 \tb 2 \ta 2 \tc 3 \td 4 \, , \nonumber \\ \nonumber \\ 
(YC)_6 &=&  i\fabe f^{adm} f^{bcn} \tn 1 \tm 1 \te 2 \tc 3 \td 4 + 
i \fauv \fbdm f^{cmu} \tv 1 \tb 2 \ta 2 \tc 3 \td 4 \nonumber \\&& 
-\fabe \fauv \fbdm f^{cmu} \tv 1 \te 2 \tc 3\td 4 -i\fabe f^{adm} f^{cmu} 
\tu 1 \tb 1 \te 2 \tc 3 \td 4 \nonumber \\&&+ 
i f^{adm} f^{bcn} f^{mnq} \tq 1 \tb 2 \ta 2 \tc 3 \td 4 
-\fabe f^{adm} f^{bcn} f^{mnq} \tq 1 \te 2 \tc 3 \td 4 \, , \nonumber \\ \nonumber \\
(YC)_7 &=& -i \fabe \fbdm f^{can} \tn 1 \tm 1 \te 2 \tc 3 \td 4 \, , \nonumber \\ \nonumber \\
(YC)_8 &=& i \fabe \fbgh \fcdg \thh 1 \ta 1 \te 2 \tc 3 \td 4 - 
\fabe  \fahk \fbgh \fcdg \tkk 1 \te 2 \tc 3 \td 4 \, , \nonumber \\ \nonumber \\
(YC)_9 &=& -\fabe \fauv \fbdm f^{cmu} \tv 1 \te 2 \tc 3 \td 4 
+ i \fabe \fbdm f^{cmu} \tu 1 \ta 1 \te 2 \tc 3 \td 4 \, .
\end{eqnarray}

\vspace{2mm}

\pagebreak


\item[{\bf 17}.] $\text W_4^{(4)}(1,1,3,3)$ \\

\vspace{-2mm}
This Cweb comprises eighteen diagrams.

\vspace{3mm}
\begin{minipage}{0.5\textwidth}
	\begin{figure}[H]
		\vspace{-2mm}
		\includegraphics[height=4cm,width=4cm]{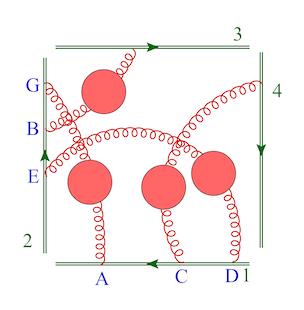}
	\end{figure}
\end{minipage} 
\hspace{-3.0cm}
\begin{minipage}{0.45\textwidth}
	\begin{tabular}{ | c | c | c |}
		\hline
		\textbf{Diagrams} & \textbf{Sequences} & \textbf{S-factors} \\ \hline
		$C_1$ & $\lbrace \lbrace DCA\rbrace,\lbrace EBG \rbrace\rbrace$ & 2 \\ 
		\hline
		$C_2$ & $\lbrace \lbrace DCA\rbrace,\lbrace BEG \rbrace\rbrace$ & 1 \\ 
		\hline
		$C_3$ & $\lbrace \lbrace DCA\rbrace,\lbrace EGB \rbrace\rbrace$ & 1 \\ 
		\hline
		$C_4$ & $\lbrace \lbrace DCA\rbrace,\lbrace GEB \rbrace\rbrace$ & 0 \\ 
		\hline
		$C_5$ & $\lbrace \lbrace DCA\rbrace,\lbrace BGE \rbrace\rbrace$ & 0 \\ 
		\hline
		$C_6$ & $\lbrace \lbrace DCA\rbrace,\lbrace GBE \rbrace\rbrace$ & 0 \\ 
		\hline
		$C_7$ & $\lbrace \lbrace CDA\rbrace,\lbrace EBG \rbrace\rbrace$ & 1 \\ 
		\hline
		$C_8$ & $\lbrace \lbrace CDA\rbrace,\lbrace BEG \rbrace\rbrace$ & 2 \\ 
		\hline
		$C_9$ & $\lbrace \lbrace CDA\rbrace,\lbrace EGB \rbrace\rbrace$ & 1 \\ 
		\hline
		$C_{10}$ & $\lbrace \lbrace CDA\rbrace,\lbrace GEB \rbrace\rbrace$ & 0 \\ 
		\hline
		$C_{11}$ & $\lbrace \lbrace CDA\rbrace,\lbrace BGE \rbrace\rbrace$ & 0 \\ 
		\hline
		$C_{12}$ & $\lbrace \lbrace CDA\rbrace,\lbrace GBE \rbrace\rbrace$ & 0 \\ 
		\hline
		$C_{13}$ & $\lbrace \lbrace DAC\rbrace,\lbrace EBG \rbrace\rbrace$ & 1 \\ 
		\hline
		$C_{14}$ & $\lbrace \lbrace DAC\rbrace,\lbrace BEG \rbrace\rbrace$ & 1 \\ 
		\hline
		$C_{15}$ & $\lbrace \lbrace DAC\rbrace,\lbrace EGB \rbrace\rbrace$ & 2 \\ 
		\hline
		$C_{16}$ & $\lbrace \lbrace DAC\rbrace,\lbrace GEB \rbrace\rbrace$ & 0 \\ 
		\hline
		$C_{17}$ & $\lbrace \lbrace DAC\rbrace,\lbrace BGE \rbrace\rbrace$ & 0 \\ 
		\hline
		$C_{18}$ & $\lbrace \lbrace DAC\rbrace,\lbrace GBE \rbrace\rbrace$ & 0 \\ 
		\hline
		\end{tabular}
	\label{tab:abcd17}	
\end{minipage}

\vspace{5mm}
\noindent The $R$, $Y$ and $D$ matrices are given by
\begin{align}
\begin{split}
&
R=\frac{1}{6} \left(
\begin{array}{cccccccccccccccccc}
0 & 0 & 0 & 0 & 0 & 0 & 0 & 0 & 0 & 0 & 0 & 0 & 0 & 0 & 0 & 0 & 0 & 0 \\
0 & 2 & -2 & 0 & 0 & 0 & 0 & -1 & 1 & 0 & 0 & 0 & 0 & -1 & 1 & 0 & 0 & 0 \\
0 & -2 & 2 & 0 & 0 & 0 & 0 & 1 & -1 & 0 & 0 & 0 & 0 & 1 & -1 & 0 & 0 & 0 \\
0 & 1 & -1 & 3 & -3 & 0 & 0 & 0 & 0 & -1 & 1 & 0 & 0 & -1 & 1 & -2 & 2 & 0 \\
0 & -1 & 1 & -3 & 3 & 0 & 0 & 1 & -1 & 2 & -2 & 0 & 0 & 0 & 0 & 1 & -1 & 0 \\
6 & -3 & -3 & -3 & -3 & 6 & -3 & 2 & 1 & 2 & 1 & -3 & -3 & 1 & 2 & 1 & 2 & -3 \\
0 & 0 & 0 & 0 & 0 & 0 & 2 & -1 & -1 & 0 & 0 & 0 & -2 & 1 & 1 & 0 & 0 & 0 \\
0 & -1 & 1 & 0 & 0 & 0 & -1 & 1 & 0 & 0 & 0 & 0 & 1 & 0 & -1 & 0 & 0 & 0 \\
0 & 1 & -1 & 0 & 0 & 0 & -1 & 0 & 1 & 0 & 0 & 0 & 1 & -1 & 0 & 0 & 0 & 0 \\
0 & 1 & -1 & 0 & 0 & 0 & -1 & 2 & -1 & 2 & -2 & 0 & 1 & -3 & 2 & -2 & 2 & 0 \\
0 & -1 & 1 & 0 & 0 & 0 & -1 & 0 & 1 & -1 & 1 & 0 & 1 & 1 & -2 & 1 & -1 & 0 \\
0 & 0 & 0 & 0 & 0 & 0 & -1 & 1 & 0 & -1 & -2 & 3 & 1 & -1 & 0 & 1 & 2 & -3 \\
0 & 0 & 0 & 0 & 0 & 0 & -2 & 1 & 1 & 0 & 0 & 0 & 2 & -1 & -1 & 0 & 0 & 0 \\
0 & -1 & 1 & 0 & 0 & 0 & 1 & 0 & -1 & 0 & 0 & 0 & -1 & 1 & 0 & 0 & 0 & 0 \\
0 & 1 & -1 & 0 & 0 & 0 & 1 & -1 & 0 & 0 & 0 & 0 & -1 & 0 & 1 & 0 & 0 & 0 \\
0 & 1 & -1 & 0 & 0 & 0 & 1 & -2 & 1 & -1 & 1 & 0 & -1 & 1 & 0 & 1 & -1 & 0 \\
0 & -1 & 1 & 0 & 0 & 0 & 1 & 2 & -3 & 2 & -2 & 0 & -1 & -1 & 2 & -2 & 2 & 0 \\
0 & 0 & 0 & 0 & 0 & 0 & 1 & 0 & -1 & 2 & 1 & -3 & -1 & 0 & 1 & -2 & -1 & 3 \\
\end{array}
\right)\,, 
\end{split}
\end{align}
\begin{align}
\begin{split}
Y=\left(
\begin{array}{cccccccccccccccccc}
-1 & 0 & 1 & 1 & 0 & -1 & 1 & 0 & -1 & 0 & 0 & 0 & 0 & 0 & 0 & -1 & 0 & 1 \\
0 & -1 & 1 & 1 & -1 & 0 & 0 & 1 & -1 & 0 & 0 & 0 & 0 & 0 & 0 & -1 & 1 & 0 \\
0 & 1 & -1 & 0 & 0 & 0 & 1 & -1 & 0 & 0 & 0 & 0 & -1 & 0 & 1 & 0 & 0 & 0 \\
0 & -1 & 1 & 0 & 0 & 0 & 1 & 0 & -1 & 0 & 0 & 0 & -1 & 1 & 0 & 0 & 0 & 0 \\
-1 & 1 & 0 & 1 & 0 & -1 & 1 & -1 & 0 & -1 & 0 & 1 & 0 & 0 & 0 & 0 & 0 & 0 \\
0 & 1 & -1 & 1 & -1 & 0 & 0 & -1 & 1 & -1 & 1 & 0 & 0 & 0 & 0 & 0 & 0 & 0 \\
0 & 0 & 0 & -1 & -1 & 0 & 0 & 0 & 0 & 0 & 0 & 1 & 0 & 0 & 0 & 0 & 0 & 1 \\
0 & \frac{1}{2} & 0 & -2 & -2 & 0 & -\frac{1}{2} & 0 & 0 & 0 & 0 & 0 & 0 & 0 & 0 & 0 & 1 & 0 \\
0 & -\frac{1}{2} & 0 & 1 & 1 & 0 & -\frac{1}{2} & 0 & 0 & 0 & 0 & 0 & 0 & 0 & 0 & 1 & 0 & 0 \\
0 & -\frac{1}{2} & 0 & 0 & 0 & 0 & -\frac{1}{2} & 0 & 0 & 0 & 0 & 0 & 0 & 0 & 1 & 0 & 0 & 0 \\
0 & \frac{1}{2} & 0 & 0 & 0 & 0 & -\frac{1}{2} & 0 & 0 & 0 & 0 & 0 & 0 & 1 & 0 & 0 & 0 & 0 \\
0 & 0 & 0 & 0 & 0 & 0 & 1 & 0 & 0 & 0 & 0 & 0 & 1 & 0 & 0 & 0 & 0 & 0 \\
0 & \frac{1}{2} & 0 & 1 & 1 & 0 & \frac{1}{2} & 0 & 0 & 0 & 1 & 0 & 0 & 0 & 0 & 0 & 0 & 0 \\
0 & -\frac{1}{2} & 0 & -2 & -2 & 0 & \frac{1}{2} & 0 & 0 & 1 & 0 & 0 & 0 & 0 & 0 & 0 & 0 & 0 \\
0 & -\frac{1}{2} & 0 & 0 & 0 & 0 & \frac{1}{2} & 0 & 1 & 0 & 0 & 0 & 0 & 0 & 0 & 0 & 0 & 0 \\
0 & \frac{1}{2} & 0 & 0 & 0 & 0 & \frac{1}{2} & 1 & 0 & 0 & 0 & 0 & 0 & 0 & 0 & 0 & 0 & 0 \\
0 & 1 & 1 & 0 & 0 & 0 & 0 & 0 & 0 & 0 & 0 & 0 & 0 & 0 & 0 & 0 & 0 & 0 \\
1 & 0 & 0 & 0 & 0 & 0 & 0 & 0 & 0 & 0 & 0 & 0 & 0 & 0 & 0 & 0 & 0 & 0 \\
\end{array}
\right)\,,\qquad
D= \D{6} \, .
\nonumber
\end{split}
\end{align}
There are six colour factors, given by
\begin{eqnarray}
(YC)_1 &=& i \fahj \fdch \fbag \tj 1 \tg 2 \td 2 \tb 3 \tc 4 \, , \nonumber \\ \nonumber \\
(YC)_2 &=& i\fahj \fdch \fbdg \tj 1 \ta 2 \tg 2 \tb 3 \tc 4 + i \fahj \fdch \fbag 
\tj 1 \tg 2 \td 2 \tb 3 \tc 4 \, , \nonumber \\ \nonumber \\
(YC)_3 &=& -i \fahj \fbdg \fdch \tj 1 \tg 2 \ta 2 \tb 3 \tc 4 -i f^{agj} 
\fbdg \fdch \ta 1 \thh 1 \tj 2 \tb 3 \tc 4  \, , \nonumber \\ \nonumber \\
(YC)_4 &=&  i \fahj \fbag \fdch \tj 1 \tg 2 \td 2 \tb 3 \tc 4 -i f^{dgl} 
\fbag \fdch  \thh 1 \ta 1 \tl 2 \tb 3 \tc  4 \, , \nonumber \\ \nonumber \\
(YC)_5 &=& -i \fbdg \fdch f^{agj} \ta 1 \thh 1 \tj 2 \tb 3 \tc 4 \, , \nonumber \\ \nonumber \\ 
(YC)_6 &=& -i f^{agj} \fbdg \fdch \ta 1 \thh 1 \tj 2 \tb 3 \tc 4+ 
i f^{dgj} \fbag \fdch  \ta 1 \thh 1 \tj 2 \tb 3 \tc  4 \, .
\end{eqnarray}

\vspace{2mm}


\item[{\bf 18}.] $\text W_{4, \rm II}^{(0,2)}(1,1,2,2)$ \\

\vspace{-2mm}
This Cweb, with two three-gluon correlators, has the same correlator and attachment
content as Cweb number 4, presented above, Eqs.~(\ref{bro1}) and (\ref{bro2}). It
has four diagrams.

\begin{minipage}{0.5\textwidth}
		\begin{figure}[H]
		\vspace{-2mm}
		\includegraphics[height=4cm,width=4cm]{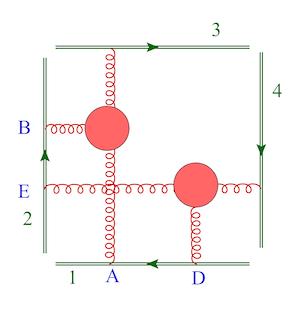}
		\end{figure}
	\end{minipage} 
	\hspace{-2.0cm}
	\begin{minipage}{0.48\textwidth}
		\begin{tabular}{ | c | c | c |}
			\hline
			\textbf{Diagrams} & \textbf{Sequences} & \textbf{S-factors} \\ \hline
			$C_1$ & $\lbrace \lbrace DA\rbrace,\lbrace EB \rbrace\rbrace$ & 1 \\ 
			\hline
			$C_2$ & $\lbrace \lbrace DA\rbrace,\lbrace BE \rbrace\rbrace$ & 0 \\ 
			\hline
			$C_3$ & $\lbrace \lbrace AD\rbrace,\lbrace EB \rbrace\rbrace$ & 0 \\ 
			\hline
			$C_4$ & $\lbrace \lbrace AD\rbrace,\lbrace BE \rbrace\rbrace$ & 1 \\ 
			\hline
		\end{tabular}
		\label{tab:abcd18}	
	\end{minipage} 
\vspace{2cm}

\noindent The $R$, $Y$ and $D$ matrices are given by
\begin{align}
\begin{split}
&
R=\left(
\begin{array}{cccc}
\frac{1}{2} & 0 & 0 & -\frac{1}{2} \\
-\frac{1}{2} & 1 & 0 & -\frac{1}{2} \\
-\frac{1}{2} & 0 & 1 & -\frac{1}{2} \\
-\frac{1}{2} & 0 & 0 & \frac{1}{2} \\
\end{array}
\right)\,, \qquad
Y=\left(
\begin{array}{cccc}
-1 & 0 & 0 & 1 \\
-1 & 0 & 1 & 0 \\
-1 & 1 & 0 & 0 \\
1 & 0 & 0 & 1 \\
\end{array}
\right)
\,,\qquad
D= \D{3} \,  ,
\end{split}
\end{align}
while the colour factors are
\begin{eqnarray}
(YC)_1 &=&i \fabc \fdef \febh \td 1 \ta 1 \thh 2 \tc 3 \tf 4 \nonumber \\ && + 
i \fabc \fdag \fdef \tg 1 \tb 2 \te 2 \tc 3 \tf 4  \, ,\nonumber \\ \nonumber \\
(YC)_2 &=&i \fabc \fdag \fdef \tg 1 \tb 2 \te 2 \tc 3 \tf 4  \, , \nonumber \\ \nonumber \\
(YC)_3 &=& i \fabc \fdef \febh \ta 1 \td 1 \thh 2 \tc 3 \tf 4 \, .
\end{eqnarray}  

\vspace{2mm}


\item[{\bf 19}.] $\text W_{4, \rm I}^{(4)}(1,2,2,3)$

\vspace{2mm}
This is the first of two Cwebs with the same correlator and attachment content. 
It has twelve diagrams.

\vspace{3mm}
\begin{minipage}{0.5\textwidth}
	\begin{figure}[H]
		\vspace{-2mm}
		\includegraphics[height=4cm,width=4cm]{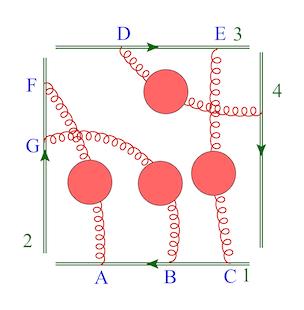}
	\end{figure}
\end{minipage} 
\hspace{-2.0cm}
\begin{minipage}{0.45\textwidth}
	\begin{tabular}{ | c | c | c |}
		\hline
		\textbf{Diagrams} & \textbf{Sequences} & \textbf{S-factors} \\ \hline
		$C_1$ & $\lbrace \lbrace CBA\rbrace,\lbrace GF \rbrace, \lbrace DE \rbrace\rbrace$ & 1 \\ 
		\hline
		$C_2$ & $\lbrace \lbrace CBA\rbrace,\lbrace GF \rbrace, \lbrace ED \rbrace\rbrace$ & 3 \\ 
		\hline
		$C_3$ & $\lbrace \lbrace CBA\rbrace,\lbrace FG \rbrace, \lbrace DE \rbrace\rbrace$ & 0 \\ 
		\hline
		$C_4$ & $\lbrace \lbrace CBA\rbrace,\lbrace FG \rbrace, \lbrace ED \rbrace\rbrace$ & 0 \\ 
		\hline
		$C_5$ & $\lbrace \lbrace BCA\rbrace,\lbrace GF \rbrace, \lbrace DE \rbrace\rbrace$ & 2 \\ 
		\hline
		$C_6$ & $\lbrace \lbrace BCA\rbrace,\lbrace GF \rbrace, \lbrace ED \rbrace\rbrace$ & 2 \\ 
		\hline
		$C_7$ & $\lbrace \lbrace BCA\rbrace,\lbrace FG \rbrace, \lbrace DE \rbrace\rbrace$ & 0 \\ 
		\hline
		$C_8$ & $\lbrace \lbrace BCA\rbrace,\lbrace FG \rbrace, \lbrace ED \rbrace\rbrace$ & 0 \\ 
		\hline
		$C_9$ & $\lbrace \lbrace BAC\rbrace,\lbrace GF \rbrace, \lbrace DE \rbrace\rbrace$ & 3 \\ 
		\hline
		$C_{10}$ & $\lbrace \lbrace BAC\rbrace,\lbrace GF \rbrace, \lbrace ED \rbrace\rbrace$ & 1 \\ 
		\hline
		$C_{11}$ & $\lbrace \lbrace BAC\rbrace,\lbrace FG \rbrace, \lbrace DE \rbrace\rbrace$ & 0 \\ 
		\hline
		$C_{12}$ & $\lbrace \lbrace BAC\rbrace,\lbrace FG \rbrace, \lbrace ED \rbrace\rbrace$ & 0 \\ 
		\hline
	\end{tabular}
	\label{tab:abcd19}	
\end{minipage} \\ \\ \\ \\ \\

\noindent The $R$, $Y$ and $D$ matrices are given by

\begin{align}
\begin{split}
&
R=\left(
\begin{array}{cccccccccccc}
\frac{1}{6} & -\frac{1}{6} & 0 & 0 & -\frac{1}{3} & \frac{1}{3} & 0 & 0 & \frac{1}{6} & -\frac{1}{6} & 0 & 0 \\
0 & 0 & 0 & 0 & 0 & 0 & 0 & 0 & 0 & 0 & 0 & 0 \\
-\frac{1}{6} & \frac{1}{6} & \frac{1}{3} & -\frac{1}{3} & -\frac{1}{3} & \frac{1}{3} & 0 & 0 & \frac{1}{2} & -\frac{1}{2} & -\frac{1}{3} & \frac{1}{3} \\
\frac{1}{6} & -\frac{1}{6} & -\frac{1}{6} & \frac{1}{6} & 0 & 0 & 0 & 0 & -\frac{1}{6} & \frac{1}{6} & \frac{1}{6} & -\frac{1}{6} \\
-\frac{1}{6} & \frac{1}{6} & 0 & 0 & \frac{1}{3} & -\frac{1}{3} & 0 & 0 & -\frac{1}{6} & \frac{1}{6} & 0 & 0 \\
\frac{1}{6} & -\frac{1}{6} & 0 & 0 & -\frac{1}{3} & \frac{1}{3} & 0 & 0 & \frac{1}{6} & -\frac{1}{6} & 0 & 0 \\
0 & 0 & -\frac{1}{6} & \frac{1}{6} & -\frac{1}{6} & \frac{1}{6} & \frac{1}{2} & -\frac{1}{2} & \frac{1}{6} & -\frac{1}{6} & -\frac{1}{3} & \frac{1}{3} \\
-\frac{1}{6} & \frac{1}{6} & \frac{1}{3} & -\frac{1}{3} & \frac{1}{6} & -\frac{1}{6} & -\frac{1}{2} & \frac{1}{2} & 0 & 0 & \frac{1}{6} & -\frac{1}{6} \\
0 & 0 & 0 & 0 & 0 & 0 & 0 & 0 & 0 & 0 & 0 & 0 \\
-\frac{1}{6} & \frac{1}{6} & 0 & 0 & \frac{1}{3} & -\frac{1}{3} & 0 & 0 & -\frac{1}{6} & \frac{1}{6} & 0 & 0 \\
\frac{1}{6} & -\frac{1}{6} & -\frac{1}{6} & \frac{1}{6} & 0 & 0 & 0 & 0 & -\frac{1}{6} & \frac{1}{6} & \frac{1}{6} & -\frac{1}{6} \\
-\frac{1}{2} & \frac{1}{2} & \frac{1}{3} & -\frac{1}{3} & \frac{1}{3} & -\frac{1}{3} & 0 & 0 & \frac{1}{6} & -\frac{1}{6} & -\frac{1}{3} & \frac{1}{3} \\
\end{array}
\right)\,, \qquad \\ \\ &
\hspace{1cm} Y=\left(
\begin{array}{cccccccccccc}
-2 & 2 & 1 & -1 & 2 & -2 & 0 & 0 & 0 & 0 & -1 & 1 \\
-1 & 1 & 0 & 0 & 2 & -2 & 0 & 0 & -1 & 1 & 0 & 0 \\
-1 & 1 & 1 & -1 & 1 & -1 & -1 & 1 & 0 & 0 & 0 & 0 \\
2 & 0 & -1 & 0 & 0 & 0 & 0 & 0 & 0 & 0 & 0 & 1 \\
-\frac{1}{2} & 0 & \frac{1}{2} & 0 & 0 & 0 & 0 & 0 & 0 & 0 & 1 & 0 \\
1 & 0 & 0 & 0 & 0 & 0 & 0 & 0 & 0 & 1 & 0 & 0 \\
0 & 0 & 0 & 0 & 0 & 0 & 0 & 0 & 1 & 0 & 0 & 0 \\
\frac{1}{2} & 0 & -\frac{1}{2} & 0 & 0 & 0 & 1 & 1 & 0 & 0 & 0 & 0 \\
-1 & 0 & 0 & 0 & 0 & 1 & 0 & 0 & 0 & 0 & 0 & 0 \\
1 & 0 & 0 & 0 & 1 & 0 & 0 & 0 & 0 & 0 & 0 & 0 \\
-\frac{1}{2} & 0 & \frac{1}{2} & 1 & 0 & 0 & 0 & 0 & 0 & 0 & 0 & 0 \\
0 & 1 & 0 & 0 & 0 & 0 & 0 & 0 & 0 & 0 & 0 & 0 \\
\end{array}
\right)
\,,\qquad
D= \D{3} \, .
\end{split}
\end{align}
The colour factors are
\begin{eqnarray}
(YC)_1 &=&-i f^{abh} f^{cbj} \fcdg \ta 1 \tj 1 \thh 2 \tg 3 \td 4 +
i f^{cbj} \fcdg f^{jam} \tm 1 \tb 2 \ta 2 \tg 3 \td 4 \nonumber \\ && 
-f^{abh} f^{cbj} \fcdg f^{jam} \tm 1 \thh 2 \tg 3 \td 4 \,  , \nonumber \\ \nonumber \\
(YC)_2 &=& -i f^{abh} f^{cbj} \fcdg \ta 1 \tj 1 \thh 2 \tg 3 \td 4+ 
i f^{cbj} \fcdg f^{jam} \tm 1 \tb 2 \ta 2 \tg 3 \td 4 \, ,  \nonumber \\ \nonumber \\
(YC)_3 &=&-i f^{abh} f^{cbj} \fcdg \ta 1 \tj 1 \thh 2 \tg 3 \td 4 \, .
\end{eqnarray} 

\vspace{2mm}

\pagebreak


\item[{\bf 20}.] $\text W_{4, \rm II}^{(4)}(1,2,2,3)$ \\

\vspace{-2mm}
Our last Cweb connecting four Wilson lines has twenty-four diagrams.

\vspace{5mm}
\begin{minipage}{0.46\textwidth}
	\begin{figure}[H]
		\includegraphics[height=4cm,width=4cm]{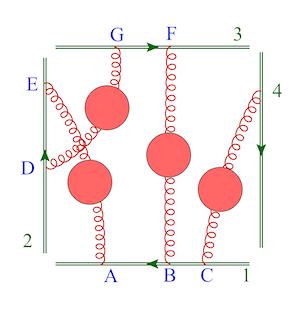}
	\end{figure}
\end{minipage} 
\hspace{-2.0cm}
\begin{minipage}{0.60\textwidth}
	\begin{tabular}{ | c | c | c |}
		\hline
		\textbf{Diagrams} & \textbf{Sequences} & \textbf{S-factors} \\ \hline
		$C_1$ & $\lbrace \lbrace CBA\rbrace,\lbrace DE \rbrace, \lbrace GF \rbrace\rbrace$ & 2 \\ 
		\hline
		$C_2$ & $\lbrace \lbrace CBA\rbrace,\lbrace DE \rbrace, \lbrace FG \rbrace\rbrace$ & 1 \\ 
		\hline
		$C_3$ & $\lbrace \lbrace CBA\rbrace,\lbrace ED \rbrace, \lbrace GF \rbrace\rbrace$ & 0 \\ 
		\hline
		$C_4$ & $\lbrace \lbrace CBA\rbrace,\lbrace ED \rbrace, \lbrace FG \rbrace\rbrace$ & 1 \\ 
		\hline
		$C_5$ & $\lbrace \lbrace BCA\rbrace,\lbrace DE \rbrace, \lbrace GF \rbrace\rbrace$ & 1 \\ 
		\hline
		$C_6$ & $\lbrace \lbrace BCA\rbrace,\lbrace DE \rbrace, \lbrace FG \rbrace\rbrace$ & 2 \\ 
		\hline
		$C_7$ & $\lbrace \lbrace BCA\rbrace,\lbrace ED \rbrace, \lbrace GF \rbrace\rbrace$ & 0 \\ 
		\hline
		$C_8$ & $\lbrace \lbrace BCA\rbrace,\lbrace ED \rbrace, \lbrace FG \rbrace\rbrace$ & 1 \\ 
		\hline
		$C_9$ & $\lbrace \lbrace CAB\rbrace,\lbrace DE \rbrace, \lbrace GF \rbrace\rbrace$ & 2 \\ 
		\hline
		$C_{10}$ & $\lbrace \lbrace CAB\rbrace,\lbrace DE \rbrace, \lbrace FG \rbrace\rbrace$ & 0 \\ 
		\hline
		$C_{11}$ & $\lbrace \lbrace CAB\rbrace,\lbrace ED \rbrace, \lbrace GF \rbrace\rbrace$ & 1 \\ 
		\hline
		$C_{12}$ & $\lbrace \lbrace CAB\rbrace,\lbrace ED \rbrace, \lbrace FG \rbrace\rbrace$ & 1 \\ 
		\hline
		$C_{13}$ & $\lbrace \lbrace ACB\rbrace,\lbrace DE \rbrace, \lbrace GF \rbrace\rbrace$ & 1 \\ 
		\hline
		$C_{14}$ & $\lbrace \lbrace ACB\rbrace,\lbrace DE \rbrace, \lbrace FG \rbrace\rbrace$ & 0 \\ 
		\hline
		$C_{15}$ & $\lbrace \lbrace ACB\rbrace,\lbrace ED \rbrace, \lbrace GF \rbrace\rbrace$ & 2 \\ 
		\hline
		$C_{16}$ & $\lbrace \lbrace ACB\rbrace,\lbrace ED \rbrace, \lbrace FG \rbrace\rbrace$ & 1 \\ 
		\hline
		$C_{17}$ & $\lbrace \lbrace BAC\rbrace,\lbrace DE \rbrace, \lbrace GF \rbrace\rbrace$ & 1 \\ 
		\hline
		$C_{18}$ & $\lbrace \lbrace BAC\rbrace,\lbrace DE \rbrace, \lbrace FG \rbrace\rbrace$ & 1 \\ 
		\hline
		$C_{19}$ & $\lbrace \lbrace BAC\rbrace,\lbrace ED \rbrace, \lbrace GF \rbrace\rbrace$ & 0 \\ 
		\hline
		$C_{20}$ & $\lbrace \lbrace BAC\rbrace,\lbrace ED \rbrace, \lbrace FG \rbrace\rbrace$ & 2 \\ 
		\hline
		$C_{21}$ & $\lbrace \lbrace ABC\rbrace,\lbrace DE \rbrace, \lbrace GF \rbrace\rbrace$ & 1 \\ 
		\hline
		$C_{22}$ & $\lbrace \lbrace ABC\rbrace,\lbrace DE \rbrace, \lbrace FG \rbrace\rbrace$ & 0 \\ 
		\hline
		$C_{23}$ & $\lbrace \lbrace ABC\rbrace,\lbrace ED \rbrace, \lbrace GF \rbrace\rbrace$ & 1 \\ 
		\hline
		$C_{24}$ & $\lbrace \lbrace ABC\rbrace,\lbrace ED \rbrace, \lbrace FG \rbrace\rbrace$ & 2 \\ 
		\hline
	\end{tabular}
	\label{tab:abcd20}	
\end{minipage}

\vspace{1cm}
\noindent The $R$, $Y$ and $D$ matrices are given by

\begin{align}
\hspace{-1.2cm}
\begin{split}
&
R=\frac{1}{12} \left(
\begin{array}{cccccccccccccccccccccccc}
2 & -1 & 0 & -1 & -1 & 0 & 0 & 1 & 0 & 0 & -1 & 1 & -1 & 0 & 0 & 1 & -1 & 1 & 0 & 0 & 1 & 0 & 1 & -2 \\
-2 & 3 & 0 & -1 & 1 & -2 & 0 & 1 & 0 & 0 & 1 & -1 & -1 & 0 & 2 & -1 & 1 & -1 & 0 & 0 & 1 & 0 & -3 & 2 \\
-4 & 1 & 6 & -3 & -1 & 2 & 0 & -1 & 2 & 0 & -3 & 1 & 1 & 0 & -2 & 1 & 5 & -3 & -6 & 4 & -3 & 0 & 5 & -2 \\
-2 & -1 & 0 & 3 & 1 & 0 & 0 & -1 & 2 & 0 & -1 & -1 & 1 & 0 & 0 & -1 & 1 & 1 & 0 & -2 & -3 & 0 & 1 & 2 \\
-2 & 1 & 0 & 1 & 3 & -2 & 0 & -1 & 0 & 0 & -1 & 1 & 1 & 0 & 2 & -3 & -1 & 1 & 0 & 0 & -1 & 0 & -1 & 2 \\
0 & -1 & 0 & 1 & -1 & 2 & 0 & -1 & 0 & 0 & 1 & -1 & 1 & 0 & -2 & 1 & 1 & -1 & 0 & 0 & -1 & 0 & 1 & 0 \\
4 & -3 & -6 & 5 & -9 & 6 & 12 & -9 & 2 & 0 & -3 & 1 & -3 & 0 & 6 & -3 & 5 & -3 & -6 & 4 & 1 & 0 & -3 & 2 \\
0 & 1 & 0 & -1 & -1 & -2 & 0 & 3 & 2 & 0 & -1 & -1 & -3 & 0 & 2 & 1 & 1 & 1 & 0 & -2 & 1 & 0 & -1 & 0 \\
0 & -1 & 0 & 1 & -1 & 0 & 0 & 1 & 2 & 0 & -1 & -1 & -1 & 0 & 0 & 1 & 1 & 1 & 0 & -2 & -1 & 0 & 1 & 0 \\
2 & -3 & 0 & 1 & 1 & -2 & 0 & 1 & -4 & 6 & 1 & -3 & -1 & 0 & 2 & -1 & -3 & 5 & 0 & -2 & 5 & -6 & -3 & 4 \\
0 & 1 & 0 & -1 & -1 & 2 & 0 & -1 & -2 & 0 & 3 & -1 & 1 & 0 & -2 & 1 & 1 & -3 & 0 & 2 & 1 & 0 & -1 & 0 \\
2 & -1 & 0 & -1 & 1 & 0 & 0 & -1 & -2 & 0 & -1 & 3 & 1 & 0 & 0 & -1 & -3 & 1 & 0 & 2 & 1 & 0 & 1 & -2 \\
0 & -1 & 0 & 1 & 1 & 2 & 0 & -3 & -2 & 0 & 1 & 1 & 3 & 0 & -2 & -1 & -1 & -1 & 0 & 2 & -1 & 0 & 1 & 0 \\
2 & -3 & 0 & 1 & -3 & 6 & 0 & -3 & 4 & -6 & -3 & 5 & -9 & 12 & 6 & -9 & 1 & -3 & 0 & 2 & 5 & -6 & -3 & 4 \\
0 & 1 & 0 & -1 & 1 & -2 & 0 & 1 & 0 & 0 & -1 & 1 & -1 & 0 & 2 & -1 & -1 & 1 & 0 & 0 & 1 & 0 & -1 & 0 \\
2 & -1 & 0 & -1 & -3 & 2 & 0 & 1 & 0 & 0 & 1 & -1 & -1 & 0 & -2 & 3 & 1 & -1 & 0 & 0 & 1 & 0 & 1 & -2 \\
-2 & 1 & 0 & 1 & -1 & 0 & 0 & 1 & 2 & 0 & 1 & -3 & -1 & 0 & 0 & 1 & 3 & -1 & 0 & -2 & -1 & 0 & -1 & 2 \\
0 & -1 & 0 & 1 & 1 & -2 & 0 & 1 & 2 & 0 & -3 & 1 & -1 & 0 & 2 & -1 & -1 & 3 & 0 & -2 & -1 & 0 & 1 & 0 \\
4 & -3 & -6 & 5 & -1 & 2 & 0 & -1 & -2 & 0 & 5 & -3 & 1 & 0 & -2 & 1 & -3 & 1 & 6 & -4 & 1 & 0 & -3 & 2 \\
0 & 1 & 0 & -1 & 1 & 0 & 0 & -1 & -2 & 0 & 1 & 1 & 1 & 0 & 0 & -1 & -1 & -1 & 0 & 2 & 1 & 0 & -1 & 0 \\
2 & 1 & 0 & -3 & -1 & 0 & 0 & 1 & -2 & 0 & 1 & 1 & -1 & 0 & 0 & 1 & -1 & -1 & 0 & 2 & 3 & 0 & -1 & -2 \\
-2 & 5 & 0 & -3 & 1 & -2 & 0 & 1 & 4 & -6 & -3 & 5 & -1 & 0 & 2 & -1 & 1 & -3 & 0 & 2 & -3 & 6 & 1 & -4 \\
2 & -3 & 0 & 1 & -1 & 2 & 0 & -1 & 0 & 0 & -1 & 1 & 1 & 0 & -2 & 1 & -1 & 1 & 0 & 0 & -1 & 0 & 3 & -2 \\
-2 & 1 & 0 & 1 & 1 & 0 & 0 & -1 & 0 & 0 & 1 & -1 & 1 & 0 & 0 & -1 & 1 & -1 & 0 & 0 & -1 & 0 & -1 & 2 \\
\end{array}
\right)\,, \qquad \\ &
\hspace{15mm} Y=\left(
\begin{array}{cccccccccccccccccccccccc}
-2 & 1 & 1 & 0 & 2 & -1 & -1 & 0 & -1 & 1 & 1 & -1 & 2 & -1 & -1 & 0 & 0 & 0 & 0 & 0 & -1 & 0 & 0 & 1 \\
-1 & 0 & 1 & 0 & 1 & 0 & -1 & 0 & 0 & 0 & 0 & 0 & 1 & 0 & -1 & 0 & 0 & 0 & 0 & 0 & -1 & 0 & 1 & 0 \\
-1 & 1 & 0 & 0 & 1 & -1 & 0 & 0 & 0 & 0 & 0 & 0 & 1 & -1 & 0 & 0 & 0 & 0 & 0 & 0 & -1 & 1 & 0 & 0 \\
-1 & 1 & 1 & -1 & 2 & -1 & -1 & 0 & -2 & 1 & 1 & 0 & 2 & -1 & -1 & 0 & -1 & 0 & 0 & 1 & 0 & 0 & 0 & 0 \\
0 & 0 & 0 & 0 & 1 & 0 & -1 & 0 & -1 & 0 & 1 & 0 & 1 & 0 & -1 & 0 & -1 & 0 & 1 & 0 & 0 & 0 & 0 & 0 \\
0 & 0 & 0 & 0 & 1 & -1 & 0 & 0 & -1 & 1 & 0 & 0 & 1 & -1 & 0 & 0 & -1 & 1 & 0 & 0 & 0 & 0 & 0 & 0 \\
0 & 0 & 0 & 0 & 0 & 0 & 0 & 0 & -1 & 1 & 1 & -1 & 1 & -1 & -1 & 1 & 0 & 0 & 0 & 0 & 0 & 0 & 0 & 0 \\
-1 & 1 & 1 & -1 & 1 & -1 & -1 & 1 & 0 & 0 & 0 & 0 & 0 & 0 & 0 & 0 & 0 & 0 & 0 & 0 & 0 & 0 & 0 & 0 \\
1 & 0 & 0 & 0 & 0 & 0 & 0 & 0 & 0 & 0 & 0 & 0 & 0 & 0 & 0 & 0 & 0 & 0 & 0 & 0 & 0 & 0 & 0 & 1 \\
0 & 1 & 0 & 0 & 0 & 0 & 0 & 0 & 0 & 0 & 0 & 0 & 0 & 0 & 0 & 0 & 0 & 0 & 0 & 0 & 0 & 0 & 1 & 0 \\
-2 & -1 & 0 & 0 & -1 & 0 & 0 & 0 & 0 & 1 & 0 & 0 & 0 & 0 & 0 & 0 & 0 & 0 & 0 & 0 & 0 & 1 & 0 & 0 \\
0 & 0 & 0 & 1 & 0 & 0 & 0 & 0 & 0 & 0 & 0 & 0 & 0 & 0 & 0 & 0 & 0 & 0 & 0 & 0 & 1 & 0 & 0 & 0 \\
1 & \frac{1}{2} & 0 & 1 & -\frac{1}{2} & 0 & 0 & 0 & 0 & 0 & 0 & 0 & 0 & 0 & 0 & 0 & 0 & 0 & 0 & 1 & 0 & 0 & 0 & 0 \\
2 & 1 & 1 & 0 & 1 & 0 & 0 & 0 & 0 & 0 & 0 & 0 & 0 & 0 & 0 & 0 & 0 & 0 & 1 & 0 & 0 & 0 & 0 & 0 \\
-2 & -\frac{1}{2} & 0 & -1 & -\frac{1}{2} & 0 & 0 & 0 & 0 & 0 & 0 & 0 & 0 & 0 & 0 & 0 & 0 & 1 & 0 & 0 & 0 & 0 & 0 & 0 \\
0 & -1 & 0 & -1 & 1 & 0 & 0 & 0 & 0 & 0 & 0 & 0 & 0 & 0 & 0 & 0 & 1 & 0 & 0 & 0 & 0 & 0 & 0 & 0 \\
0 & 0 & 0 & 0 & 1 & 0 & 0 & 0 & 0 & 0 & 0 & 0 & 0 & 0 & 0 & 1 & 0 & 0 & 0 & 0 & 0 & 0 & 0 & 0 \\
-1 & -\frac{1}{2} & 0 & 0 & -\frac{1}{2} & 0 & 0 & 0 & 0 & 0 & 0 & 0 & 0 & 0 & 1 & 0 & 0 & 0 & 0 & 0 & 0 & 0 & 0 & 0 \\
2 & \frac{3}{2} & 0 & 1 & -\frac{1}{2} & 0 & 0 & 0 & 0 & 0 & 0 & 0 & 1 & 0 & 0 & 0 & 0 & 0 & 0 & 0 & 0 & 0 & 0 & 0 \\
0 & 1 & 0 & 1 & -1 & 0 & 0 & 0 & 0 & 0 & 0 & 1 & 0 & 0 & 0 & 0 & 0 & 0 & 0 & 0 & 0 & 0 & 0 & 0 \\
2 & \frac{1}{2} & 0 & 1 & \frac{1}{2} & 0 & 0 & 0 & 0 & 0 & 1 & 0 & 0 & 0 & 0 & 0 & 0 & 0 & 0 & 0 & 0 & 0 & 0 & 0 \\
-1 & -\frac{1}{2} & 0 & -1 & \frac{1}{2} & 0 & 0 & 0 & 1 & 0 & 0 & 0 & 0 & 0 & 0 & 0 & 0 & 0 & 0 & 0 & 0 & 0 & 0 & 0 \\
-2 & -\frac{3}{2} & 0 & -1 & \frac{1}{2} & 0 & 0 & 1 & 0 & 0 & 0 & 0 & 0 & 0 & 0 & 0 & 0 & 0 & 0 & 0 & 0 & 0 & 0 & 0 \\
1 & \frac{1}{2} & 0 & 0 & \frac{1}{2} & 1 & 0 & 0 & 0 & 0 & 0 & 0 & 0 & 0 & 0 & 0 & 0 & 0 & 0 & 0 & 0 & 0 & 0 & 0 \\
\end{array}
\right)
\,,\qquad \\ \\ &
\hspace{5mm} D= \D{8}  \, .
\end{split}
\end{align}
The eight colour structures are
\begin{eqnarray}
(YC)_1 &=& - i\fadh \fbcm \fdbj \ta 1 \tm 1 \thh 2 \tj 3 \tc 4 -
i \facp \fadh \fdbj \tb 1 \tp 1 \thh 2 \tj 3 \tc 4 \nonumber \\&& +
i \fbcm \fdbj \fman \tn 1 \td 2 \ta 2 \tj 3 \tc 4\nonumber -
i \fadh \fbcm \fman \tn 1 \thh 2 \tb 3 \td 3 \tc 4 \, , \\ \nonumber \\
(YC)_2 &=& -i \fadh \fbcm \fman\tn 1 \thh 2 \tb 3 \td 3 \tc 4  \, , \nonumber \\ \nonumber \\
(YC)_3 &=& i \fbcm \fdbj \fman  \tn 1 \ta 2 \td 2 \tj 3 \tc 4 \, , \nonumber \\ \nonumber \\
(YC)_4 &=& -i\fadh \fbcm \fdbj \ta 1 \tm 1 \thh 2 \tj 3 \tc 4-
i \facp \fadh \fdbj \tb 1 \tp 1 \thh 2 \tj 3 \tc 4 \nonumber \\&& + 
i \facp \fdbj \fpbq \tq 1 \td 2 \ta 2 \tj 3 \tc 4 -
i \facp \fadh \fpbq \tq 1 \thh 2 \tb 3 \td 3 \tc 4 \, , \nonumber \\ \nonumber \\
(YC)_5 &=& -i \facp \fadh \fpbq \tq 1 \thh 2 \tb 3 \td 3 \tc 4  \, , \nonumber \\ \nonumber \\
(YC)_6 &=& i \facp \fdbj \fpbq \tq 1 \ta 2 \td 2 \tj 3 \tc 4 \, , \nonumber \\ \nonumber \\
 (YC)_7 &=& i \fadh \facp \fdbj \tb 1 \tp 1 \thh 2 \tj 3 \tc 4 \, ,\nonumber \\ \nonumber \\
 (YC)_8 &=& -i \fadh \fbcm \fdbj \ta 1 \tm 1 \thh 2 \tj 3 \tc 4 \, .
\end{eqnarray} 


\end{itemize}


\vspace{4mm}


\subsection{Cwebs connecting five Wilson lines}

\vspace{4mm}


\begin{itemize}


\item[{\bf 1}.] $\text W_5^{(1,0,1)}(1,1,1,1,2)$ \\

\vspace{-2mm}
A simple two-diagram Cweb,

\vspace{-2mm}
\begin{minipage}{0.5\textwidth}
	\begin{figure}[H]
		\includegraphics[height=4cm,width=4cm]{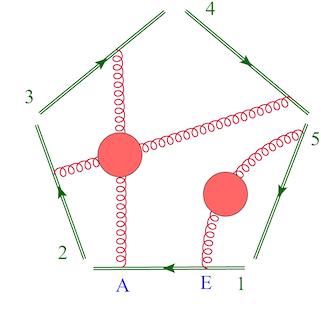}
	\end{figure}
\end{minipage} 
\hspace{-2.0cm}
\begin{minipage}{0.45\textwidth}
	\begin{tabular}{ | c | c | c |}
		\hline
		\textbf{Diagrams} & \textbf{Sequences} & \textbf{S-factors} \\ \hline
		$C_1$ & $\lbrace \lbrace EA\rbrace\rbrace$ & 1 \\ 
		\hline
		$C_2$ & $\lbrace \lbrace AE\rbrace\rbrace$ & 1 \\ 
		\hline
	\end{tabular}
	\label{tab:abcd21}	
\end{minipage} 

\vspace{2mm}
\noindent The $R$, $Y$ and $D$ matrices are given by
\begin{align}
\begin{split}
&
R=\displaystyle{\left(
		\begin{array}{cc}
		\frac{1}{2} & -\frac{1}{2} \\
		-\frac{1}{2} & \frac{1}{2} \\
		\end{array}
		\right) } \, , \qquad
Y=\left(
	\begin{array}{cc}
	-1 & 1 \\
	1 & 1 \\
	\end{array}
	\right) \, , \qquad
D= \D{1} \, ,
\end{split}
\end{align} 
and the single colour factor is
\begin{eqnarray}
  (YC)_1 &=&  i \fabg \fcdg \feah \thh 1 \tb 2 \tc 3 \td 4 \te 5 \, .
\end{eqnarray}

\vspace{2mm}


\item[{\bf 2}.] $\text W_5^{(0,2)}(1,1,1,1,2)$ \\

\vspace{-2mm}
Another simple two-diagram Cweb,

\begin{minipage}{0.5\textwidth}
		\begin{figure}[H]
		\includegraphics[height=4cm,width=4cm]{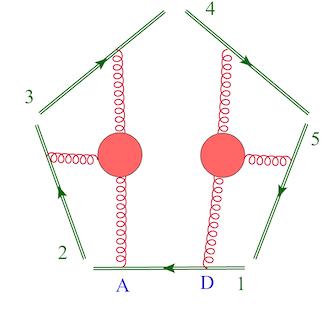}
		\end{figure}
	\end{minipage} 
	\hspace{-2.0cm}
	\begin{minipage}{0.45\textwidth}
		\begin{tabular}{ | c | c | c |}
			\hline
			\textbf{Diagrams} & \textbf{Sequences} & \textbf{S-factors} \\ \hline
			$C_1$ & $\lbrace \lbrace DA\rbrace\rbrace$ & 1 \\ 
			\hline
			$C_2$ & $\lbrace \lbrace AD\rbrace\rbrace$ & 1 \\ 
			\hline
		\end{tabular}
		\label{tab:abcd22}	
	\end{minipage} 
	
\vspace{2mm}
\noindent The $R$, $Y$ and $D$ matrices are given by
\begin{align}
\begin{split}
		&
		R=\displaystyle{\left(
			\begin{array}{cc}
			\frac{1}{2} & -\frac{1}{2} \\
			-\frac{1}{2} & \frac{1}{2} \\
			\end{array}
			\right) }\,, \qquad
		Y=\left(
		\begin{array}{cc}
		-1 & 1 \\
		1 & 1 \\
		\end{array}
		\right)
		\,,\qquad
		D= \D{1} \,  ,
\end{split}
\end{align} 
and there is a single colour factor,
\begin{eqnarray}
  (YC)_1 &=& -i \fabc \fdef \fdah \thh 1 \tb 2 \tc 3 \te 4 \tf 5 \, .
\end{eqnarray} 

\vspace{2mm}


\item[{\bf 3}.] $\text W_5^{(2,1)}(1,1,1,1,3)$ \\

\vspace{-2mm}
A six-diagram Cweb,

\begin{minipage}{0.5\textwidth}
		\begin{figure}[H]
		\includegraphics[height=4cm,width=4cm]{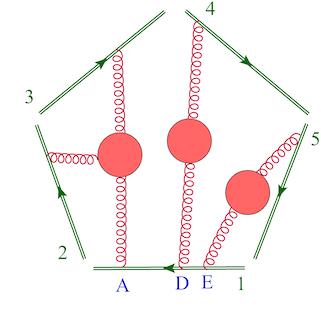}
		\end{figure}
	\end{minipage} 
	\hspace{-2.0cm}
	\begin{minipage}{0.45\textwidth}
		\begin{tabular}{ | c | c | c |}
			\hline
			\textbf{Diagrams} & \textbf{Sequences} & \textbf{S-factors} \\ \hline
			$C_1$ & $\lbrace \lbrace EDA\rbrace\rbrace$ & 1 \\ 
			\hline
			$C_2$ & $\lbrace \lbrace DEA\rbrace\rbrace$ & 1 \\ 
			\hline
				$C_3$ & $\lbrace \lbrace EAD\rbrace\rbrace$ & 1 \\ 
			\hline	
			$C_4$ & $\lbrace \lbrace AED\rbrace\rbrace$ & 1 \\ 
			\hline	
			$C_5$ & $\lbrace \lbrace DAE\rbrace\rbrace$ & 1 \\ 
			\hline
			$C_6$ & $\lbrace \lbrace ADE\rbrace\rbrace$ & 1 \\ 
			\hline
		\end{tabular}
		\label{tab:abcd23}	
	\end{minipage} 

\vspace{3mm}
\noindent where the $R$, $Y$ and $D$ matrices are given by

\begin{align}
\begin{split}
&
R=\displaystyle{\left(
		\begin{array}{cccccc}
		\frac{1}{3} & -\frac{1}{6} & -\frac{1}{6} & -\frac{1}{6} & -\frac{1}{6} & \frac{1}{3} \\
		-\frac{1}{6} & \frac{1}{3} & -\frac{1}{6} & \frac{1}{3} & -\frac{1}{6} & -\frac{1}{6} \\
		-\frac{1}{6} & -\frac{1}{6} & \frac{1}{3} & -\frac{1}{6} & \frac{1}{3} & -\frac{1}{6} \\
		-\frac{1}{6} & \frac{1}{3} & -\frac{1}{6} & \frac{1}{3} & -\frac{1}{6} & -\frac{1}{6} \\
		-\frac{1}{6} & -\frac{1}{6} & \frac{1}{3} & -\frac{1}{6} & \frac{1}{3} & -\frac{1}{6} \\
		\frac{1}{3} & -\frac{1}{6} & -\frac{1}{6} & -\frac{1}{6} & -\frac{1}{6} & \frac{1}{3} \\
		\end{array}
		\right) }\,, \quad
Y=\left(
		\begin{array}{cccccc}
		1 & -1 & 0 & -1 & 0 & 1 \\
		0 & -1 & 1 & -1 & 1 & 0 \\
		-1 & 0 & 0 & 0 & 0 & 1 \\
		1 & 1 & 0 & 0 & 1 & 0 \\
		0 & -1 & 0 & 1 & 0 & 0 \\
		1 & 1 & 1 & 0 & 0 & 0 \\
		\end{array}
	\right) \,, \quad
D= \D{2} \, ,
\end{split}
\end{align} 

\vspace{2mm}
and there are two colour structures,
\begin{eqnarray}
	(YC)_1 &=&  i \fabc \fdeg \fgah \thh 1 \tb 2 \tc 3 \td 4 \te 5 \, , \nonumber \\ \nonumber \\
	(YC)_2 &=& -i \fabc \faej \fdjm \tm 1 \tb 2 \tc 3 \td 4 \te 5 \, .
\end{eqnarray} 

\vspace{2mm}


\item[{\bf 4}.] $\text W_{5 ,\, {\rm I}}^{(2,1)}(1,1,1,2,2)$ \\

\vspace{-2mm}
A four-diagram Cweb, one of two with this set  of correlators and attachments.

\begin{minipage}{0.5\textwidth}
		\begin{figure}[H]
		\includegraphics[height=4cm,width=4cm]{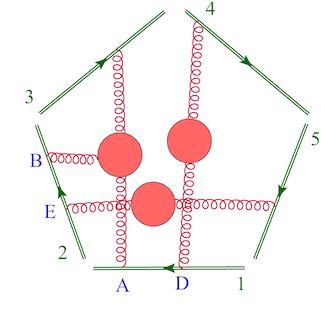}
		\end{figure}
	\end{minipage} 
	\hspace{-2.0cm}
\begin{minipage}{0.46\textwidth}
		\begin{tabular}{ | c | c | c |}
			\hline
			\textbf{Diagrams} & \textbf{Sequences} & \textbf{S-factors} \\ \hline
			$C_1$ & $\lbrace \lbrace DA\rbrace , \lbrace EB \rbrace \rbrace$ & 2 \\ 
			\hline
			$C_2$ & $\lbrace \lbrace DA\rbrace , \lbrace BE \rbrace \rbrace$ & 1 \\ 
			\hline
			$C_3$ & $\lbrace \lbrace AD\rbrace , \lbrace EB \rbrace \rbrace$ & 1 \\ 
			\hline
			$C_4$ & $\lbrace \lbrace AD\rbrace , \lbrace BE \rbrace \rbrace$ & 2 \\ 
			\hline
			\end{tabular}
		\label{tab:abcd24}	
\end{minipage} 

\vspace{2mm} 
\noindent The $R$, $Y$ and $D$ matrices are given by
	
\vspace{-2mm}		
\begin{align}
\begin{split}
		&
R=\displaystyle{\left(
		\begin{array}{cccc}
		\frac{1}{6} & -\frac{1}{6} & -\frac{1}{6} & \frac{1}{6} \\
		-\frac{1}{3} & \frac{1}{3} & \frac{1}{3} & -\frac{1}{3} \\
		-\frac{1}{3} & \frac{1}{3} & \frac{1}{3} & -\frac{1}{3} \\
		\frac{1}{6} & -\frac{1}{6} & -\frac{1}{6} & \frac{1}{6} \\
		\end{array}
		\right) }\,, \qquad
Y=\left(
		\begin{array}{cccc}
		1 & -1 & -1 & 1 \\
		-1 & 0 & 0 & 1 \\
		2 & 0 & 1 & 0 \\
		2 & 1 & 0 & 0 \\
		\end{array}
		\right)
		\,,\qquad
D= \D{1} \, ,
\end{split}
\end{align} 
The single colour factor is
\begin{eqnarray}
  (YC)_1 &=&  - i \fabc \fadh \fbeg \thh 1 \tg 2 \tc 3 \td 4 \te 5 \, .
\end{eqnarray} 

\vspace{2mm}


\item[{\bf 5}.] $\text W_{5 ,\, {\rm II}}^{(2,1)}(1,1,1,2,2)$ \\

\vspace{-2mm}
The second Cweb of the set, also with four diagrams.

\begin{minipage}{0.5\textwidth}
		\begin{figure}[H]
		\includegraphics[height=4cm,width=4cm]{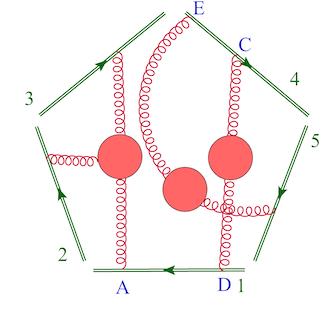}
		\end{figure}
	\end{minipage} 
	\hspace{-2.0cm}
	\begin{minipage}{0.45\textwidth}
		\begin{tabular}{ | c | c | c |}
			\hline
			\textbf{Diagrams} & \textbf{Sequences} & \textbf{S-factors} \\ \hline
			$C_1$ & $\lbrace \lbrace DA\rbrace , \lbrace EC \rbrace \rbrace$ & 1 \\ 
			\hline
			$C_2$ & $\lbrace \lbrace DA\rbrace  , \lbrace CE \rbrace \rbrace$ & 2 \\ 
			\hline
			$C_3$ & $\lbrace \lbrace AD\rbrace  , \lbrace EC \rbrace \rbrace$ & 2 \\ 
			\hline
			$C_4$ & $\lbrace \lbrace AD\rbrace  , \lbrace CE \rbrace \rbrace$ & 1 \\ 
			\hline
		\end{tabular}
		\label{tab:abcd25}	
	\end{minipage}

\vspace{2mm} 
\noindent The $R$, $Y$ and $D$ matrices are given by

\vspace{-2mm}
\begin{align}
\begin{split}
&
R=\displaystyle{\left(
			\begin{array}{cccc}
			\frac{1}{3} & -\frac{1}{3} & -\frac{1}{3} & \frac{1}{3} \\
			-\frac{1}{6} & \frac{1}{6} & \frac{1}{6} & -\frac{1}{6} \\
			-\frac{1}{6} & \frac{1}{6} & \frac{1}{6} & -\frac{1}{6} \\
			\frac{1}{3} & -\frac{1}{3} & -\frac{1}{3} & \frac{1}{3} \\
			\end{array}
			\right) }\,, \qquad
Y=\left(
		\begin{array}{cccc}
		1 & -1 & -1 & 1 \\
		-1 & 0 & 0 & 1 \\
		\frac{1}{2} & 0 & 1 & 0 \\
		\frac{1}{2} & 1 & 0 & 0 \\
		\end{array}
		\right)
		\,,\qquad
D= \D{1} \, .
\end{split}
\end{align}
\vspace{2mm}
The single colour factor is
\begin{eqnarray}
	(YC)_1 &=&  -i \fabc \fadh \fdeg \thh 1 \tb 2 \tc 3 \tg 4 \te 5  \, .
\end{eqnarray} 

\vspace{2mm}


\item[{\bf 6}.] $\text W_5^{(4)}(1,1,1,2,2)$ \\

\vspace{-2mm}
A Cweb with eight diagrams

\begin{minipage}{0.5\textwidth}
		\begin{figure}[H]
		\includegraphics[height=4cm,width=4cm]{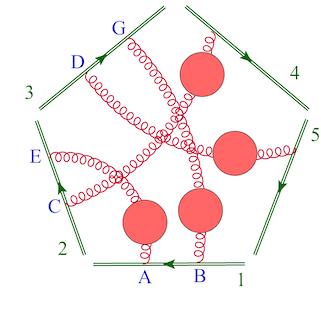}
		\end{figure}
	\end{minipage} 
	\hspace{-2.0cm}
	\begin{minipage}{0.55\textwidth}
		\begin{tabular}{ | c | c | c |}
			\hline
			\textbf{Diagrams} & \textbf{Sequences} & \textbf{S-factors} \\ \hline
			$C_1$ & $\lbrace \lbrace BA\rbrace,  \lbrace CE \rbrace, \lbrace DG\rbrace \rbrace$ & 3 \\ 
			\hline
			$C_2$ & $\lbrace \lbrace BA\rbrace,  \lbrace CE \rbrace, \lbrace GD\rbrace \rbrace$ & 4 \\ 
			\hline
			$C_3$ & $\lbrace \lbrace BA\rbrace,  \lbrace EC \rbrace, \lbrace DG\rbrace \rbrace$ & 1 \\ 
			\hline
			$C_4$ & $\lbrace \lbrace BA\rbrace,  \lbrace EC \rbrace, \lbrace GD\rbrace \rbrace$ & 2 \\ 
			\hline
			$C_5$ & $\lbrace \lbrace AB\rbrace,  \lbrace CE \rbrace, \lbrace DG\rbrace \rbrace$ & 2 \\ 
			\hline
			$C_6$ & $\lbrace \lbrace AB\rbrace,  \lbrace CE \rbrace, \lbrace GD\rbrace \rbrace$ & 1 \\ 
			\hline
			$C_7$ & $\lbrace \lbrace AB\rbrace,  \lbrace EC \rbrace, \lbrace DG\rbrace \rbrace$ & 4 \\ 
			\hline
			$C_8$ & $\lbrace \lbrace AB\rbrace,  \lbrace EC \rbrace, \lbrace GD\rbrace \rbrace$ & 3 \\ 
			\hline
		\end{tabular}
		\label{tab:abcd26}	
	\end{minipage} 

\vspace{6mm} 
\noindent The $R$, $Y$ and $D$ matrices are given by

\begin{align}
\begin{split}
&
R=\displaystyle\left(
		\begin{array}{cccccccc}
		\frac{1}{12} & -\frac{1}{12} & -\frac{1}{12} & \frac{1}{12} & -\frac{1}{12} & \frac{1}{12} & \frac{1}{12} & -\frac{1}{12} \\
		-\frac{1}{12} & \frac{1}{12} & \frac{1}{12} & -\frac{1}{12} & \frac{1}{12} & -\frac{1}{12} & -\frac{1}{12} & \frac{1}{12} \\
		-\frac{1}{4} & \frac{1}{4} & \frac{1}{4} & -\frac{1}{4} & \frac{1}{4} & -\frac{1}{4} & -\frac{1}{4} & \frac{1}{4} \\
		\frac{1}{12} & -\frac{1}{12} & -\frac{1}{12} & \frac{1}{12} & -\frac{1}{12} & \frac{1}{12} & \frac{1}{12} & -\frac{1}{12} \\
		-\frac{1}{12} & \frac{1}{12} & \frac{1}{12} & -\frac{1}{12} & \frac{1}{12} & -\frac{1}{12} & -\frac{1}{12} & \frac{1}{12} \\
		\frac{1}{4} & -\frac{1}{4} & -\frac{1}{4} & \frac{1}{4} & -\frac{1}{4} & \frac{1}{4} & \frac{1}{4} & -\frac{1}{4} \\
		\frac{1}{12} & -\frac{1}{12} & -\frac{1}{12} & \frac{1}{12} & -\frac{1}{12} & \frac{1}{12} & \frac{1}{12} & -\frac{1}{12} \\
		-\frac{1}{12} & \frac{1}{12} & \frac{1}{12} & -\frac{1}{12} & \frac{1}{12} & -\frac{1}{12} & -\frac{1}{12} & \frac{1}{12} \\
		\end{array}
		\right)\,, \qquad \\ &
Y=\left(
		\begin{array}{cccccccc}
		-1 & 1 & 1 & -1 & 1 & -1 & -1 & 1 \\
		1 & 0 & 0 & 0 & 0 & 0 & 0 & 1 \\
		-1 & 0 & 0 & 0 & 0 & 0 & 1 & 0 \\
		-3 & 0 & 0 & 0 & 0 & 1 & 0 & 0 \\
		1 & 0 & 0 & 0 & 1 & 0 & 0 & 0 \\
		-1 & 0 & 0 & 1 & 0 & 0 & 0 & 0 \\
		3 & 0 & 1 & 0 & 0 & 0 & 0 & 0 \\
		1 & 1 & 0 & 0 & 0 & 0 & 0 & 0 \\
		\end{array}
		\right)
		\,,\qquad
		\\ & \\ &
D= \D{1} \,  .
\end{split}
\end{align}

\vspace{2mm}
While large, this Cweb has only one exponentiated colour factor,
\begin{eqnarray}
	(YC)_1 &=&  i \fach \fabg f^{bdj} \tg 1 \thh 2 \tj 3 \tc 4 \td 5 \,  . 
\end{eqnarray} 

\vspace{2mm}

\pagebreak


\item[{\bf 7}.] $\text W_5^{(4)}(1,1,1,2,3)$ \\

\vspace{-2mm}
A Cweb with twelve diagrams,

\vspace{2mm}
\begin{minipage}{0.5\textwidth}
		\begin{figure}[H]
		\includegraphics[height=4cm,width=4cm]{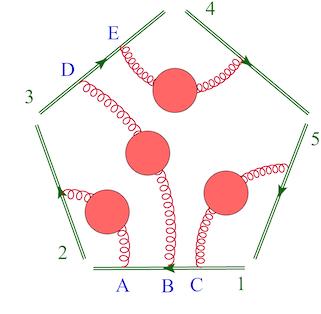}
		\end{figure}
	\end{minipage} 
	\hspace{-2.0cm}
	\begin{minipage}{0.45\textwidth}
	\begin{tabular}{ |c|c|c| } 
		\hline
		\textbf{Diagrams}  & \textbf{Sequences}  & \textbf{S-factor}  \\ 
		\hline
		$C_1$ & \{\{CBA\},\{DE\}\} & 2 \\ 
		\hline
		$C_2$ & \{\{CBA\},\{ED\}\} & 1 \\ 
		\hline
		$C_3$ & \{\{BCA\},\{DE\}\} & 1 \\ 
		\hline
		$C_4$ & \{\{BCA\},\{ED\}\} & 1 \\ 
		\hline
		$C_5$ & \{\{CAB\},\{DE\}\} & 2 \\ 
		\hline
		$C_6$ & \{\{CAB\},\{ED\}\} & 1 \\ 
		\hline
		$C_7$ & \{\{ACB\},\{DE\}\} & 2 \\ 
		\hline
		$C_8$ & \{\{ACB\},\{ED\}\} & 1 \\ 
		\hline
		$C_9$ & \{\{BAC\},\{DE\}\} & 1 \\ 
		\hline
		$C_{10}$ & \{\{BAC\},\{ED\}\} & 1 \\ 
		\hline
		$C_{11}$ & \{\{ABC\},\{DE\}\} & 2 \\ 
		\hline
		$C_{12}$ & \{\{ABC\},\{ED\}\} & 1 \\ 
		\hline
		\end{tabular}
		\label{tab:abcd27}	
	\end{minipage} 

\vspace{2mm}
with $R$, $Y$ and $D$ matrices given by

\vspace{-2mm}
\begin{align}
\begin{split}
&
R=\frac{1}{12} \left(
		\begin{array}{cccccccccccc}
		2 & -2 & -1 & 1 & -1 & 1 & -1 & 1 & -1 & 1 & 2 & -2 \\
		-2 & 2 & 1 & -1 & 1 & -1 & 1 & -1 & 1 & -1 & -2 & 2 \\
		-2 & 2 & 3 & -3 & -1 & 1 & 3 & -3 & -1 & 1 & -2 & 2 \\
		0 & 0 & -1 & 1 & 1 & -1 & -1 & 1 & 1 & -1 & 0 & 0 \\
		0 & 0 & -1 & 1 & 1 & -1 & -1 & 1 & 1 & -1 & 0 & 0 \\
		2 & -2 & 1 & -1 & -3 & 3 & 1 & -1 & -3 & 3 & 2 & -2 \\
		0 & 0 & 1 & -1 & -1 & 1 & 1 & -1 & -1 & 1 & 0 & 0 \\
		2 & -2 & -3 & 3 & 1 & -1 & -3 & 3 & 1 & -1 & 2 & -2 \\
		-2 & 2 & -1 & 1 & 3 & -3 & -1 & 1 & 3 & -3 & -2 & 2 \\
		0 & 0 & 1 & -1 & -1 & 1 & 1 & -1 & -1 & 1 & 0 & 0 \\
		2 & -2 & -1 & 1 & -1 & 1 & -1 & 1 & -1 & 1 & 2 & -2 \\
		-2 & 2 & 1 & -1 & 1 & -1 & 1 & -1 & 1 & -1 & -2 & 2 \\
		\end{array}
		\right)\,, \qquad \\ &
\hspace{1cm} Y=\left(
		\begin{array}{cccccccccccc}
		-1 & 1 & 1 & -1 & 0 & 0 & 1 & -1 & 0 & 0 & -1 & 1 \\
		0 & 0 & 1 & -1 & -1 & 1 & 1 & -1 & -1 & 1 & 0 & 0 \\
		1 & 0 & 0 & 0 & 0 & 0 & 0 & 0 & 0 & 0 & 0 & 1 \\
		-1 & 0 & 0 & 0 & 0 & 0 & 0 & 0 & 0 & 0 & 1 & 0 \\
		-\frac{1}{2} & 0 & -\frac{1}{2} & 0 & 0 & 0 & 0 & 0 & 0 & 1 & 0 & 0 \\
		2 & 0 & 1 & 0 & 0 & 0 & 0 & 0 & 1 & 0 & 0 & 0 \\
		0 & 0 & 1 & 0 & 0 & 0 & 0 & 1 & 0 & 0 & 0 & 0 \\
		-\frac{1}{2} & 0 & -\frac{1}{2} & 0 & 0 & 0 & 1 & 0 & 0 & 0 & 0 & 0 \\
		-2 & 0 & -1 & 0 & 0 & 1 & 0 & 0 & 0 & 0 & 0 & 0 \\
		\frac{1}{2} & 0 & \frac{1}{2} & 0 & 1 & 0 & 0 & 0 & 0 & 0 & 0 & 0 \\
		\frac{1}{2} & 0 & \frac{1}{2} & 1 & 0 & 0 & 0 & 0 & 0 & 0 & 0 & 0 \\
		1 & 1 & 0 & 0 & 0 & 0 & 0 & 0 & 0 & 0 & 0 & 0 \\
		\end{array}
		\right)	\,,\qquad
D= \D{2} \, .
\end{split}
\end{align}
and two colour factors,
\begin{eqnarray}
	(YC)_1 &=&  i \fach \fbdm f^{bhv} \tv 1 \ta 2 \tm 3 \td 4 \tc 5 \nonumber \\ && 
		 -i \fabg \fbdm f^{cgu} \tu 1 \ta 2 \tm 3 \td 4 \tc 5 \, , \nonumber  \\ \nonumber\\
	(YC)_2&=&i \fach \fbdm f^{bhv} \tv 1 \ta 2 \tm 3 \td 4 \tc 5 \, .
\end{eqnarray} 

\vspace{2mm}


\item[{\bf 8}.] $\text W_{5}^{(4)}(1,1,1,1,4)$ \\

\vspace{-2mm}
Last, but not least, a Cweb with twenty-four diagrams

\vspace{1cm}
\begin{minipage}{0.5\textwidth}
		\begin{figure}[H]
			\includegraphics[height=4cm,width=4cm]{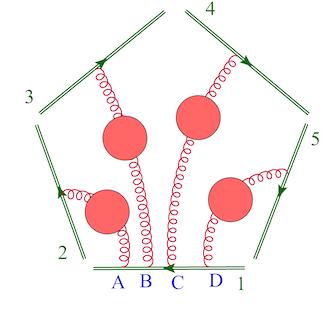}
		\end{figure}
	\end{minipage} 
	\hspace{-2.0cm}
	\begin{minipage}{0.45\textwidth}
		\begin{tabular}{ |c|c|c| } 
			\hline
			\textbf{Diagrams}  & \textbf{Sequences}  & \textbf{S-factor}  \\ 
			\hline
			$C_1$ & \{ABCD\} & 1 \\ 
			\hline
			$C_2$ & \{DCBA\} & 1 \\ 
			\hline
			$C_3$ & \{CDBA\} & 1 \\ 
			\hline
			$C_4$ & \{DBCA\} & 1 \\ 
			\hline
			$C_5$ & \{BDCA\} & 1 \\ 
			\hline
			$C_6$ & \{CBDA\} & 1 \\ 
			\hline
			$C_7$ & \{BCDA\} & 1 \\ 
			\hline
			$C_8$ & \{DCAB\} & 1 \\ 
			\hline
			$C_9$ & \{CDAB\} & 1 \\ 
			\hline
			$C_{10}$ & \{DACB\} & 1 \\ 
			\hline
			$C_{11}$ & \{ADCB\} & 1 \\ 
			\hline
			$C_{12}$ & \{CADB\} & 1 \\ 
			\hline
			$C_{13}$ & \{ACDB\} & 1\\ 
			\hline
			$C_{14}$ & \{DBAC\} & 1 \\ 
			\hline
			$C_{15}$ & \{BDAC\} & 1 \\ 
			\hline
			$C_{16}$ & \{DABC\} & 1 \\ 
			\hline
			$C_{17}$ & \{ADBC\} & 1 \\ 
			\hline
			$C_{18}$ & \{BADC\} & 1 \\ 
			\hline
			$C_{19}$ & \{ABDC\} & 1 \\ 
			\hline
			$C_{20}$ & \{CBAD\} & 1 \\ 
			\hline
			$C_{21}$ & \{BCAD\} & 1 \\ 
			\hline
			$C_{22}$ & \{CABD\} & 1 \\ 
			\hline
			$C_{23}$ & \{ACBD\} & 1 \\ 
			\hline
			$C_{24}$ & \{BACD\} & 1 \\ 
			\hline
			\end{tabular}
		\label{tab:abcd28}	
	\end{minipage} 

\vspace{5mm}
with $R$, $Y$ and $D$ matrices given by

\begin{align}
\begin{split}
\hspace{-1cm}
&
R=\frac{1}{12} \left(
		\begin{array}{cccccccccccccccccccccccc}
		3 & -1 & -1 & -1 & -1 & 1 & -1 & 1 & -1 & -1 & -1 & 1 & -1 & 1 & 1 & 1 & -1 & 1 & -1 & 1 & 1 & 1 & 1 & -3 \\
		-1 & 3 & -1 & 1 & -1 & -1 & 1 & -1 & -1 & 1 & -1 & -1 & -1 & 1 & 1 & 1 & 1 & -3 & -1 & 1 & 1 & 1 & -1 & 1 \\
		-1 & -1 & 3 & -1 & 1 & -1 & -1 & 1 & 1 & 1 & -1 & 1 & -1 & 1 & -1 & -1 & -1 & 1 & 1 & -1 & 1 & -3 & 1 & 1 \\
		-1 & 1 & -1 & 3 & -1 & -1 & -1 & 1 & 1 & 1 & 1 & -3 & 1 & -1 & -1 & 1 & -1 & -1 & 1 & -1 & -1 & 1 & 1 & 1 \\
		-1 & -1 & 1 & -1 & 3 & -1 & 1 & -1 & -1 & 1 & 1 & 1 & 1 & -1 & 1 & -3 & 1 & 1 & -1 & 1 & -1 & -1 & -1 & 1 \\
		1 & -1 & -1 & -1 & -1 & 3 & 1 & -1 & 1 & -3 & 1 & 1 & 1 & -1 & -1 & 1 & 1 & 1 & 1 & -1 & -1 & 1 & -1 & -1 \\
		-1 & 1 & -1 & -1 & -1 & 1 & 3 & -1 & -1 & -1 & -1 & 1 & 1 & 1 & -1 & 1 & 1 & -1 & 1 & 1 & -1 & 1 & -3 & 1 \\
		1 & -1 & -1 & 1 & -1 & -1 & -1 & 3 & -1 & 1 & -1 & -1 & 1 & 1 & -1 & 1 & -3 & 1 & 1 & 1 & -1 & 1 & 1 & -1 \\
		-1 & 1 & 1 & 1 & -1 & 1 & -1 & -1 & 3 & -1 & 1 & -1 & -1 & -1 & -1 & 1 & 1 & -1 & 1 & -3 & 1 & -1 & 1 & 1 \\
		-1 & 1 & 1 & 1 & 1 & -3 & -1 & 1 & -1 & 3 & -1 & -1 & -1 & 1 & 1 & -1 & -1 & -1 & -1 & 1 & 1 & -1 & 1 & 1 \\
		1 & -1 & -1 & 1 & 1 & 1 & -1 & -1 & 1 & -1 & 3 & -1 & 1 & -3 & 1 & -1 & 1 & 1 & -1 & -1 & -1 & 1 & 1 & -1 \\
		1 & -1 & 1 & -3 & 1 & 1 & 1 & -1 & -1 & -1 & -1 & 3 & -1 & 1 & 1 & -1 & 1 & 1 & -1 & 1 & 1 & -1 & -1 & -1 \\
		-1 & -1 & -1 & 1 & 1 & -1 & 1 & 1 & -1 & 1 & 1 & -1 & 3 & -1 & -1 & -1 & -1 & 1 & 1 & 1 & -3 & 1 & -1 & 1 \\
		-1 & 1 & 1 & -1 & -1 & -1 & 1 & 1 & -1 & 1 & -3 & 1 & -1 & 3 & -1 & 1 & -1 & -1 & 1 & 1 & 1 & -1 & -1 & 1 \\
		1 & 1 & -1 & 1 & 1 & -1 & -1 & -1 & -1 & 1 & 1 & -1 & -1 & -1 & 3 & -1 & 1 & -1 & -3 & 1 & 1 & 1 & 1 & -1 \\
		1 & 1 & -1 & 1 & -3 & 1 & -1 & 1 & 1 & -1 & -1 & -1 & -1 & 1 & -1 & 3 & -1 & -1 & 1 & -1 & 1 & 1 & 1 & -1 \\
		-1 & 1 & 1 & -1 & 1 & 1 & 1 & -3 & 1 & -1 & 1 & 1 & -1 & -1 & 1 & -1 & 3 & -1 & -1 & -1 & 1 & -1 & -1 & 1 \\
		1 & -3 & 1 & -1 & 1 & 1 & -1 & 1 & 1 & -1 & 1 & 1 & 1 & -1 & -1 & -1 & -1 & 3 & 1 & -1 & -1 & -1 & 1 & -1 \\
		-1 & -1 & 1 & -1 & -1 & 1 & 1 & 1 & 1 & -1 & -1 & 1 & 1 & 1 & -3 & 1 & -1 & 1 & 3 & -1 & -1 & -1 & -1 & 1 \\
		1 & -1 & -1 & -1 & 1 & -1 & 1 & 1 & -3 & 1 & -1 & 1 & 1 & 1 & 1 & -1 & -1 & 1 & -1 & 3 & -1 & 1 & -1 & -1 \\
		1 & 1 & 1 & -1 & -1 & 1 & -1 & -1 & 1 & -1 & -1 & 1 & -3 & 1 & 1 & 1 & 1 & -1 & -1 & -1 & 3 & -1 & 1 & -1 \\
		1 & 1 & -3 & 1 & -1 & 1 & 1 & -1 & -1 & -1 & 1 & -1 & 1 & -1 & 1 & 1 & 1 & -1 & -1 & 1 & -1 & 3 & -1 & -1 \\
		1 & -1 & 1 & 1 & 1 & -1 & -3 & 1 & 1 & 1 & 1 & -1 & -1 & -1 & 1 & -1 & -1 & 1 & -1 & -1 & 1 & -1 & 3 & -1 \\
		-3 & 1 & 1 & 1 & 1 & -1 & 1 & -1 & 1 & 1 & 1 & -1 & 1 & -1 & -1 & -1 & 1 & -1 & 1 & -1 & -1 & -1 & -1 & 3 \\
		\end{array}
		\right)\,, \\ & \\ &
Y=\left(
		\begin{array}{cccccccccccccccccccccccc}
		-1 & 1 & 0 & 1 & 0 & -1 & 0 & 0 & 0 & 1 & 0 & -1 & 0 & 0 & 0 & 0 & 0 & -1 & 0 & 0 & 0 & 0 & 0 & 1 \\
		0 & 0 & 0 & 1 & 0 & -1 & -1 & 1 & 0 & 1 & 0 & -1 & 0 & 0 & 0 & 0 & -1 & 0 & 0 & 0 & 0 & 0 & 1 & 0 \\
		0 & 1 & -1 & 1 & -1 & 0 & 0 & 0 & 0 & 0 & 0 & -1 & 0 & 0 & 0 & 1 & 0 & -1 & 0 & 0 & 0 & 1 & 0 & 0 \\
		0 & 1 & 0 & 0 & -1 & 0 & 0 & 0 & 0 & 0 & -1 & 0 & -1 & 1 & 0 & 1 & 0 & -1 & 0 & 0 & 1 & 0 & 0 & 0 \\
		0 & 0 & 0 & 0 & 0 & -1 & 0 & 1 & -1 & 1 & -1 & 0 & 0 & 1 & 0 & 0 & -1 & 0 & 0 & 1 & 0 & 0 & 0 & 0 \\
		0 & 0 & 0 & 0 & -1 & 0 & 0 & 1 & 0 & 0 & -1 & 0 & 0 & 1 & -1 & 1 & -1 & 0 & 1 & 0 & 0 & 0 & 0 & 0 \\
		1 & 0 & 0 & 0 & 0 & 0 & 0 & 0 & 0 & 0 & 0 & 0 & 0 & 0 & 0 & 0 & 0 & 0 & 0 & 0 & 0 & 0 & 0 & 1 \\
		0 & 0 & 0 & 0 & 0 & 0 & 1 & 0 & 0 & 0 & 0 & 0 & 0 & 0 & 0 & 0 & 0 & 0 & 0 & 0 & 0 & 0 & 1 & 0 \\
		0 & 0 & 1 & 0 & 0 & 0 & 0 & 0 & 0 & 0 & 0 & 0 & 0 & 0 & 0 & 0 & 0 & 0 & 0 & 0 & 0 & 1 & 0 & 0 \\
		1 & -\frac{1}{3} & \frac{1}{3} & \frac{5}{3} & \frac{4}{3} & 0 & 1 & 0 & 0 & 0 & 0 & 0 & 0 & 0 & 0 & 0 & 0 & 0 & 0 & 0 & 1 & 0 & 0 & 0 \\
		-2 & -\frac{1}{3} & -\frac{2}{3} & -\frac{4}{3} & -\frac{5}{3} & 0 & -1 & 0 & 0 & 0 & 0 & 0 & 0 & 0 & 0 & 0 & 0 & 0 & 0 & 1 & 0 & 0 & 0 & 0 \\
		0 & \frac{2}{3} & -\frac{2}{3} & -\frac{1}{3} & \frac{1}{3} & 0 & -1 & 0 & 0 & 0 & 0 & 0 & 0 & 0 & 0 & 0 & 0 & 0 & 1 & 0 & 0 & 0 & 0 & 0 \\
		0 & 1 & 0 & 0 & 0 & 0 & 0 & 0 & 0 & 0 & 0 & 0 & 0 & 0 & 0 & 0 & 0 & 1 & 0 & 0 & 0 & 0 & 0 & 0 \\
		2 & 0 & 1 & 2 & 1 & 0 & 1 & 0 & 0 & 0 & 0 & 0 & 0 & 0 & 0 & 0 & 1 & 0 & 0 & 0 & 0 & 0 & 0 & 0 \\
		0 & 0 & 0 & 0 & 1 & 0 & 0 & 0 & 0 & 0 & 0 & 0 & 0 & 0 & 0 & 1 & 0 & 0 & 0 & 0 & 0 & 0 & 0 & 0 \\
		0 & -\frac{2}{3} & \frac{2}{3} & \frac{1}{3} & -\frac{1}{3} & 0 & 1 & 0 & 0 & 0 & 0 & 0 & 0 & 0 & 1 & 0 & 0 & 0 & 0 & 0 & 0 & 0 & 0 & 0 \\
		-1 & -\frac{2}{3} & -\frac{4}{3} & -\frac{2}{3} & -\frac{1}{3} & 0 & -1 & 0 & 0 & 0 & 0 & 0 & 0 & 1 & 0 & 0 & 0 & 0 & 0 & 0 & 0 & 0 & 0 & 0 \\
		-1 & \frac{1}{3} & -\frac{1}{3} & -\frac{5}{3} & -\frac{4}{3} & 0 & -1 & 0 & 0 & 0 & 0 & 0 & 1 & 0 & 0 & 0 & 0 & 0 & 0 & 0 & 0 & 0 & 0 & 0 \\
		0 & 0 & 0 & 1 & 0 & 0 & 0 & 0 & 0 & 0 & 0 & 1 & 0 & 0 & 0 & 0 & 0 & 0 & 0 & 0 & 0 & 0 & 0 & 0 \\
		1 & \frac{2}{3} & \frac{4}{3} & \frac{2}{3} & \frac{1}{3} & 0 & 1 & 0 & 0 & 0 & 1 & 0 & 0 & 0 & 0 & 0 & 0 & 0 & 0 & 0 & 0 & 0 & 0 & 0 \\
		-1 & -1 & -1 & -1 & -1 & 0 & 0 & 0 & 0 & 1 & 0 & 0 & 0 & 0 & 0 & 0 & 0 & 0 & 0 & 0 & 0 & 0 & 0 & 0 \\
		2 & \frac{1}{3} & \frac{2}{3} & \frac{4}{3} & \frac{5}{3} & 0 & 1 & 0 & 1 & 0 & 0 & 0 & 0 & 0 & 0 & 0 & 0 & 0 & 0 & 0 & 0 & 0 & 0 & 0 \\
		-2 & 0 & -1 & -2 & -1 & 0 & -1 & 1 & 0 & 0 & 0 & 0 & 0 & 0 & 0 & 0 & 0 & 0 & 0 & 0 & 0 & 0 & 0 & 0 \\
		1 & 1 & 1 & 1 & 1 & 1 & 0 & 0 & 0 & 0 & 0 & 0 & 0 & 0 & 0 & 0 & 0 & 0 & 0 & 0 & 0 & 0 & 0 & 0 \\
		\end{array}
		\right) , \quad D= \D{6} \, ,
\end{split}
\end{align}
so that only six independent colour factors are present, given by
\begin{eqnarray}
(YC)_1 &=&  i \fahj \fbgh \fcdg \tj 1 \ta 2 \tb 3 \tc 4 \td 5 \, , \nonumber \\ \nonumber \\
(YC)_2&=& -i f^{agk} \fcdg f^{kbm} \tm 1 \ta 2 \tb 3 \tc 4 \td 5 \, ,\nonumber \\ \nonumber \\
(YC)_3&=&- i \fahj \fbdg f^{gch} \tj 1 \ta 2 \tb 3 \tc 4 \td 5 \, ,\nonumber \\ \nonumber \\
(YC)_4&=& -i \fahj \fbdg f^{gch} \tj 1 \ta 2 \tb 3 \tc 4 \td 5  \nonumber \\ && 
+i f^{ack} \fbdg f^{gkm} \tm 1 \ta 2 \tb 3 \tc 4 \td 5 \, ,\nonumber \\ \nonumber \\
(YC)_5&=& i \fadh f^{bjk} f^{chj} \tkk 1 \ta 2 \tb 3 \tc 4 \td 5 \, ,\nonumber \\ \nonumber \\
(YC)_6&=& i \fadh f^{bhl} f^{clm} \tm 1 \ta 2 \tb 3 \tc 4 \td 5 \, .
\end{eqnarray}  


\end{itemize}


\noindent This completes our listing of all Cwebs with a perturbative  expansion 
starting at  ${\cal O} (g^8)$, and connecting four and five Wilson lines.

\bibliographystyle{JHEP}
\bibliography{mybib}


\end{document}